\documentclass[sn-mathphys-num]{sn-jnl}
\usepackage{graphicx}%
\usepackage{epsfig}
\usepackage{amsmath,amssymb,amsfonts}%
\usepackage{amsthm}%
\usepackage{mathrsfs}%
\usepackage[title]{appendix}%
\usepackage{xcolor}%
\usepackage{textcomp}%
\usepackage{manyfoot}%
\usepackage{booktabs}%
\usepackage{algorithm}%
\usepackage{algorithmicx}%
\usepackage{algpseudocode}%
\usepackage{listings}%

\begin{document}

\title{From Low- to High-Frequency QPOs around the Non-Rotating Hairy Horndeski Black Hole: Microquasar GRS 1915+105}

\author{Orhan Donmez}\email{orhan.donmez@aum.edu.kw}
\affil{College of Engineering and Technology, American University
  of the Middle East, Egaila 54200, Kuwait.}

\abstract{
  Research on the Horndeski black hole, associated with the scalar hairy parameter, offers insights into enigmatic cosmic phenomena such as dark matter. Additionally, the numerical study of the GRS 1915+105 source, which exhibits continuous variability in X-ray observations, along with its physical properties and mechanisms behind Quasi-periodic oscillations (QPOs) frequencies, can contribute to observational studies. Motivated by this, we examine the variations in physical mechanisms around the non-rotating Horndeski black hole with Bondi-Hoyle-Lyttleton (BHL) accretion related to the scalar hair parameter and the resulting QPO frequencies. Numerical simulations have shown the formation of a shock cone around the black hole. With a decrease in the scalar hair parameter, the shock cone  opening angle narrows due to the influence of the scalar field potential, and the stagnation point within the cone moves closer to the black hole horizon. With the changing scalar hair parameter, the simultaneous formation of the shock cone and bow shock is observed. Due to the intense increase in scalar potential, both the shock cone and bow shock disappeared, and a cavity surrounding the black hole forms in the area where the shock cone was. Additionally, QPO oscillations induced by the physical mechanisms observed in relation to the hair parameter are revealed through numerical simulations. A broad range of QPO frequencies is observed, from low to high frequencies, with resonance states like 3:2 occurring. The QPO frequencies determined numerically are compared with the observational results of the GRS 1915+105 source, demonstrating a match between the observations and numerical findings. From this, it is concluded that the shock cone, bow shock, and cavity are suitable physical mechanisms for generating QPOs for the GRS 1915+105 source. Lastly, we define the potential range of the spin parameter for the GRS 1915+105 source based on the agreement between observational and numerical results. It has also been found that for most of the QPOs obtained from numerical calculations to be consistent with observations, $h/M$ should be greater than $-0.5$.
}


\maketitle

%%%%%%%%%%%%%%%%%%%%%%%%%%%%%%%%%%%%%%%%%%%%%%%%%%%%%%%%%%%%%%%%%%%%%%%
%%%%%%%%%%%%%%%%%%%%%%%%%%%%%%%%%%%%%%%%%%%%%%%%%%%%%%%%%%%%%%%%%%%%%%%
%%%%%%%%%%%%%%%%%%%%%%%%%%%%%%%%%%%%%%%%%%%%%%%%%%%%%%%%%%%%%%%%%%%%%%%
%%%%%%%%%%%%%%%%%%%%%%%%%%%%%%%%%%%%%%%%%%%%%%%%%%%%%%%%%%%%%%%%%%%%%%%
%%%%%%%%%%%%%%%%%%%%%%%%%%%%%%%%%%%%%%%%%%%%%%%%%%%%%%%%%%%%%%%%%%%%%%%
%%%%%%%%%%%%%%%%%%%%%%%%%%%%%%%%%%%%%%%%%%%%%%%%%%%%%%%%%%%%%%%%%%%%%%%

\section{Introduction}
\label{Introduction}

Quasi-periodic oscillations (QPOs) describe the rapid oscillations in the $X-$ray data obtained from various astrophysical systems such as Active Galactic Nuclei (AGN), low-mass and high-mass $X-$ray binary systems, and ultraluminous $X-$ray systems \cite{Pasham2018ALQ, Smith_2021, Singh2022MNRAS}. The low-mass $X-$ray binary (LMXB) and high-mass $X-$ray binary (HMXB) systems generally harbor black holes at their centers and undergo continuous mass accretion from the companion star to the black hole \cite{charles2003optical, Walter_2015}. It is believed that these oscillations arise from the interaction of the matter accreting towards the black hole. As a result of this interaction, QPOs with oscillation frequencies ranging from a few milliHertz to a few hundred Hz occur \cite{Fragile_2016}. In AGNs, QPOs are revealed through different observations in a wide spectral range spanning from radio to gamma-ray regions of the electromagnetic spectrum \cite{Ackermann2015ApJ, Zhang_2021}. However, this situation is observed in systems containing stellar-mass black holes, seen in X-ray spectral analyses.

The frequency range of low-frequency QPOs (LFQPOs) observed from different sources varies from a few milliHertz to several tens of Hertz. Understanding these frequencies may be possible through comprehending the physical structure formed by the matter accreted around the black holes, thereby enabling an understanding of physical properties such as mass and angular momentum of the black holes at the centers of the observed sources. Results from XTE J1550-564 and other sources indicate three types of LFQPOs, namely $A$, $B$, and $C$ \cite{Homan2001ApJS, belloni2011black, Belloni2024MNRAS}. Type-A QPOs, observed infrequently in the black hole binary systems, appear in approximately 10 sources, typically during the transition from the hard to the soft state. Type-B QPOs, observed in a few black hole binary systems, are more associated with the physical mechanisms of astrophysical jets. Type-C QPOs, the most common in black hole binary systems, are generally observed in X-ray spectral analyses \cite{Motta_2012, Motta2016AN, Liu_2021}. GRS 1915+105, an exemplary source of Type-C QPOs, has been extensively observed and studied both theoretically and numerically to understand the physical properties of the black hole and the mechanisms occurring around it \cite{ Belloni2024MNRAS, Belloni2013MNRAS, Misra2020ApJ, wang20242022, Chauhan2024MNRAS}. It is thought that QPOs occur due to dynamic changes in the matter flow near the black hole horizon. Studying this source is crucial for uncovering the interactions between black hole matter, accretion mechanisms, the effects of strong gravity, and the impacts of modified gravities.

High-frequency QPOs (HFQPOs) occur due to rapid luminosity changes as matter falls into the black hole, with oscillation frequencies ranging from the low twenties to several hundred Hz, often displaying stable ratios, such as $3:2$ \cite{Pasham_2015, Varniere_2018}. The physical mechanisms behind these QPOs are not fully understood, yet they are known to arise from interactions near the black hole horizon within a strong gravitational field. HFQPOs have been observed in various sources with central black holes, including GRO J1655-40, GRS 1915+105, XTE J1550-564, H1743-322, 4U 1630-47, XTE J1650-500, XTE J1859+226, IGR J17091-3624, GX 339-4, and XTE J1752-223 \cite{wang20242022, BelloniMNRAS2012, Strohmayer2001ApJ, Remillard2006ARA&A, Varniere_2018, Majumder_2022}. Notably, GRS 1915+105 has exhibited numerous HFQPOs, ranging from $63$ to $71$ Hz \cite{BelloniMNRAS2012, Sreehari_2020}. Power spectrum analyses of these sources have shown frequencies varying from $27$ to $450$ Hz \cite{Varniere_2018, Ingram2019}. HFQPOs occur less frequently than LFQPOs, contributing to their intrigue.

To understand the physical mechanisms behind the observed HFQPOs and LFQPOs, it is necessary to comprehend the outcomes of the interaction between matter and the black hole in a strong gravitational field. Therefore, studying the different mechanisms of matter falling towards the black hole is crucial in X-ray and gravitational wave astronomy \cite{DonmezMPLA24, DonmezEPJC24}. One such mechanism is the Bondi-Hoyle-Lyttleton (BHL) accretion, which predicts the fall of matter onto the black hole with a homogeneous distribution and at a constant asymptotic speed \cite{Bondi1, Bondi1952MNRAS, Edgar1}. As a result, the process of matter entering the gravitational field of the black hole from upstream and falling towards it, forming distinct structures on the downstream side, has been observed \cite{Donmez6, Zanotti1, Penner2, Donmez5, Koyuncu1, DonmezUniverse2022, CruzOsorio2023JCAP, Donmez2024arXiv240216707D}. The BHL mechanism has been widely applied to a variety of astrophysical systems, with hydrodynamic equations numerically solved in both Newtonian \cite{Ruffert2, Foglizzo1, Ohsugi1} and relativistic regimes \cite{CruzOsorio2023JCAP, Donmez2024arXiv240216707D, Wenrui1, Donmez20, Donmez_2024ES} to uncover the causes of electromagnetic emissions in astrophysical systems and to investigate potential QPO frequencies.

Theoretical studies and observational results from various sources indicate that HFQPOs and LFQPOs originate from physical mechanisms occurring in strong gravitational fields \cite{Smith_2021, Ingram2019, Thomson2014PhDT}. These QPOs result from sudden transitions from a hard state to a soft state near the event horizon of the black hole \cite{Motta2016AN}. To explain these observational results, alternative relativistic theories must be considered.

Although general relativity is a pivotal theory explaining gravitational interactions, it fails to provide a detailed explanation for certain phenomena, such as the early development of the universe, its accelerating expansion, and the existence of dark matter and dark energy. However, the scalar field is thought to offer solutions to these problems \cite{Comelli_2003, BERTACCA_2007, Chung_2007, de_la_Macorra_2008, Errehymy2024EPJC, Maurya2024ptep}. Horndeski gravity depends on the hair parameter, which defines a scalar field around the black hole. Consequently, the spacetime of the  Horndeski black hole is modified due to the interaction between the scalar field and gravity \cite{BERTACCA_2007, Arbuzov_2022}, significantly changing the nature, formation mechanisms, and resulting physical mechanisms of matter falling towards the black hole \cite{martens2020dark}. The potential created by the scalar field around the black hole could represent the field of dark matter. In some accepted theories, the scalar field is the predominant notion for dark matter, instead of particles like WIMPs or axions \cite{Hu_2000, Hui_2017}. In cosmological simulations, dark matter has been defined as a scalar field, and its effects on galaxy formation and dynamics have been studied. Therefore, some observational phenomena have been attempted to be explained by the results obtained from these models, positioning the scalar field of dark matter as an important concept in unraveling the mysteries of the universe \cite{Wang_2023}.

In this study, we model the physical mechanism created by matter falling towards the non-rotating Horndeski black hole through the BHL mechanism, particularly in the strong gravitational field around the black hole, by numerically solving the general relativistic hydrodynamic (GRH) equations. We simulate the matter falling towards the black hole from the upstream region in two dimensions on the equatorial plane, resulting in the formation of a shock mechanism on the downstream side. 
The main purpose of this paper is to reveal the effects of the interaction of hydrodynamic parameters of the matter with black holes in the strong gravitational fields on the dynamics of the resulting shock cone and QPOs, and to identify the modes that produce continuous QPO frequencies through long-term simulations. Therefore, the effects of magnetic fields and viscosity have been neglected.
We reveal how the dynamic structure of the physical mechanism around the non-rotating Horndeski black hole changes with respect to the scalar field parameter of the Horndeski black hole, demonstrate how the physical mechanism undergoes significant changes or completely vanishes in a strong scalar field, and discuss the new physical mechanisms formed. We also explore how the physical mechanism around the black hole varies depending on the asymptotic speed. Finally, we examine the QPOs generated by the physical mechanism depending on the hair parameter and asymptotic speed, and their consistency with observational results from some astrophysical sources.

In this paper, we numerically reveal the formation of physical structures such as shock cones, bow shocks, and cavities around the non-rotating hairy Horndeski black hole with BHL accretion, dependent on the scalar hair parameter, and identify both low- and high-frequency QPOs. The paper is organized as follows: In Section \ref{Equations}, the GRH equations and the spacetime metric of the non-rotating hairy Horndeski black hole are presented, summarizing the black hole's physical properties. The initial and boundary conditions necessary for the formation of physical mechanisms around the black hole through the BHL mechanism are given in Section \ref{InitBoun}. In Section \ref{Result1}, the characteristics of the shock cone, bow shock, and cavity formed as a result of numerical calculations, and their dependence on the scalar hair parameter, are discussed in detail. Additionally, the effect of the asymptotic velocity on such formations is briefly summarized. The physical mechanisms identified in Section \ref{Result1}, and the resulting QPO frequencies for each model, are discussed separately, calculating the genuine modes and their linear and nonlinear coupling frequencies in Section \ref{QPOs1}. In Section \ref{QPOs4}, the numerically calculated QPO frequencies are compared in detail with the observed LFQPOs and HFQPOs for the source GRS 1915+105, discussing the similarities and differences. Based on the agreement of the numerically derived QPOs with observations, Section \ref{physical_mech} proposes physical mechanisms that could generate the QPOs in the source GRS 1915+105. We conclude our results in Section \ref{conc}.

%%%%%%%%%%%%%%%%%%%%%%%%%%%%%%%%%%%%%%%%%%%%%%%%%%%%%%%%%%%%%%%%%%%%%%%
%%%%%%%%%%%%%%%%%%%%%%%%%%%%%%%%%%%%%%%%%%%%%%%%%%%%%%%%%%%%%%%%%%%%%%%

\section{Equations}
\label{Equations}

\subsection{General Relativistic Hydrodynamic Equations}
\label{GRHE1}

The General Relativistic Hydrodynamics (GRH) equations are instrumental in understanding phenomena such as mass accretion around compact objects like black holes. They reveal the disk structures around compact objects in X-ray binaries and help understand the nature of emitted X-rays. Unlike Newtonian hydrodynamics, which overlooks strong gravitational interactions near the  black hole horizon, the GRH equations elucidate the physical mechanisms in regions close to the black hole horizon where such interactions are significant. With broad applications in astrophysics and cosmology, these equations are crucial for explaining observational results. Therefore, for numerical solutions, the GRH equations are transformed from covariant to conservative form using the 3+1 formalism \cite{Font2000LRR, Donmez1, Donmez2}. Ignoring the effects of magnetic fields and viscosity, they describe the behavior of a perfect fluid stress-energy tensor on the equatorial plane. GRH equations are,

\begin{eqnarray}
  \frac{\partial U}{\partial t} + \frac{\partial F^r}{\partial r} + \frac{\partial F^{\phi}}{\partial \phi}
  = S,
\label{GRHE2}
\end{eqnarray}

\noindent
where $U$ denotes the conserved variables, $F^r$ the fluxes in the $r$ direction, $F^{\phi}$ the fluxes in the $\phi$ direction, and $S$ the sources for the conserved variables. These are determined by the primitive variables, the 3-metric defining the curvature geometry around the black hole, and other fluid variables to be defined later.  These variables include,

\begin{eqnarray}
  U =
  \begin{pmatrix}
    D \\
    S_j \\
    \tau
  \end{pmatrix}
  =
  \begin{pmatrix}
    \sqrt{\gamma}W\rho \\
    \sqrt{\gamma}\eta\rho W^2 v_j\\
    \sqrt{\gamma}(\eta\rho W^2 - P - W \rho)
    \end{pmatrix},
\label{GREq5}
\end{eqnarray}

\noindent and fluxes are

\begin{eqnarray}
  \vec{F}^i =
  \begin{pmatrix}
    \alpha\left(v^i - \frac{1}{\alpha\beta^i}\right)D \\
    \alpha\left(\left(v^i - \frac{1}{\alpha\beta^i}\right)S_j + \sqrt{\gamma}P\delta^i_j\right)\\
    \alpha\left(\left(v^i - \frac{1}{\alpha\beta^i}\right)\tau  + \sqrt{\gamma}P v^i\right)
    \end{pmatrix},
\label{GREq6}
\end{eqnarray}

\noindent and the  source is 

\begin{eqnarray}
  \vec{S} =
  \begin{pmatrix}
    0 \\
    \alpha\sqrt{\gamma}T^{ab}g_{bc}\Gamma^c_{aj} \\
    \alpha\sqrt{\gamma}\left(T^{a0}\partial_{a}\alpha - \alpha T^{ab}\Gamma^0_{ab}\right)
   \end{pmatrix}.
\label{GREq7}
\end{eqnarray}

\noindent
Here, $\Gamma^c_{ab}$ represents the Christoffel symbol, $v^i = u^i/W + \beta^i$ denotes the three-velocity of the fluid, $\eta = 1 + \epsilon + P/\rho$ is the enthalpy, and $W = (1 - \gamma_{ij}v^i v^j)^{1/2}$ is the Lorentz factor. The variables $\rho$, $\epsilon$, $u^{a}$, $p$, $\alpha$, $\beta^i$, $\eta$, $\gamma_{ij}$, $g^{ab}$, and $\gamma$ represent the rest-mass density, internal energy, four-velocity of the fluid, fluid pressure, lapse function, shift vector, specific enthalpy, 3-metric, four-metric of the spacetime curvature, and determinant of the three-metric, respectively. The indices $i$, $j$, and $k$ range from 1 to 3, representing spatial dimensions, while $a$, $b$, and $c$ range from 0 to 3, representing spacetime dimensions.

%%%%%%%%%%%%%%%%%%%%%%%%%%%%%%%%%%%%%%%%%%%%%%%%%%%%%%%%%%%%%%%%%%%%%%%%%%%%%%%%%%%%%

\subsection{Non-Rotating Black Hole Space-Time Metric in Horndeski gravity}
\label{Horndeski_gravity}

The numerical solutions of the GRH equations are crucial for understanding the physical phenomena resulting from the interaction between a black hole and a disk in a strong gravitational field. Comparing these results with observational data helps elucidate the nature of certain astrophysical events. For example, we can reveal the properties and mechanisms of QPO oscillations in strong gravitational fields, thereby pinpointing the sources in the observational data. This process allows different gravitational theories to explain the reasons for previously undetermined X-ray oscillations. In this context, alongside the Schwarzschild, Kerr, EGB, and Hartle-Thorne black holes used in our previous studies \cite{Donmez6, Donmez5, Donmez2024arXiv240216707D, Donmez20, Donmez_EGB_Rot, Donmez2023arXiv231013847D}, we have investigated scenarios around the non-rotating Horndeski black hole. Horndeski gravity, defined in terms of the four-metric determinant, the Ricci scalar \cite{Horndeski1974IJTP, Heydari-Fard2023arXiv}, and the scalar field determinant function, yields a static spacetime solution with spherical symmetry upon applying the scalar field canonical operation and a finite energy condition \cite{Badia2017EPJC, Esteban2021arXiv}.

\begin{eqnarray}
  ds^2 = -f(r)dt^2 + \frac{dr^2}{f(r)}+r^2\left(d\theta^2 + \rm{Sin}^2\theta d\phi^2\right),
\label{Horn1}
\end{eqnarray}

\noindent
where, $f(r) = 1 - \frac{2M}{r} + \frac{h}{r} \ln\left(\frac{2}{2M}\right)$, where $h$ is the scalar hair parameter with the dimension of length, and $M$ is the mass of the black hole. Horndeski gravity is characterized by second-order field equations, derived from the most general scalar-tensor theory. In this framework, the hair parameter $h/M$ represents a scalar field term that modifies spacetime. This modification influences the scalar potential in spacetime, which in turn alters the physical processes occurring around the black hole.

\begin{figure*}
  \vspace{1cm}
  \center
     \psfig{file=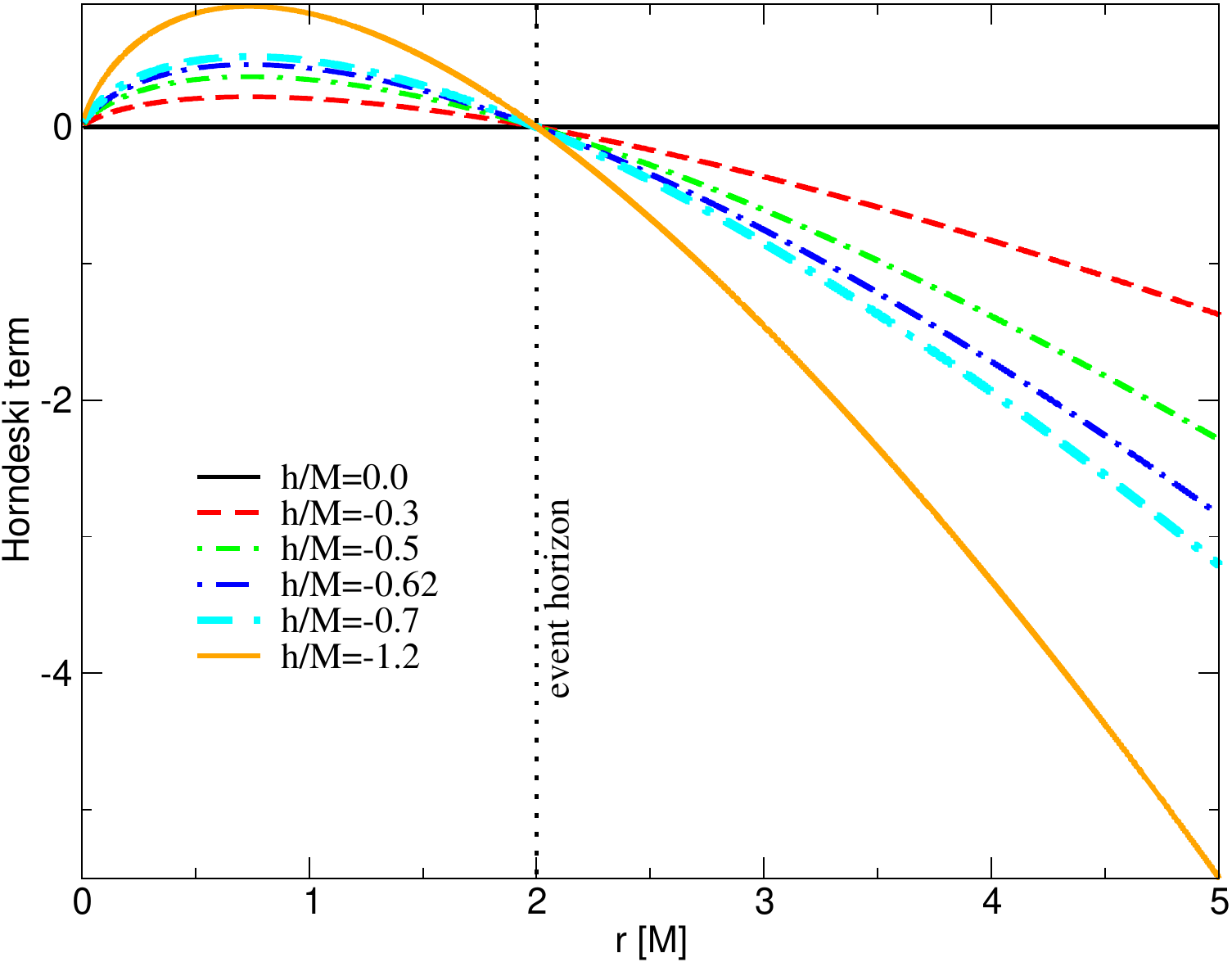,width=12.0cm}
     \caption{The variation of the Horndeski term ($hr\ln\left(\frac{r}{2M}\right)$) with respect to $r$ is illustrated. In various Horndeski hairy parameter scenarios, we observe that the event horizon consistently forms at $R_g=2M$. Simultaneously, we gain insights into how the Horndeski modification, namely the scalar field, affects spacetime. The impact of the scalar field intensifies with increasing $r$. This suggests that in regions where the scalar field, matter pressure, or gravitational forces are weaker, the dynamic structure of the physical mechanisms would be significantly influenced.
}
\vspace{1cm}
\label{Horndeski_EH}
\end{figure*}

The equation defining the non-rotating black hole in Horndeski gravity, expressed in Boyer-Lindquist coordinates, is presented in Ref.\cite{Walia2022EPJC},

\begin{eqnarray}
  ds^2 = -\left(\frac{\bigtriangleup}{r^2}\right)dt^2 + \frac{r^2}{\bigtriangleup}dr^2
  +r^2  d\theta^2 + \rm{Sin}^2\theta r^2 d\phi^2,
\label{Horn2}
\end{eqnarray}

\noindent
where $\Delta = r^2 - 2Mr + hr \ln\left(\frac{r}{2M}\right)$. Here, the ratio $h/M$ lies within the interval $-2 \leq h/M \leq 0$. Horndeski gravity admits a spherically symmetric hairy black hole solution. As $h/M$ approaches $0$, the Horndeski metric, as given in Eq.\ref{Horn2}, converges to the Schwarzschild solution \cite{Kerr1963PhRvL}. The horizons of the Horndeski black hole are determined by setting $\Delta = 0$, revealing two distinct horizons. The event horizon consistently appears at the Schwarzschild radius $R_s = 2M$, as shown in Fig.\ref{Horndeski_EH}, while the Cauchy horizon changes with $h/M$, decreasing as $h/M$ increases \cite{Rayimbaev2023EPJC}. This aspect is overlooked in some calculations because it pertains to the singularity within the black hole event horizon.

Figure \ref{Horndeski_EH} also illustrates how the scalar field effect varies with $r$. An increase in the scalar field intensity is observed as $\lvert h \rvert$ increases. The negative value imparted to spacetime by the scalar hair parameter could signify the presence of exotic, non-baryonic matter or a non-standard form of energy, potentially related to dark matter or energy, or a repulsive gravitational field. This field significantly influences the physical mechanisms formed, affecting the flow rate and direction of matter falling towards the black hole due to BHL accretion. The scalar field induces changes in the total forces acting on the matter, such as gravitational and pressure forces, thus impacting the accretion process and the potential physical mechanisms. This effect becomes more pronounced with increasing $\lvert h \rvert$.

It is important to determine the lapse function $\alpha$ and the shift vectors $\beta_i$ for solving the GRH equations numerically. These variables are derived from the relationship between the four-metric $g_{ab}$ and the three-metric $\gamma_{ij}$, as detailed in Ref.\cite{Misner1977},

\begin{eqnarray}
  \left( {\begin{array}{cc}
   g_{tt} & g_{ti} \\
   g_{it} & \gamma_{ij} \\
  \end{array} } \right)
=  
  \left( {\begin{array}{cc}
   (\beta_k\beta^k - \alpha^2) & \beta_k \\
   \beta_i & \gamma_{ij} \\
  \end{array} } \right),
\label{Horn3}
\end{eqnarray}

\noindent
where the lapse function and the shift vectors  for Horndeski black hole are

\begin{eqnarray}
  \alpha = \sqrt{\frac{\bigtriangleup}{r^2}},
\label{Horn4}
\end{eqnarray}

\noindent
and  

\begin{eqnarray}
  \beta_r = 0, \;\;\;  \beta_{\theta}= 0, \;\;\;  \beta_{\phi}= 0.
\label{Horn5}
\end{eqnarray}

%%%%%%%%%%%%%%%%%%%%%%%%%%%%%%%%%%%%%%%%%%%%%%%%%%%%%%%%%%%%%%%%%%%%%%%

%%%%%%%%%%%%%%%%%%%%%%%%%%%%%%%%%%%%%%%%%%%%%%%%%%%%%%%%%%%%%%%%%%%%%%%
%%%%%%%%%%%%%%%%%%%%%%%%%%%%%%%%%%%%%%%%%%%%%%%%%%%%%%%%%%%%%%%%%%%%%%%
\section{ Numerical and Physical Setup}
\label{InitBoun}

Understanding the physical properties of matter-black hole interactions around the non-rotating Horndeski black hole, which depends on the scalar hair parameter, may help elucidate the characteristics of observed black holes of different masses and contribute to understanding the physical mechanisms behind QPOs in strong gravitational fields. Thus, we model the matter-black hole interactions using BHL accretion, which is influenced by the velocity at which matter approaches the black hole, the thermodynamic properties of the matter, and the strong gravitational force exerted by the black hole. BHL accretion enables the calculation of the rest-mass accretion rate around the black hole, aiding in understanding the dynamic structure of various physical mechanisms like shock waves, shock cones, and bow shocks. This dynamic structure helps to clarify observational results from systems with stellar and massive black holes at their centers.

In modeling shock cones and other potential mechanisms around the black hole via BHL accretion, the numerical domain is initially populated with negligible values. Matter is then introduced from the upstream region towards the non-rotating Horndeski black hole. The parameters for the matter falling towards the black hole include rest-mass density $\rho = 1$, sound speed $C_s/c = 0.1$, and adiabatic index $\Gamma = 4/3$. The gas pressure is calculated using the ideal gas equation of state $P = (\Gamma-1)\rho\epsilon$, ensuring consistency with the initial values in the numerical calculations. In addition to these parameters, the radial and azimuthal components of the velocity of matter falling towards the black hole are,

\begin{eqnarray}
  V^r = \sqrt{\gamma^{rr}}V_{\infty}\rm{cos}(\phi) \;\;\; and  \;\;\;  V^{\phi} = -\sqrt{\gamma^{\phi \phi}}V_{\infty}\rm{sin}(\phi).
\label{Asymptotic_veloc}
\end{eqnarray}

\noindent Here, $V_{\infty}$ represents the asymptotic velocity, the velocity of the gas as it moves through empty space. As seen in Table \ref{Inital_Con}, different values of this velocity are used in the simulations, as our previous work \cite{Donmez5, Donmez2024arXiv240216707D, Donmez20, Donmez_EGB_Rot} shows that the asymptotic velocity affects the mechanisms around the black holes. We examine the mechanisms due to Horndeski gravity interaction with matter and highlight the asymptotic velocity impact on these mechanisms.

In this article, we are investigating the physical mechanisms that can occur around the non-rotating Horndeski black hole in the equatorial plane. The horizon of the Horndeski black hole is at $\rm{R_s}=2M$. Therefore, we placed the inner boundary of the computational domain as close to the black hole horizon as possible, i.e., at $\rm{r_{\text{min}}}=2.3M$. This way, in the strong gravitational field, we will be able to reveal the QPO oscillations that can occur as a result of the interaction between matter and the black hole. Additionally, to prevent non-physical oscillations that could come from the outer boundary, we placed the outer boundary far enough from the strong gravitational field at $\rm{r_{\text{max}}}=100M$. The boundary is set for $\phi \in [0, \pi]$.

In the equatorial plane, for modeling the matter flow and physical mechanisms towards the black hole, we used a resolution of $1024$ in the radial direction and $512$ in the azimuthal direction. Test trials have shown that increasing the resolution does not change the physical mechanisms and QPO oscillations. To provide sufficient time for the matter to reach a steady-state and for the accurate determination of QPOs, we set the maximum simulation time to $\rm{t_{\text{max}}}=35000M$. Although most models reach the steady-state within $\rm{\tau_{\text{ss}}}\sim 3000M$, as indicated in Table \ref{Inital_Con}, we believe the additional time is ample for QPO formation.

Finally, when setting up the model, we aim to prevent matter and oscillations from propagating into the computational domain from the inner and outer boundaries, as such backflows can eliminate possible physical oscillations and alter the structure of the shock cone. Therefore, we implement an outflow boundary condition at the inner boundary, close to the black hole horizon, ensuring that matter reaching this boundary falls into the black hole. At the outer boundary, specifically in the upstream region where matter is continuously injected, we maintain the previously mentioned values; meanwhile, in the downstream region, we apply the outflow boundary condition to prevent the propagation of non-physical oscillations into the computational domain. For the azimuthal direction, we employ periodic boundary conditions.

\begin{table}
%\scriptsize
\footnotesize
\caption{The initial model adopted for the numerical simulations for the non-rotating black hole in Horndeski gravity, along with some outcomes from the numerical results, includes several key parameters. These are: $Model$, representing the name of the model; $h/M$, the scalar hair parameter in the Horndeski metric; $V_{\infty}/c$, the asymptotic velocity of gas injected from the outer boundary; $r_{stag}/M$, the position of the stagnation point; $\theta_{sh}/rad$, the shock cone opening angle; $\tau_{ss}$, the time required to reach the steady-state; and $PM$, the physical mechanism that has been formed.
}
 \label{Inital_Con}
%\begin{center}
%\vspace*{2cm}
  \begin{tabular}{ccccccc}
    \hline
    \hline

    $Model$               & $h/M$ & $V_{\infty}/c$ & $r_{stag}/M$ & $\theta_{sh}/rad$
    &$\tau_{ss}/M$ & $PM$ \\ 
    \hline
    $H02A$          & $0.0$    & $0.2$ & $26.9$ & $1.0524$ & $3500$ & $Shock \; Cone$\\
    $H02B$                         & $-0.2$ & $0.2$ & $20.4$ & $0.979$  & $3300$ & $Shock \; Cone$\\    
    $H02C$                        & $-0.3$ & $0.2$ & $17.7$ & $0.9295$ & $3200$ & $Shock \; Cone$\\
    $H02D$                         & $-0.4$ & $0.2$ & $15.84$ & $0.856$ & $17000$ & $Cone\; and\; bow\; shock$\\   
    $H02E$           & $-0.5$ & $0.2$ & $14.1$ & $0.807$  & $7800$ & $Shock \; Cone$\\
    $H02F$                         & $-0.62$& $0.2$ & $12.6$ & $0.708$  & $6000$ & $Shock \; Cone$\\    
    $H02G$                         & $-0.7$ & $0.2$ & $11.8$ & $0.635$  & $3100$ & $Shock \; Cone$\\
    $H02H$                         & $-0.9$ & $0.2$ & $10.12$ & $0.463$ & $2600$ & $Shock \; Cone$\\       
    $H02I$                         &$-1.14834$&$0.2$& $52.67$& $1.961$  & $1500$ & $Cavity$\\
    $H02L$                         &$-1.2$  &$0.2$  & $79.03$& $1.420$  & $1300$ & $Cavity$\\
    \hline
    $H04A$         & $0.0$ & $0.4$ & $12.03$ & $0.807$ & $850$ & $Shock \; Cone$\\
    $H04B$           & $-0.5$& $0.4$ & $9.16$  & $0.758$ & $1000$ & $Shock \; Cone$\\
    $H04C$                        & $-1.0$& $0.4$ & $7.26$ & $0.561$ & $14500$ & $ Cone\; and\; bow\; shock$\\
    \hline
    \hline
  \end{tabular}
%\end{center}
%  \tablenotetext{}{}
%\vskip -0.8truecm
\end{table}

The numerical code we employ here, we solve the GRH equations using the high-resolution shock capturing schemes. Detailed technical explanations of the code and initial test results are provided in Refs.\cite{Donmez1} and \cite{Donmez2}. Subsequently, using the same code, various physical phenomena resulting from matter-black hole interactions in different accretion scenarios around Schwarzschild and Kerr black holes have been numerically calculated \cite{Donmez6, Donmez5, Koyuncu1}. Additionally, instabilities resulting from perturbations of torus structures have been examined to determine whether QPO formation occurs. In recent years, the same code has been modified for alternative gravities to contribute to identifying instabilities, shock wave formation, and QPO behaviors \cite{Donmez2024arXiv240216707D, Donmez20, Donmez_EGB_Rot, Donmez2023arXiv231013847D}.

%%%%%%%%%%%%%%%%%%%%%%%%%%%%%%%%%%%%%%%%%%%%%%%%%%%%%%%%%%%%%%%%%%%%%%%%%%%%%%%%%%%%%%%%%%%%%%%%%%%
\section{Numerical Results }
\label{Result1}

The numerical simulation of BHL accretion explores how matter flows towards the black hole, forming a disk due to the strong gravitational field of the black hole, whether in an interstellar medium or the black hole binary system. We focus on a two-dimensional simulation in the equatorial plane, driven by the inflow speeds from the upstream region as specified in Eq.~\ref{Asymptotic_veloc}. This simulation elucidates the physical mechanisms and properties of accreted matter, facilitating the analysis of system stability and matter behavior at the steady-state around the black hole. Through these dynamics, we can probe the instabilities in the physical mechanisms and identify QPO frequencies via PSD analysis.

\begin{figure*}
  \vspace{1cm}
  \center
  \psfig{file=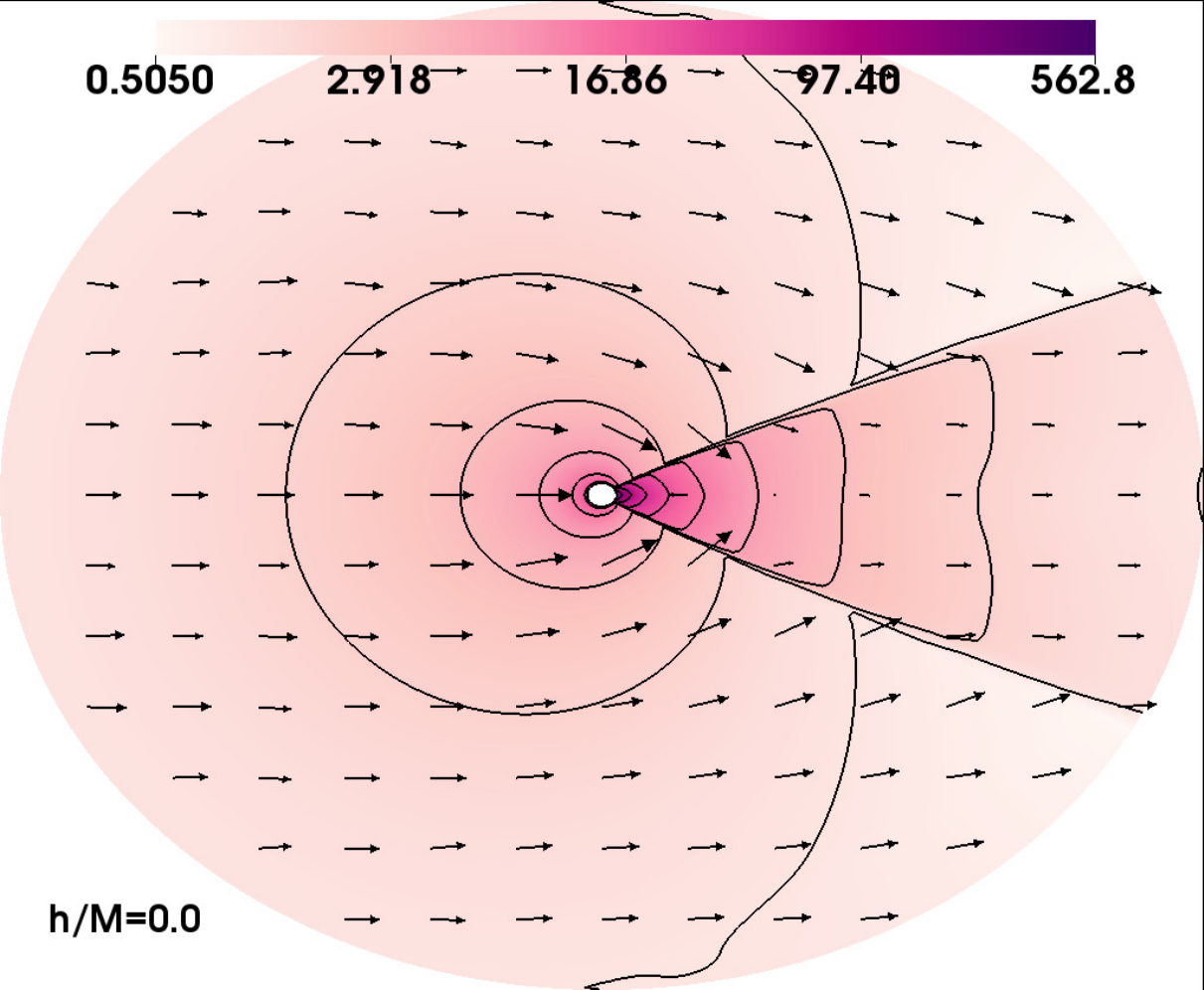,width=6.0cm}
     \psfig{file=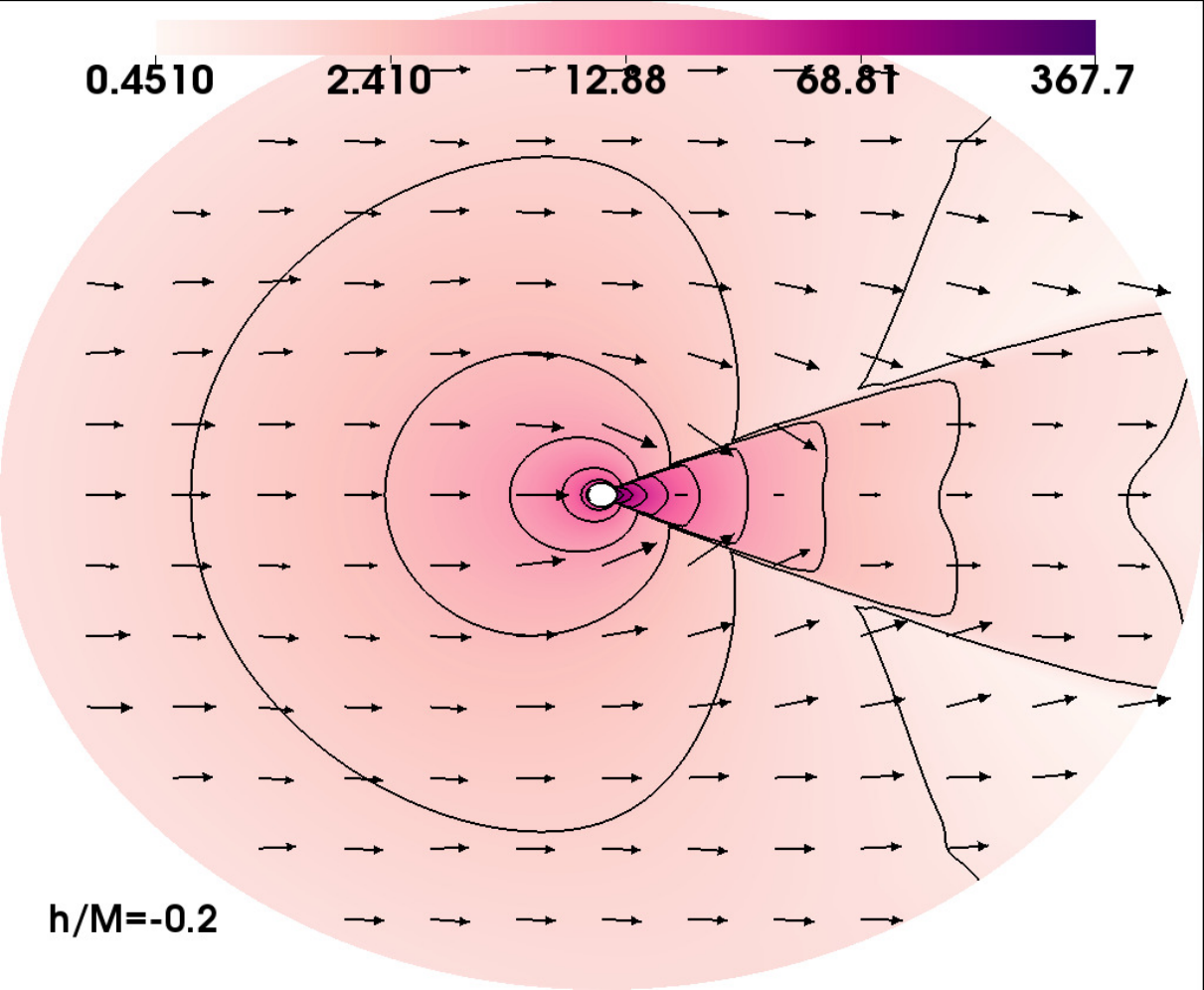,width=6.0cm}  
     \psfig{file=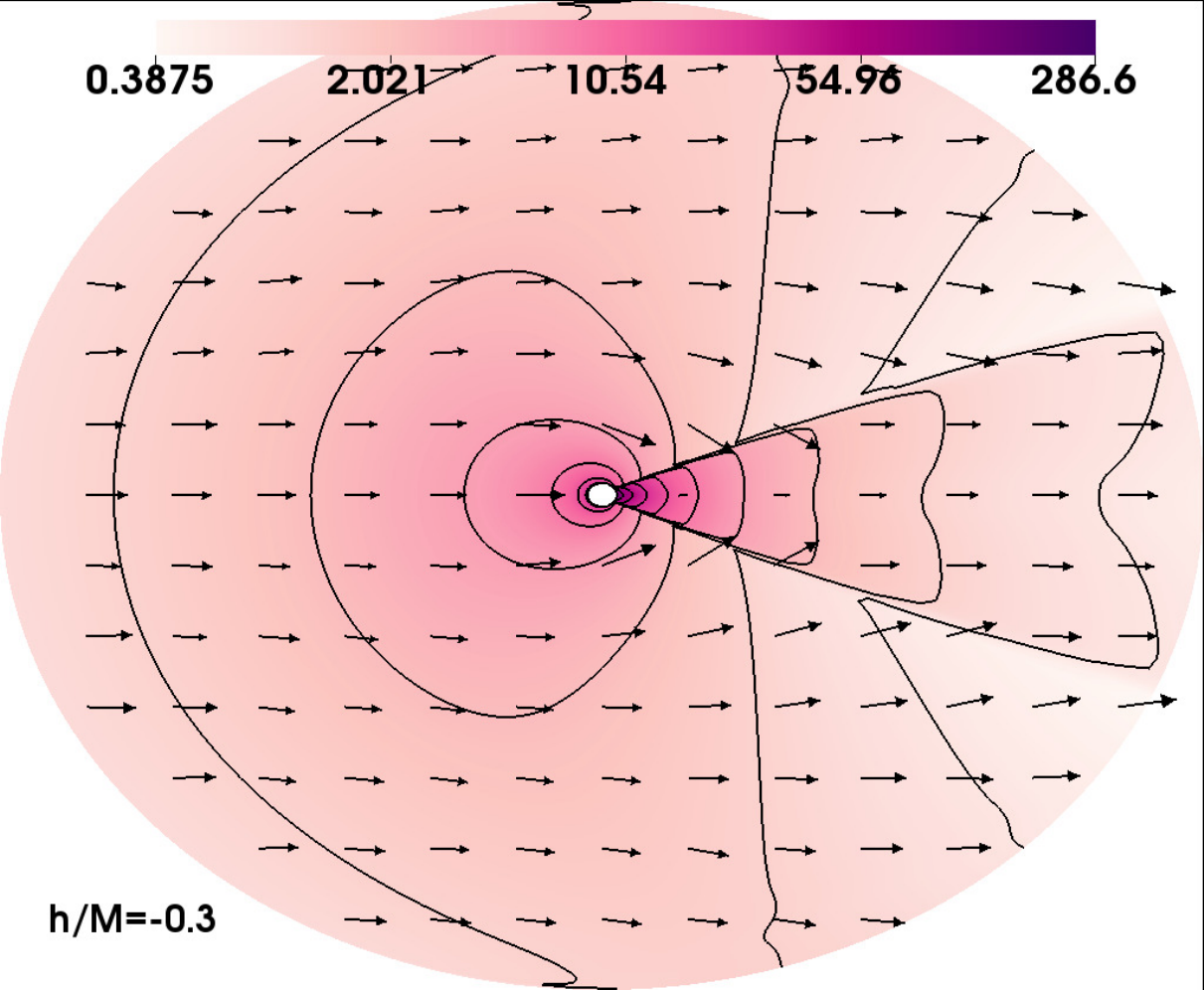,width=6.0cm}
     \psfig{file=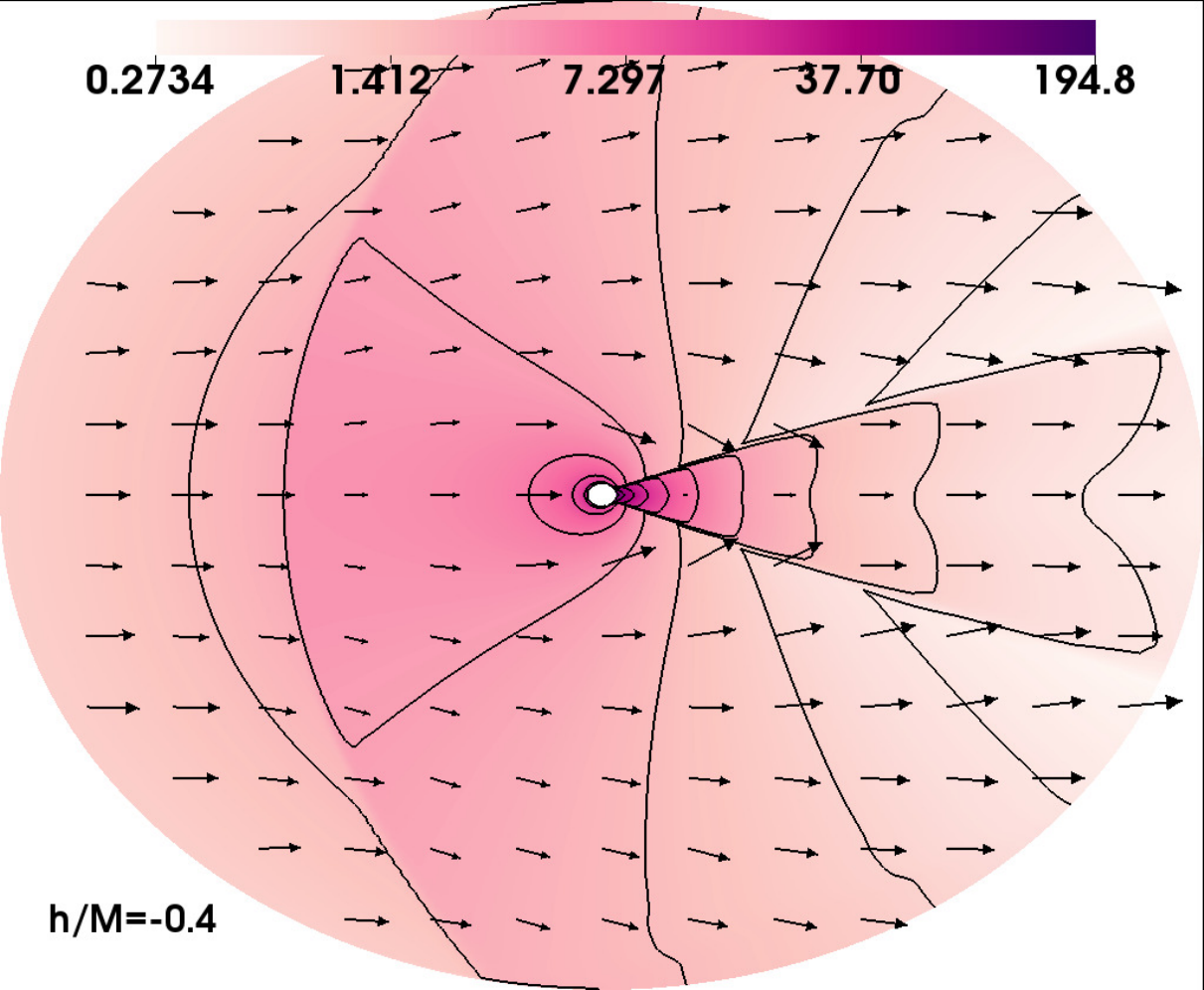,width=6.0cm}
     \psfig{file=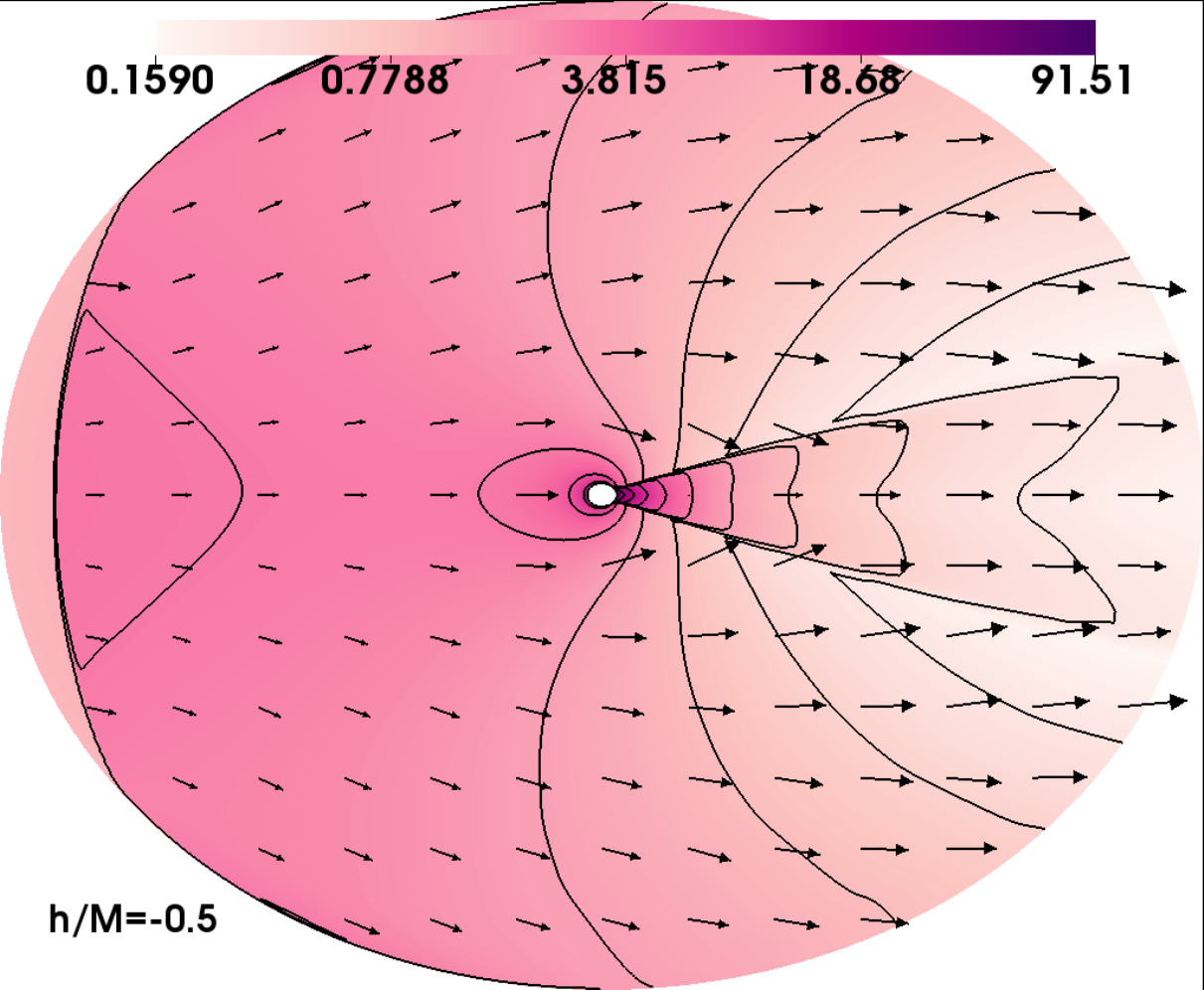,width=6.0cm}
     \psfig{file=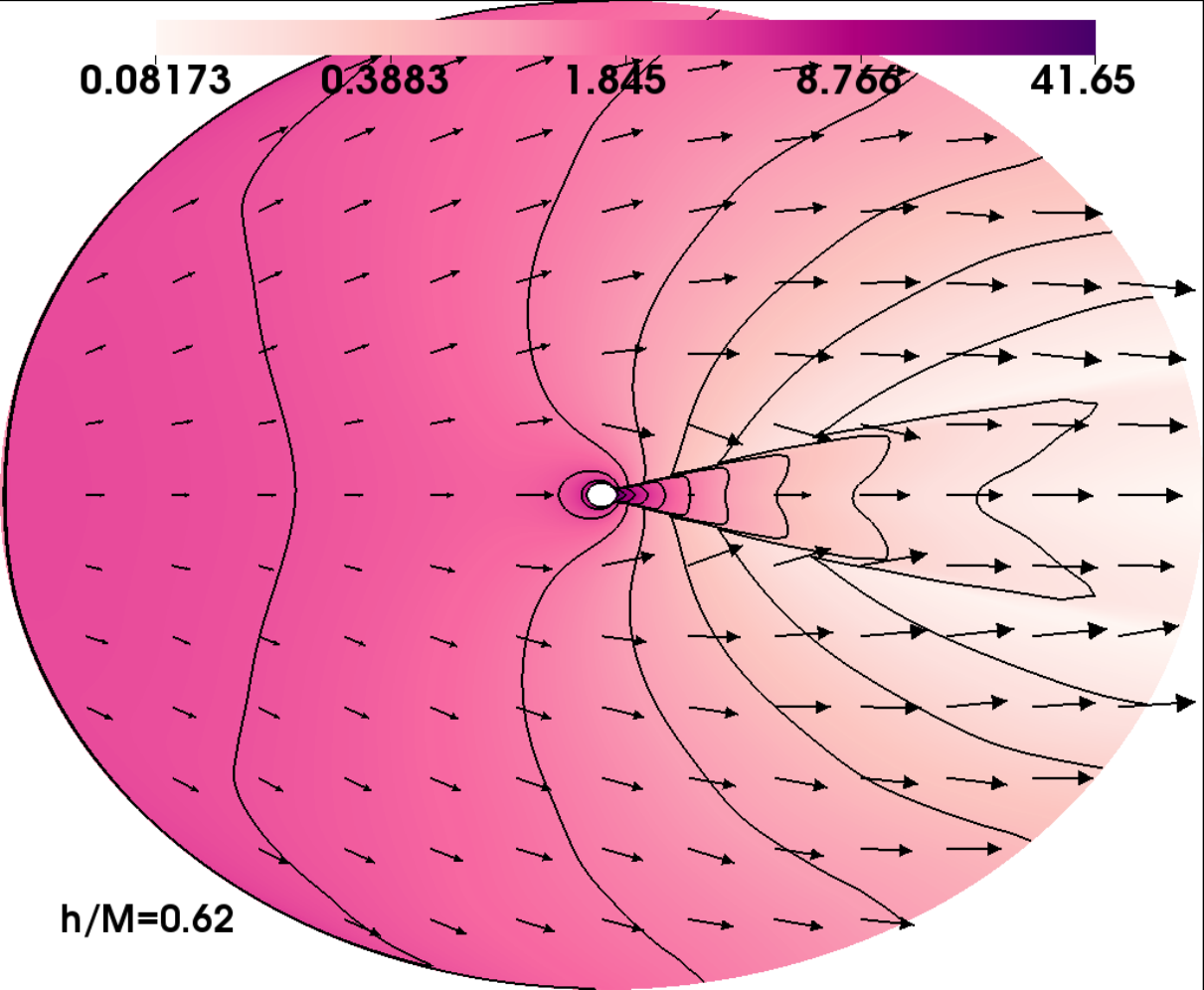,width=6.0cm}     
     \caption{
  On the equatorial plane, for $V_{\infty}/c=0.2$, the physical structure of the rest-mass density formed by BHL accretion is shown in both color and contour plots logarithmically. Additionally, to show the flow pattern and how the matter is accreted, the vector field of velocity is superimposed on these graphs. This illustrates how the physical mechanism around the non-rotating Horndeski black hole changes according to the scalar hair parameter. As the scalar field parameter decreases, significant changes are observed in the dynamic structure of the shock cone, as well as in the flow speed and direction of the matter. The domain is $[X_{\rm{min}},Y_{\rm{min}}]$ $\Rightarrow$ $[X_{\rm{max}},Y_{\rm{max}}] =$ $[-100M,-100M]$ $\Rightarrow$ $[100M,100M]$.
}
\vspace{1cm}
\label{colV02_1}
\end{figure*}

\begin{figure*}
  \vspace{1cm}
  \center
    \psfig{file=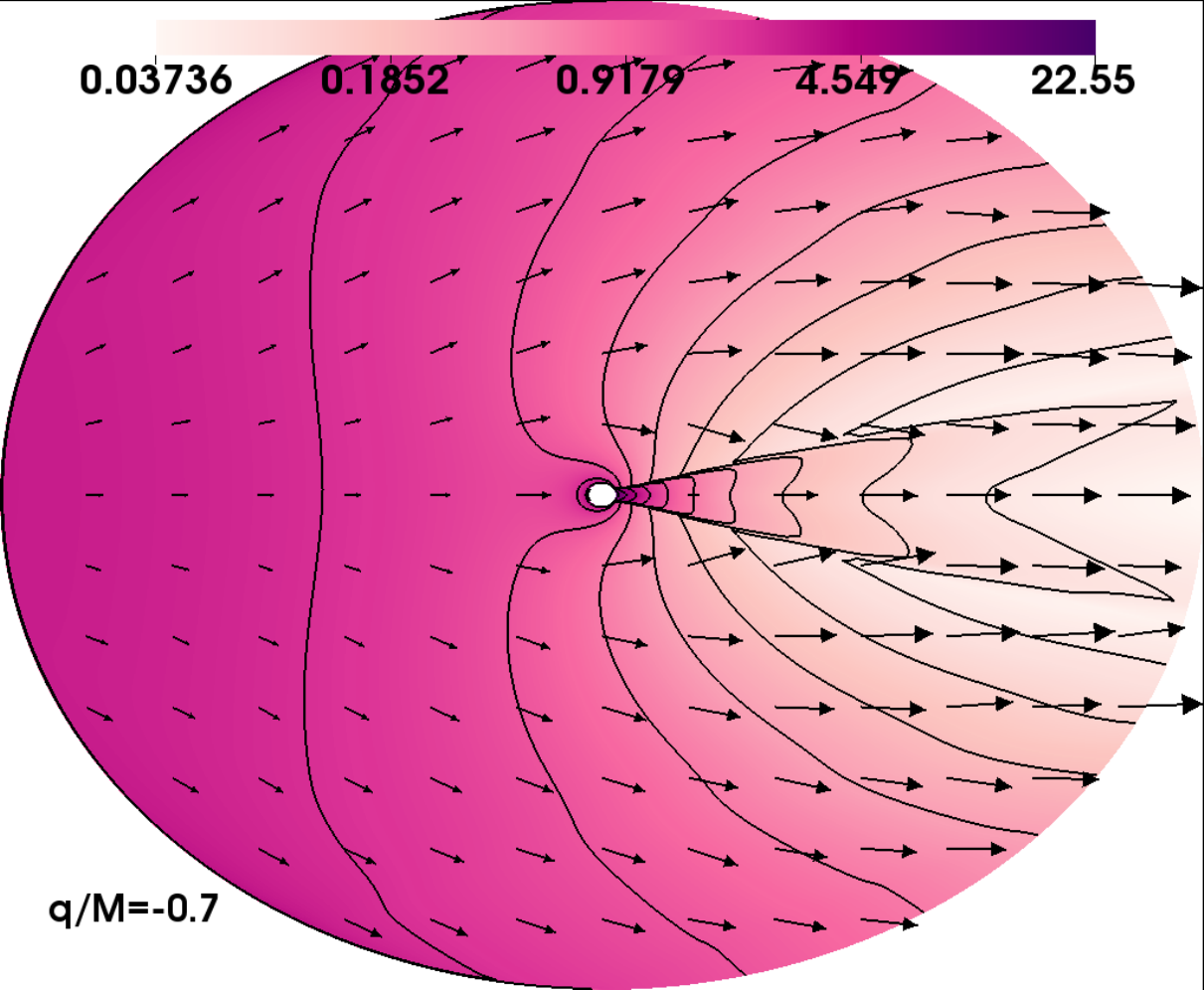,width=6.0cm}
    \psfig{file=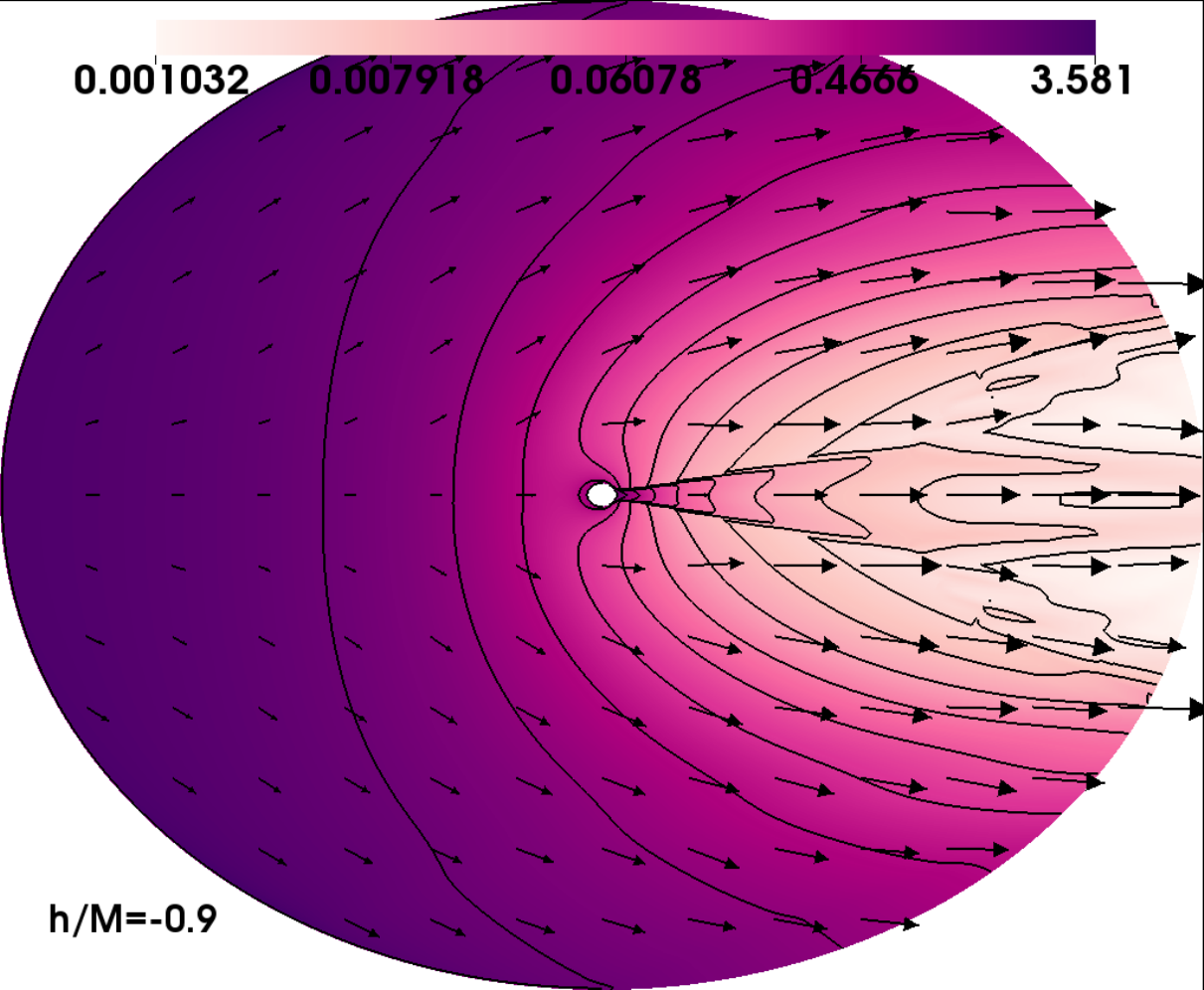,width=6.0cm}
    \psfig{file=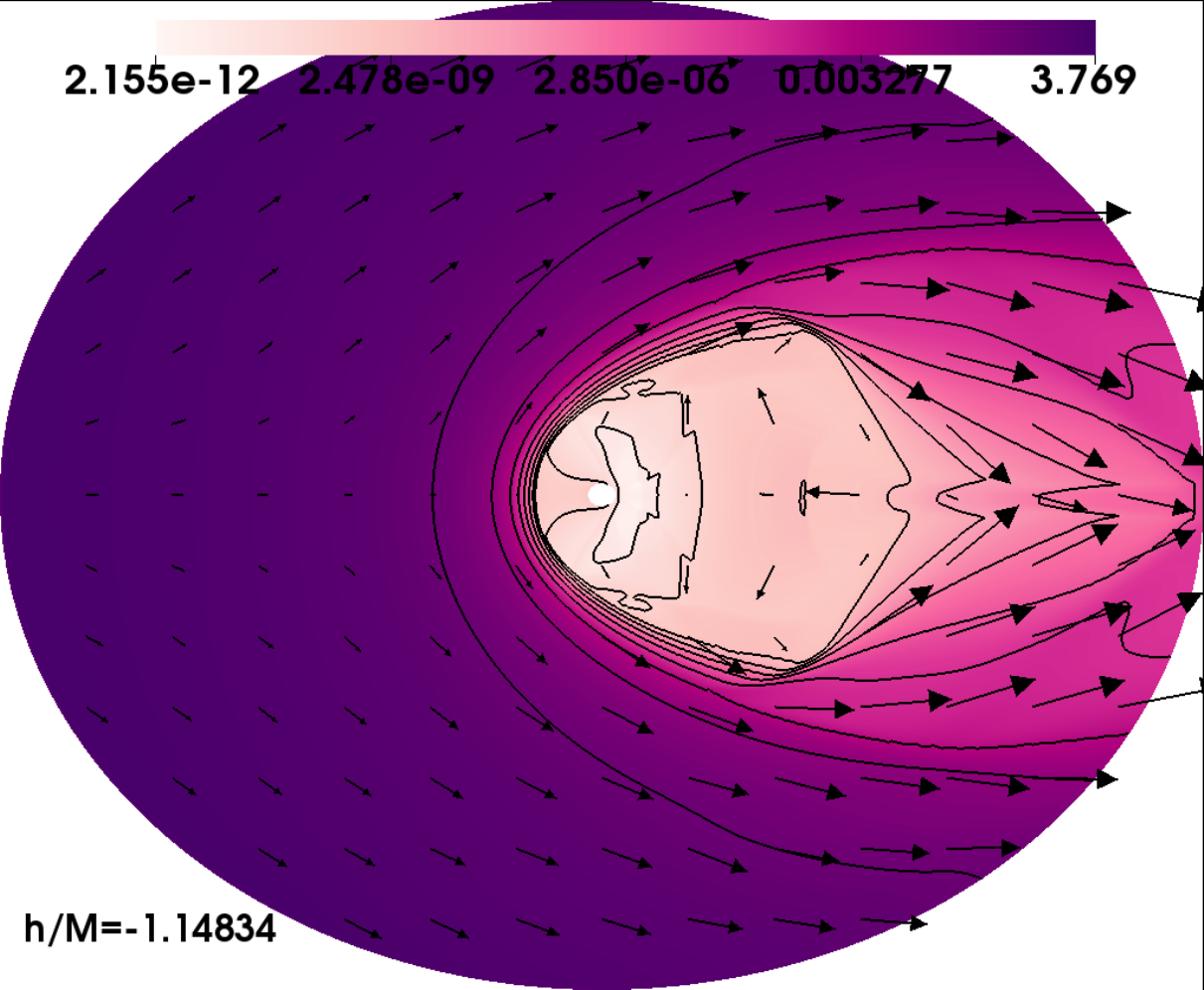,width=6.0cm}
    \psfig{file=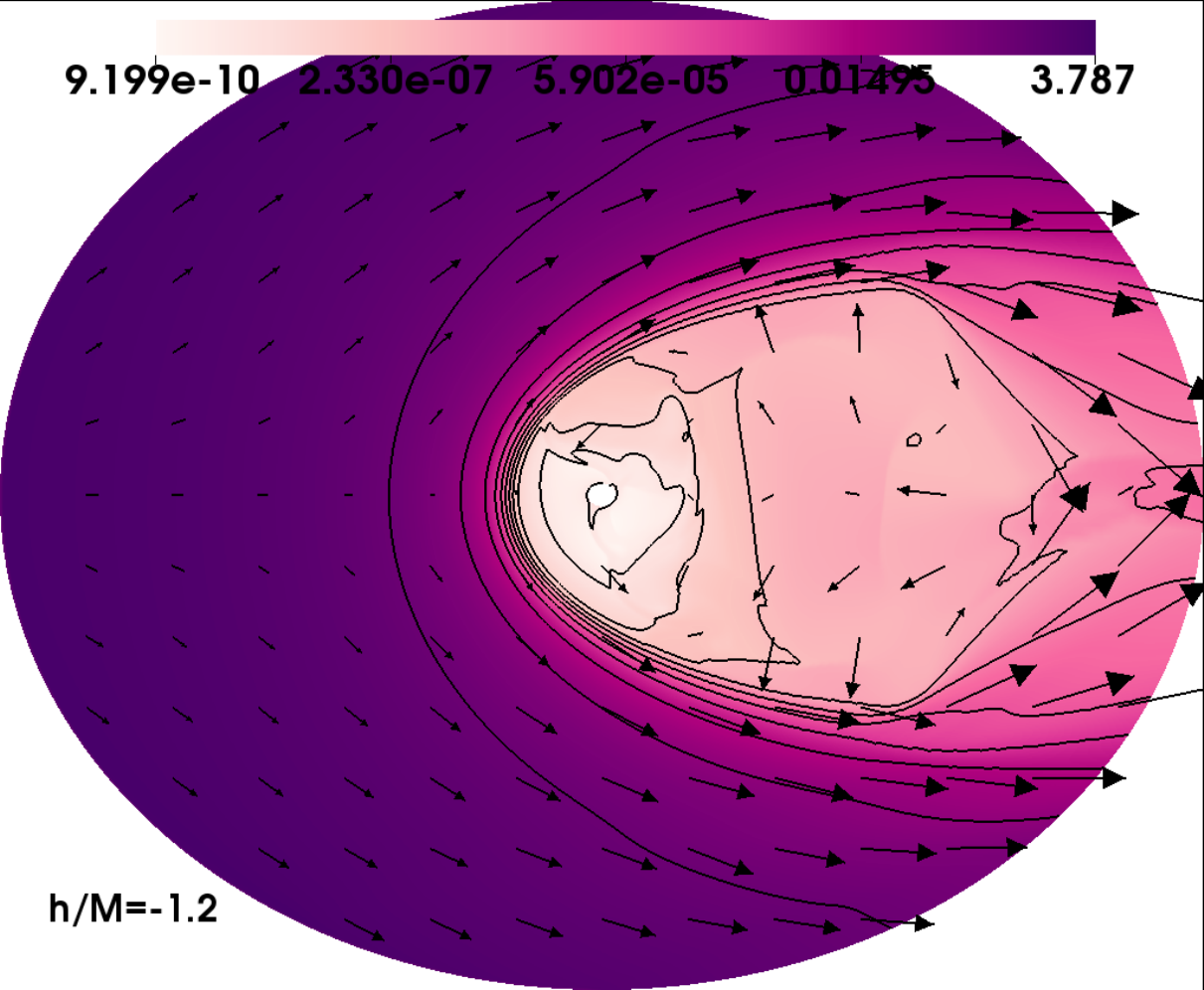,width=6.0cm}    
    \caption{Same as Fig.\ref{colV02_1}, but this figure shows the change in the physical structure under a smaller scalar field parameter. It has been observed that the matter trapped inside the shock cone is completely ejected out of the computational domain, turning the shock cone into a cavity.}
\vspace{1cm}
\label{colV02_2}
\end{figure*}

\begin{figure*}
  \vspace{1cm}
  \center
  \psfig{file=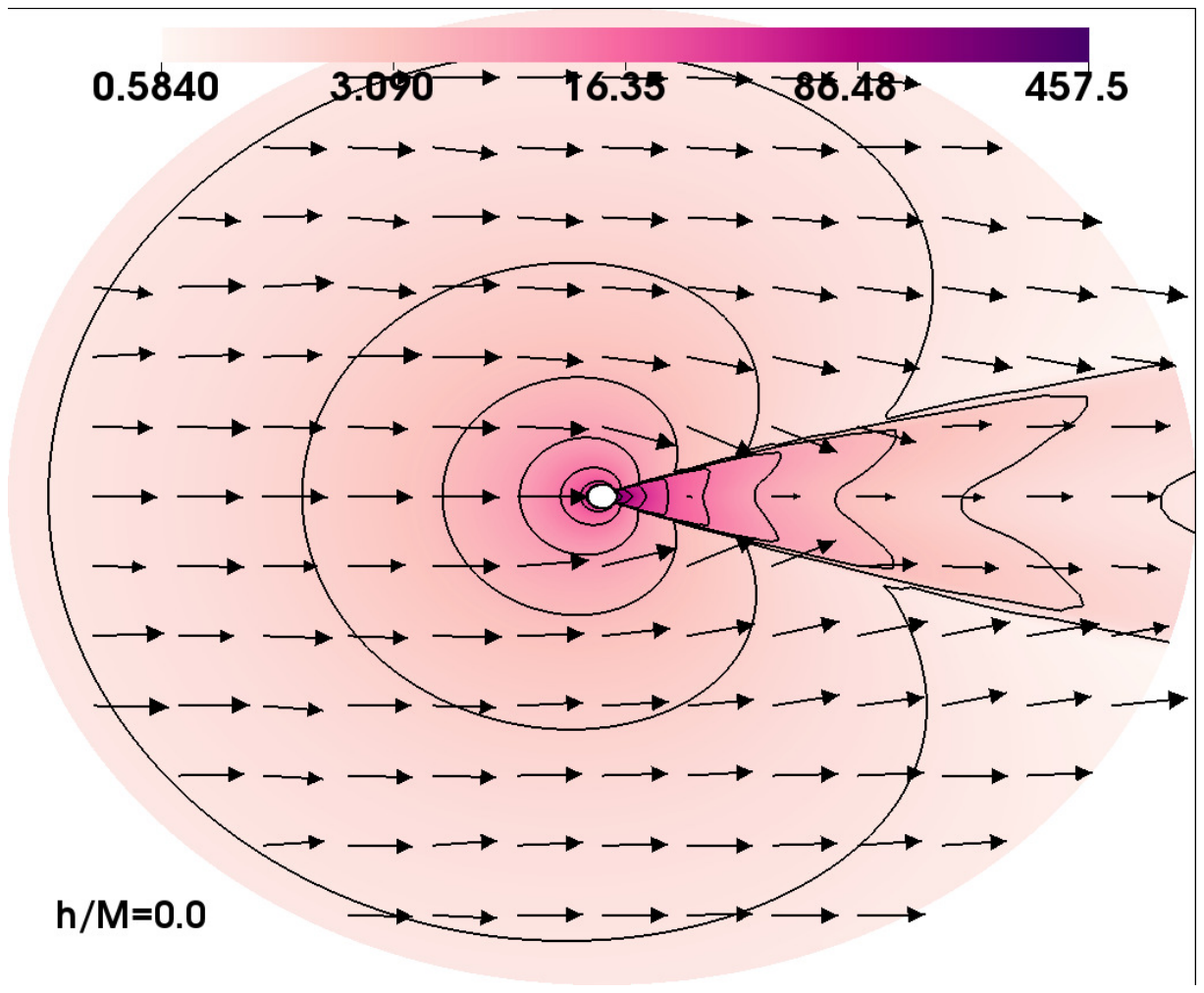,width=9.0cm, height=7.0cm}
  \psfig{file=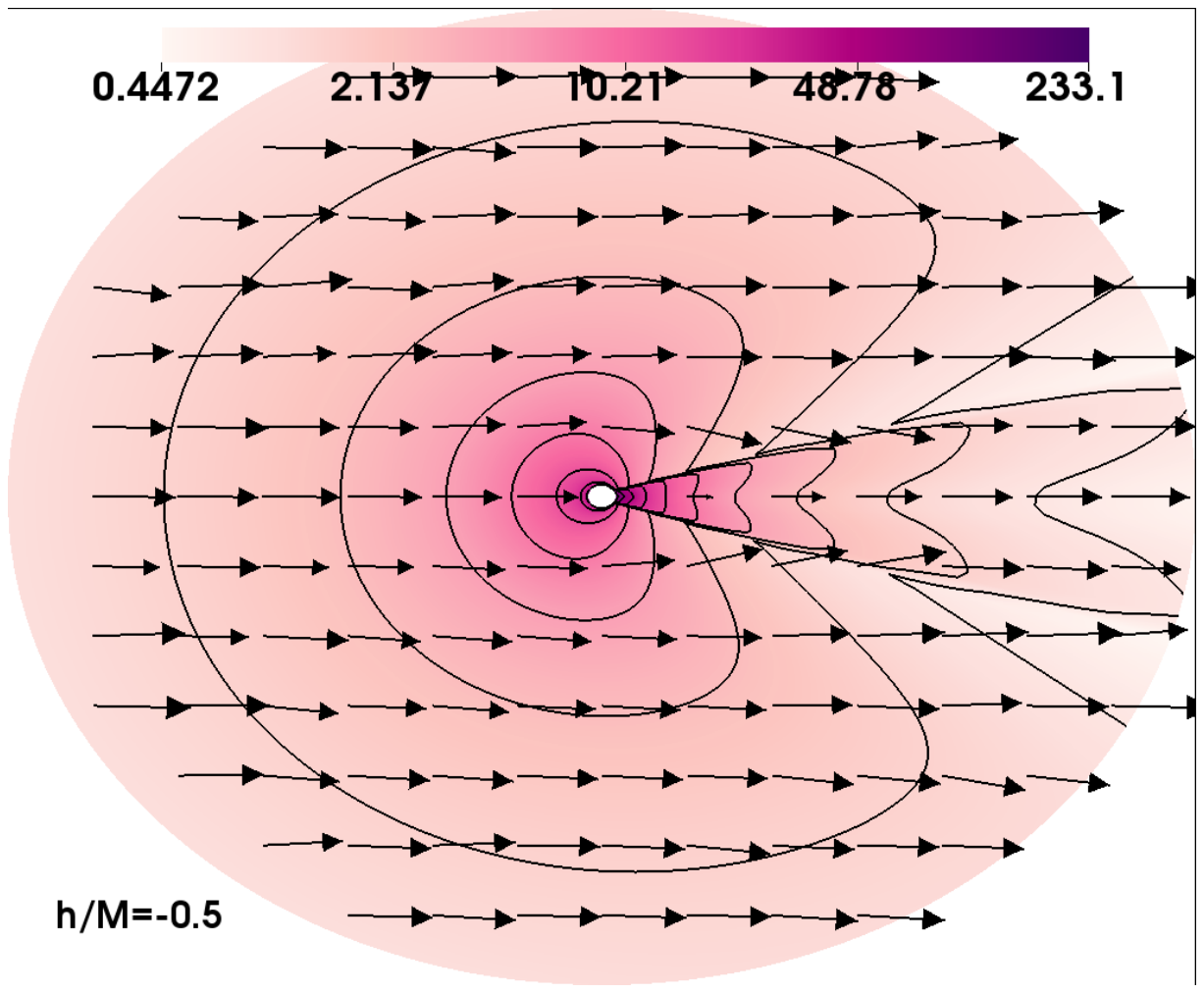,width=9.0cm, height=7.0cm}
  \psfig{file=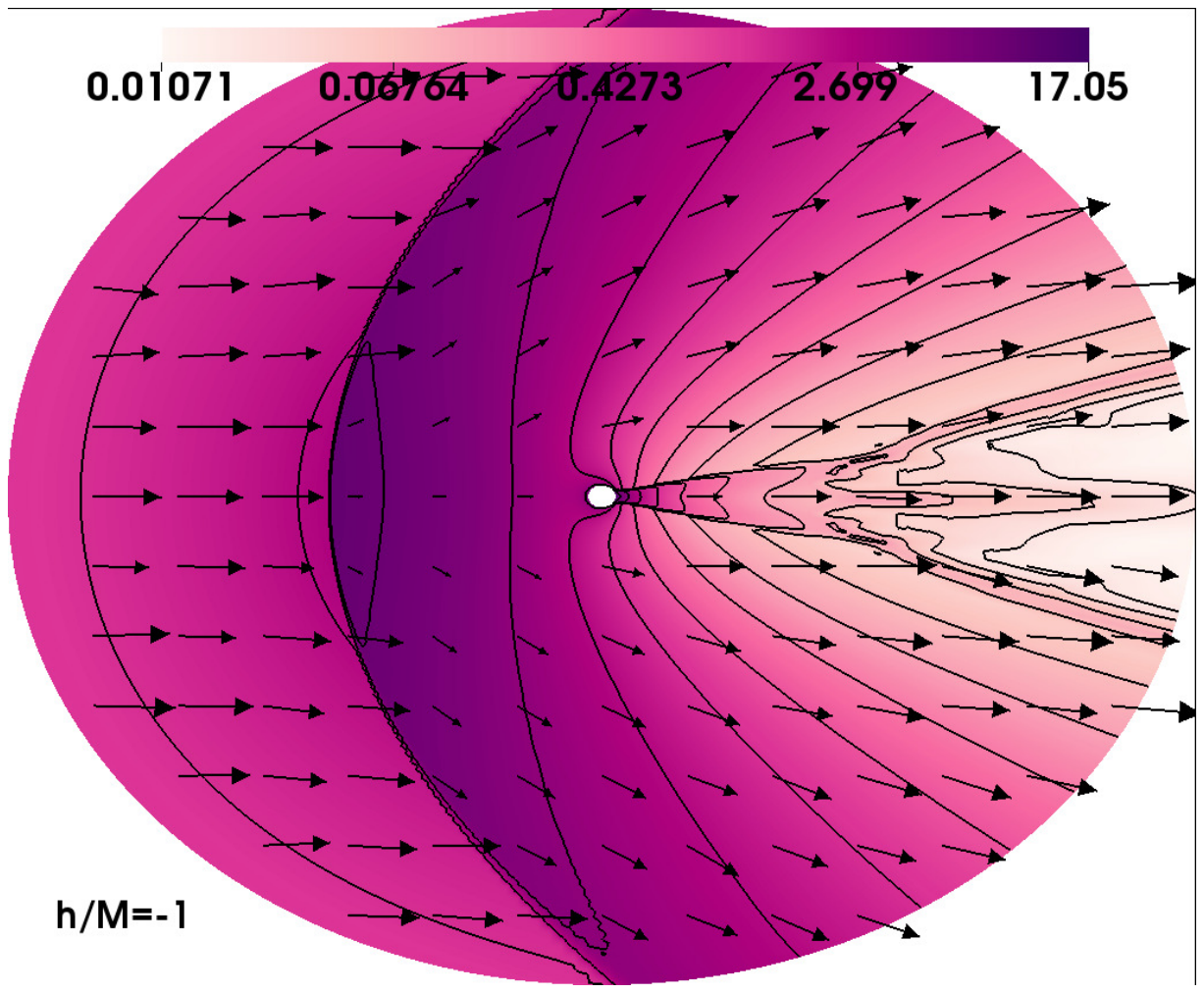,width=9.0cm, height=7.0cm}
    \caption{Unlike Figs.\ref{colV02_1} and \ref{colV02_2}, this graph shows the changes in the physical mechanism around the black hole for three different values of the scalar hair parameter, in the case where the asymptotic speed is $V_{\infty}/c=0.4$.
}
\vspace{1cm}
\label{colV04_1}
\end{figure*}

Figs. \ref{colV02_1} and \ref{colV02_2} show how conditions like shock cones, bow shocks, or cavities around the black hole depend on the parameters, under the asymptotic speed $V_{\infty}/c=0.2$, which appears in the radial and angular velocities in Eq. \ref{Asymptotic_veloc}. The asymptotic speed, determining the velocity magnitude of matter entering the computational domain, influences the infall speed of matter, the formed physical mechanisms, and all related phenomena such as QPOs. This influence has been identified in the literature for certain specific speeds, discussing how these speeds can affect the physical mechanisms and their instabilities \cite{Zanotti1, Donmez5, CruzOsorio2023JCAP, Donmez2024arXiv240216707D}. In this paper, we have modeled the impact of this parameter on the physical mechanisms around the non-rotating hairy Horndeski black hole at $V_{\infty}/c=0.4$ for three different scalar hair parameters. The results of these models are presented in Fig. \ref{colV04_1}, and the differences and similarities arising in both asymptotic speed conditions are discussed in detail later.

Figs. \ref{colV02_1} and \ref{colV02_2} depict the stable flow system around the black hole, established long after reaching the steady-state, around $t=34000M$. They illustrate the evolution of the stable flow system around the black hole and the dynamic structures of the formed physical mechanisms, and how different physical mechanisms emerge depending on the scalar hair parameter. To discuss these scenarios in detail, logarithmic color and contour plots of the rest-mass density are provided, along with vector plots to show the flow behavior of the matter. In the top row of Fig. \ref{colV02_1}, the structure of the shock cone formed in the $h/M=0$ (Schwarzschild case) and $h/M=-0.2$ cases is revealed. Even in cases where the scalar field parameter is small, the geometric structures of the contour lines change, leading to a decrease in the density of matter trapped in the cone. As seen from the vector plots, this decrease is due to an increase in the flow speed of the matter trapped in the cone away from the black hole. This phenomenon can also be explained by the stagnation point, as shown in Table \ref{Inital_Con}, moving closer to the black hole as $h/M$ decreases. Furthermore, as Table \ref{Inital_Con} shows, the opening angle of the cone decreases with decreasing $h/M$. As expected, and as discussed in detail in the following section, the QPO frequencies are significantly affected by these changes.

The scalar hair parameters define a new potential around the black hole in addition to the Schwarzschild field, which alters the spacetime. This change is related to the potential barriers created by the hair parameters. Due to the scalar field around the black hole, as the intensity of the potential barrier increases, it not only prevents the direct fall of matter into the black hole but also slows down the velocity of the falling matter. The behavior of the scalar field term is presented in Fig. \ref{Horndeski_EH} and discussed in Section \ref{Horndeski_gravity}. As seen in the middle and bottom rows of Fig. \ref{colV02_1}, as $h/M$ decreases, the flow pattern changes significantly, and the structure of the shock cone is altered. This change can also be explained by the variation of the stagnation point and the opening angle of the cone with $h/M$, as listed in Table \ref{Inital_Con}. With the decreasing value of $h/M$, there is a narrowing in the opening angle of the shock cone, a decrease in the density of the matter inside the cone, and a slowdown in the velocity of the matter falling from the upstream side. Meanwhile, the strong potential created by the scalar field has changed the flow velocity, density, and amount of the matter, resulting in the formation of the bow shock at $h/M=-0.4$ and $h/M=-0.5$. This bow shock has been completely pushed out of the computational domain when $h/M$ reaches values smaller than these. As seen at $h/M=-0.62$ in Fig. \ref{colV02_1}, the bow shock has disappeared, and the shock cone, as the ejection velocity of the matter inside it increases and the stagnation point approaches the horizon of the black hole, has begun to evolve towards disappearance. Finally, when comparing $h/M=0.0$ and $h/M=-0.62$, the effect of the scalar hair parameters is very significantly observed. In the case of $h/M=0.0$, the matter density is much lower upstream than downstream, while at $h/M=-0.62$, this situation has reversed. This reversal is entirely due to the increase in the intensity of the potential barrier created by the scalar field, which results in the scattering of matter falling into the black hole through the BHL mechanism into the surrounding area of the black hole, with the scattered matter escaping from the strong gravitational field and falling out.

Fig. \ref{colV02_2} shows that at lower values of the scalar hair parameter, the flow pattern and physical mechanisms around the black hole undergo significant changes. In the top row of Fig. \ref{colV02_2}, even as the intensity of the scalar field increases, the shock cone is still observed on the downstream side. However, it is evident that the amount of matter inside the cone is decreasing, and the cone is on the verge of disappearing. Conversely, an increase in the amount of matter on the upstream side is observed, and due to the pressure forces created by this matter, instead of directly falling into the black hole through the BHL mechanism, the matter is now scattering towards points far from the event horizon of the black hole.

In the bottom row of Fig. \ref{colV02_2}, the increasing intensity of the scalar field indicates that the shock cone on the downstream side has completely disappeared, and a cavity has formed in the region where the cone was initially present, ejecting matter outwards. At $h/M=-1.2$, the cavity's area has expanded, moving towards completely covering the downstream side. The observations in Fig. \ref{colV02_2} suggest that as $h/M$ decreases, the flow velocity of the matter on the upstream side gradually decreases, while a significant increase is observed on the downstream side. Cavities around the black holes not only cause QPO frequencies by trapping oscillation modes; they also affect the phase structure of the black hole system, leading to changes in physical parameters such as pressure and energy \cite{Tzikas2021PhLB}.

The velocity vector plots in Figs. \ref{colV02_1} and \ref{colV02_2} illustrate that as $\lvert h/M \rvert$ increases, the speed of the matter falling towards the black hole from the upstream side due to BHL accretion decreases, and concurrently, the matter on the downstream side tends to move further away from the black hole. This effect is attributed entirely to the scalar field. The scalar field, by exerting an additional influence, reduces the influx of matter towards the black hole as $\lvert h/M \rvert$ increases. As shown in these figures, with the increase of $\lvert h/M \rvert$, the BHL accretion becomes completely ineffective. At the same time, the matter within the shock cone on the downstream side is being pushed away from the black hole due to the closeness of the stagnation point to the black hole horizon and the influence of the scalar field. The reason for this, as stated in the analytical study for a test particle in Ref.\cite{Rayimbaev2021}, is entirely due to the increase in the matter's energy and angular momentum. Particularly, the matter inside the cone, due to these increases, has overcome the strong gravitational field, been expelled outward, and eventually, the shock cone has transformed into a cone cavity. This phenomenon is observed in Fig. \ref{colV02_2}. Additionally, the effect of the scalar field is clearly visible in the contour lines of the rest-mass density seen in Figs. \ref{colV02_1} and \ref{colV02_2}. Initially, except for the location of the shock cone, these lines, which exhibited spherical symmetry, have transformed into a more elliptical shape with the intensification of the scalar field, eventually becoming completely distinct contour lines.

Although the main purpose of this study is to examine the changes in the physical structure of the shock cone formed around the black hole with scalar hair parameters and BHL accretion, analyzing the case with an asymptotic velocity of $V_{\infty}/c=0.4$ is important for revealing the effect of physical parameters related to the formation of the QPOs obtained in this study. Our preliminary results for the $V_{\infty}/c=0.4$ case can inform future studies of such structures. Moreover, the choice of asymptotic velocity is related to the placement of the outer boundary of the computational domain. Setting the outer boundary further away changes the roles of the physical mechanisms created by different asymptotic velocities. However, ultimately, for a certain asymptotic velocity, the same physical mechanisms and QPOs occur. In summary, based on our previous experience and the studies we have conducted, the most appropriate asymptotic velocity for explaining the QPO observations at $r_{max}=100M$ is $V_{\infty}/c=0.2$ \cite{Donmez6, Donmez2024arXiv240216707D, Donmez20, Donmez_EGB_Rot}.

In Fig. \ref{colV04_1}, for $V_{\infty}/c=0.4$ with hair parameters $h/M=0$, $h/M=-0.5$, and $h/M=-1$, we demonstrate the effect of the scalar hair parameter on the matter falling towards the black hole and on the formation of the shock cone. Although the behavior observed here shows similarities to Figs. \ref{colV02_1} and \ref{colV02_2} for $V_{\infty}/c=0.2$, especially in vector graphics, it is evident that the flow speed of the matter significantly increases in the case of $V_{\infty}/c=0.4$. On the other hand, when comparing $h/M=-0.5$ in Figs. \ref{colV02_1} and \ref{colV04_1}, a bow shock has formed and been pushed outwards for $V_{\infty}/c=0.2$, whereas it has not yet formed for $V_{\infty}/c=0.4$. As expected, for $V_{\infty}/c=0.4$, the energy of the accreted matter is higher, hence more matter overcomes the scalar field's potential, preventing the formation of a bow shock. However, for $h/M=-1$, although it occurs later than in the $V_{\infty}/c=0.2$ case, a bow shock has formed with much greater intensity compared to $V_{\infty}/c=0.2$. Generally, for $V_{\infty}/c=0.4$, the effect of $h/M$ on the shock cone, bow shock, and potential cavity formation shows much similarity with $V_{\infty}/c=0.2$. If we consider values smaller than $h/M=-1$ for $V_{\infty}/c=0.4$, cavity formation is expected. The onset times differ, with later occurrences for $V_{\infty}/c=0.4$ due to the higher initial energy. However, the main difference between these two models, observed in our numerical calculations, is in the QPO calculations, with significant differences in QPO frequencies discussed in Section \ref{QPOs1}. Revisiting the $h/M=-1$ model in Fig. \ref{colV04_1}, the formed bow shock appears more intense. While the velocity of the matter reaching the bow shock in the upstream region is constant, these materials are pushed back due to the pressure force created by the bow shock. Therefore, as seen in the vector graphic, within the bow shock, the matter moves very slowly and tends to scatter towards the edges. Consequently, this situation complicates the maintenance of the dynamic structure of the shock cone, leading to the matter within the shock cone being pushed away from the black hole, evolving the shock cone towards cavity formation.

As shown in Table \ref{Inital_Con}, the time it takes for the matter falling into the black hole and for the resulting physical mechanisms to reach the steady-state varies from model to model. When comparing all models, the cases with $h/M=0$ and $h/M=-0.5$ in Fig. \ref{colV04_1} reach the steady-state very quickly. This rapid establishment of a stable structure is due to both the large asymptotic velocity and the formation of the shock cone around the black hole. On the other hand, in the cases of $h/M=-0.4$ and $h/M=-1$ in Figs. \ref{colV02_1} and \ref{colV04_1}, where both the shock cone and the bow shock are observed, it takes a significant amount of time to reach the steady-state. In scenarios where a cavity is formed, the potential created by the scalar field dominates compared to the energy of the accreted matter, allowing the system to quickly reach the steady-state. This occurs because the scalar field obstructs the direct fall of matter into the black hole.

%%%%%%%%%%%%%%%%%%%%%%%%%%%%%%%%%%%%%%%%%%%%%%%%%%%%%%%%%%%%%%%%%%%%%%%%%%%%%%%%%%%

\begin{figure*}
  \vspace{1cm}
  \center
     \psfig{file=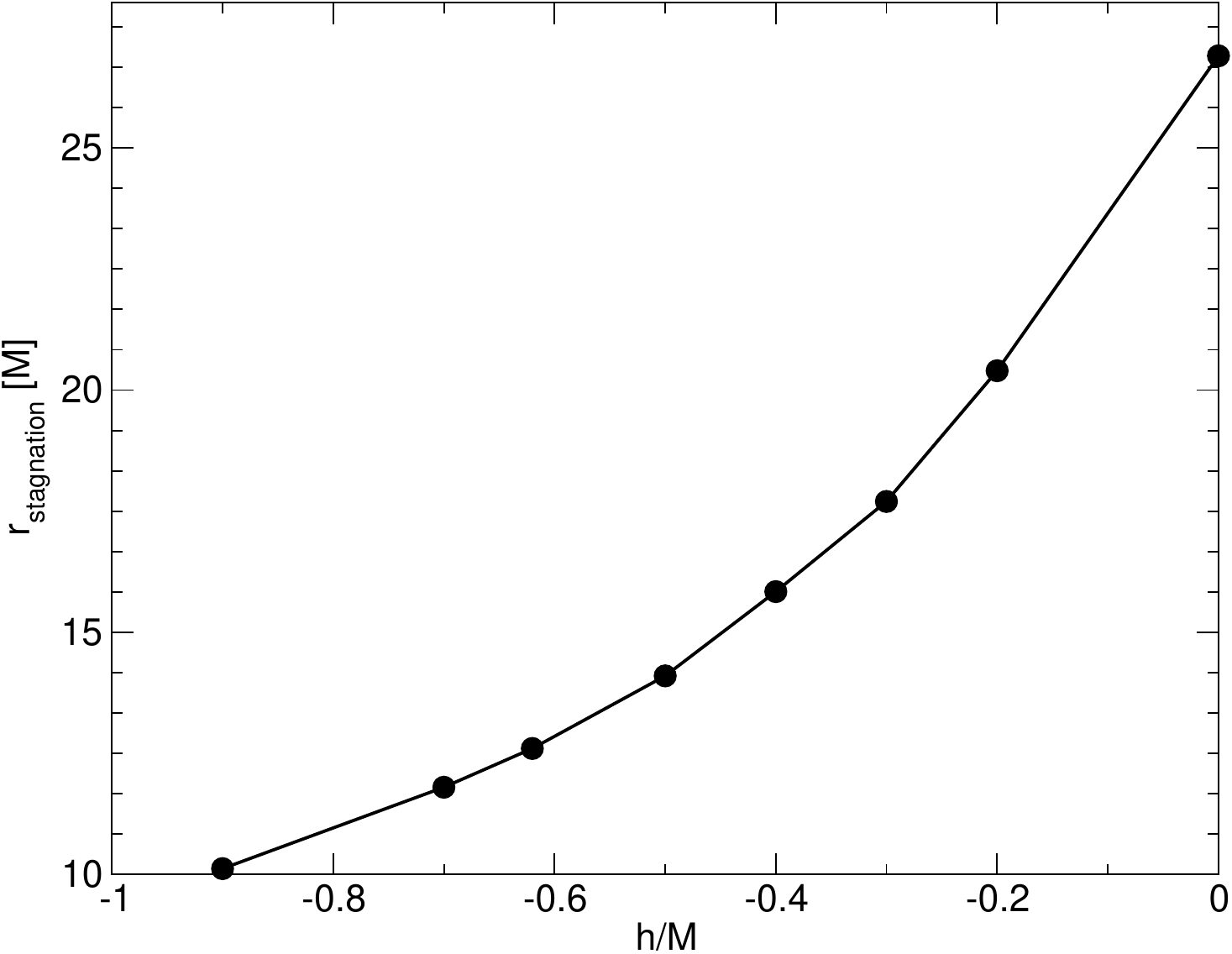,width=12.0cm}
     \caption{The variation of the stagnation point as a function of the scalar hair parameter $h/M$ is significant. As the value of $h/M$ decreases, the intensity of the scalar field increases, causing the matter to be pushed away from the black hole. Consequently, the stagnation point approaches the horizon of the black hole.
}
\vspace{1cm}
\label{Stag_1}
\end{figure*}

\begin{figure*}
  \vspace{1cm}
  \center
     \psfig{file=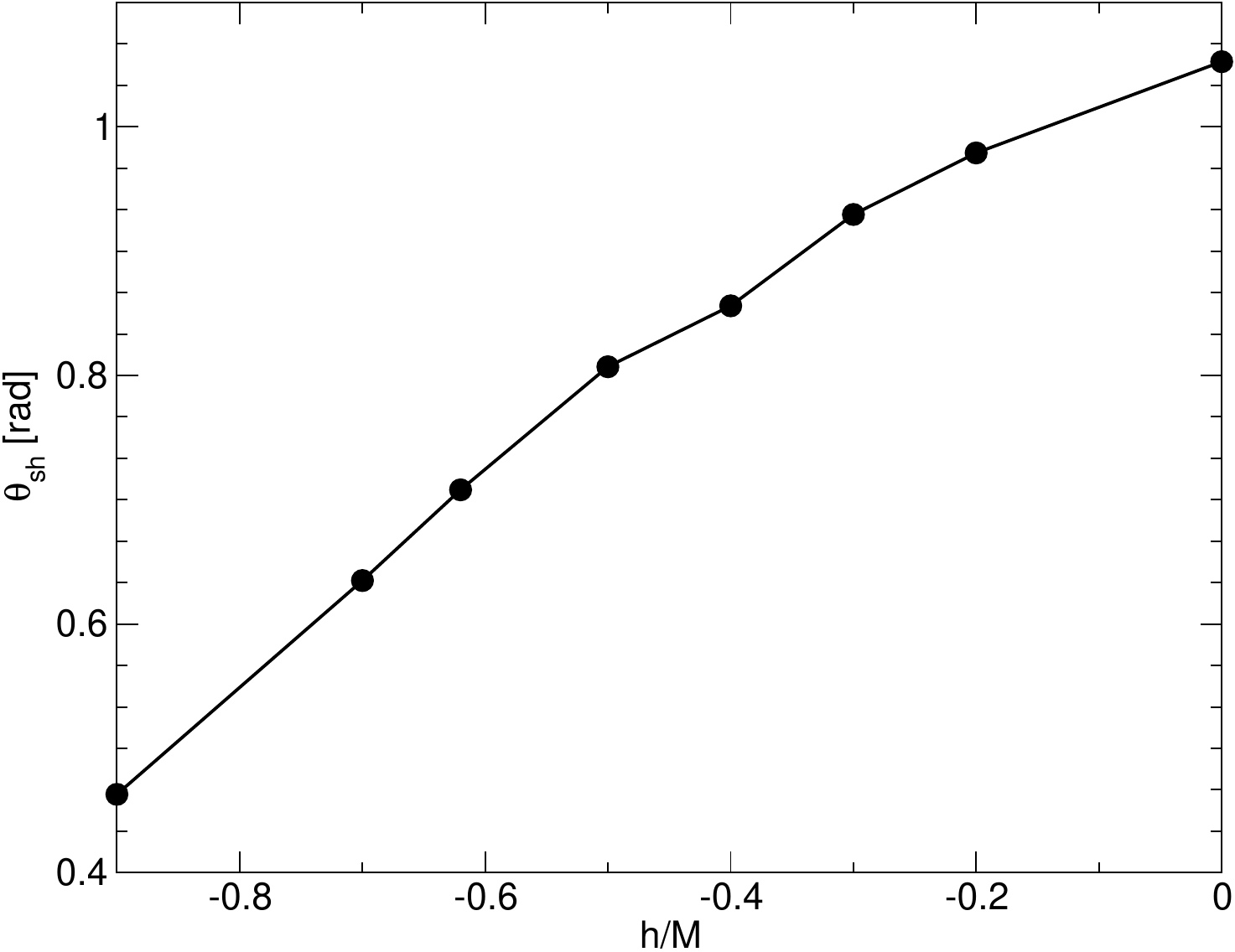,width=12.0cm}
     \caption{The opening angle of the shock cone changes with respect to $h/M$. As $h/M$ decreases, the opening angle diminishes. This reduction occurs because, as the intensity of the scalar field increases, the matter is pushed outward. Consequently, the matter inside the cone cannot maintain the structure of the cone, leading to the bilateral compression of the cone’s boundaries by the pressure of the gas sent via BHL accretion. This compression causes the shock locations, which define the boundaries of the shock cone, to approach each other.
}
\vspace{1cm}
\label{open_angl_1}
\end{figure*}

As previously discussed, the scalar hair parameter modifies the Schwarzschild solution and, consequently, alters the gravitational potential around the black hole. Thus, changes occur in the behavior of matter falling towards the black hole compared to the general assumptions of general relativity. Numerical solutions have shown that the scalar hair parameter significantly affects the dynamic structure of the shock cone formed around the black hole. To demonstrate this effect more clearly, for each model, we present the changes in the stagnation point and the opening angle of the shock cone, as detailed in Table \ref{Inital_Con}, and depicted in Figs. \ref{Stag_1} and \ref{open_angl_1} as functions of $h/M$. Fig. \ref{Stag_1} specifically illustrates the change in the stagnation point for the formed shock cones as a function of $h/M$. As shown, the stagnation point decreases exponentially with decreasing $h/M$, indicating that this point approaches the horizon of the black hole. This change affects many behaviors, from the instability of the shock cone to the induced QPOs within it. The function representing this change has been determined by applying a fit.

On the other hand, the opening angle of the shock cone ($\theta_{sh}$) is presented in Fig. \ref{open_angl_1} as a function of $h/M$. While the behavior of the stagnation point is exponential with respect to $h/M$, as seen in Fig. \ref{Stag_1}, the behavior of the opening angle in Fig. \ref{open_angl_1} is also exponential but inversely related. The empirical formula for the stagnation point of the shock cone is $r_{stag} = 20.01 e^{2.03 \frac{h}{M}} + 6.92$, while for the shock opening angle, it is $\theta_{sh} = -0.1901 e^{-1.6634 \frac{h}{M}} + 1.2431$. Although these functions may vary with different asymptotic speeds, $V_{\infty}/c$, asymptotic sound speed $C_{s,\infty}/c$, and adiabatic index, $\Gamma$ \cite{Donmez2024arXiv240216707D, Yang2019CoTPh, Yalinewich2018MNRAS}, their general behavior remains almost the same, with only the initial values and the intensity of the exponential functions showing variation. Additionally, it has been shown numerically in our previous work \cite{Donmez2024arXiv240216707D} in Figs. 2 and 3, and theoretically in \cite{Yalinewich2018MNRAS}, that the stagnation point and shock opening angle do not depend on the spin parameter of the black hole.

%%%%%%%%%%%%%%%%%%%%%%%%%%%%%%%%%%%%%%%%%%%%%%%%%%%%%%%%%%%%%%%%%%%%%%%%%%%%%%%%%%%
%%%%%%%%%%%%%%%%%%%%%%%%%%%%%%%%%%%%%%%%%%%%%%%%%%%%%%%%%%%%%%%%%%%%%%%%%%%%%%%%%%%
%%%%%%%%%%%%%%%%%%%%%%%%%%%%%%%%%%%%%%%%%%%%%%%%%%%%%%%%%%%%%%%%%%%%%%%%%%%%%%%%%%%
%%%%%%%%%%%%%%%%%%%%%%%%%%%%%%%%%%%%%%%%%%%%%%%%%%%%%%%%%%%%%%%%%%%%%%%%%%%%%%%%%%%

\section{Computing QPOs in Numerical Simulatons}
\label{QPOs1}

Here, for the $V_{\infty}/c=0.2$ and $V_{\infty}/c=0.4$ scenarios, we demonstrate how the shock cone, bow shock, and cavity mechanisms, formed under different scalar hair parameters, trap and excite QPO modes, and how these modes change depending on the parameter. We discuss how the fundamental modes and their nonlinear couplings generate new QPOs based on the shock mechanism around the black hole. We identify the circumstances under which specific frequencies are formed, ranging from LFQPOs to HFQPOs. The results of the PSD analyses indicate that the QPO formations around the non-rotating Horndeski black hole significantly vary with the $h/M$ parameter. Furthermore, a summary of the fundamental oscillation modes generated for each model is provided in Table \ref{geniue_mode}. This section details the reasons for the formation of these fundamental modes, the impact of physical mechanisms on their generation, and how these modes are produced, based on the analyses we have discussed.

\begin{table}
\footnotesize
\caption{The frequencies of the first three genuine modes are measured close to the black hole horizon from the evolution of the mass accretion rate for different models. Here, 'Model' denotes the model name, and $h/M$ represents the hair parameter. The terms $f_1$, $f_2$, and $f_3$ denote the genuine modes, while 'Remarks' indicate the physical mechanism created around the black hole. Since determining which of the numerically observed frequencies correspond to $f_{sh}$, $f_{eh}$, or $f_h$ is challenging, they are referred to as $f_1$, $f_2$, and $f_3$ here. However, in cases where there is only one genuine mode, it is presumed that $f_1 = f_{sh}$.
}
 \label{geniue_mode}
%\begin{center}
  \begin{tabular}{cccccc}
    \hline
    \hline
    $Model$ & $h/M$ & $f_1(\rm{Hz})$ & $f_2(\rm{Hz})$ & $f_3(\rm{Hz})$ & $Remarks$ \\
    \hline
    $H02A$ & $0.0$     &  $8$   & $24$  & $30$ &  $Cone$ \\
    $H02B$ & $-0.2$    &  $30.5$& $NO$  & $NO$ &  $Cone$ \\    
    $H02C$ & $-0.3$    &  $12.4$& $26$  & $50$ &  $Cone$ \\
    $H02D$ & $-0.4$    &  $5.1$ & $8.1$ & $10.6$ &  $Cone \; and\; bow$  \\
    $H02E$ & $-0.5$    &  $52$  & $NO$  & $NO$ &  $Cone$ \\
    $H02F$ & $-0.62$   &  $22$  & $26$  & $32$ &  $Cone$ \\
    $H02G$ & $-0.7$    &  $31$  & $NO$  & $NO$ &  $Cone$ \\
    $H02H$ & $-0.9$    &  $NO$  & $NO$  & $NO$ &  $Cone$ \\
    $H02I$ & $-1.14834$&  $16.4$& $NO$  & $NO$ &  $Cavity$ \\
    $H02L$ & $-1.2$    &  $8.3$& $NO$  & $NO$ &  $Cavity$ \\
    \hline
    $H04A$ & $0.0$     &  $19$ & $27$  & $31.5$ &  $Cone$ \\
    $H04B$ & $-0.5$    &  $NO$& $NO$  & $NO$ &  $Cone$ \\
    $H04C$ & $-1.0$    &  $NO$& $NO$  & $NO$ &  $Cone \; and\; bow$ \\
    \hline
    \hline
  \end{tabular}
%\end{center}
\end{table}
%

%%%%%%%%%%%%%%%%%%%%%%%%%%%%%%%%%%%%%%%%%%%%%%%%%%%%%%%%%%%%%%%%%%%%%%%%%%%%%%%%%%%
\subsection{The Case of $V_{\infty}/c=0.2$}
\label{QPOs2}

The observational results have shown that the mass accretion rate, calculated at the inner radius of the disk, is strongly connected to the QPOs formed around the black hole \cite{Liu_2021, VarniERe:2015kda}. Therefore, the mass accretion rate has been calculated at $r=2.3M$, and data showing its variation over time have been obtained. Subsequently, by conducting a PSD analysis of this mass accretion rate, the QPOs arising from the physical mechanism around the black hole have been numerically calculated. This analysis reveals the oscillation characteristics and frequencies of the shock cone. Due to the geometrized units defined in the introduction, the unit of the obtained frequencies is in terms of the black hole mass. However, for a better understanding and comparison with the literature, these frequencies are also expressed in Hz assuming the black hole mass of $M=10M_{\odot}$. For this conversion, the formula   $f(\rm{Hz}) = f(M) \times 2.03 \times 10^5 \times \left(\frac{M_{\odot}}{M}\right)$ is used. Furthermore, in this article, and in other studies utilizing the data provided here, the frequencies for sources with different black hole masses can be calculated in the PSD analyses by multiplying the frequency axis by the factor $10M_{\odot}/M$, where $M$ is the mass of the black hole used to compute the frequency.

In this paper, the effects of viscosity and magnetic fields are not considered, so the oscillation modes resulting from these physical parameters do not occur in our study. However, the structures of the shock cone, bow shock, and cavity formed around the black hole, the confinement of pressure modes, and their excitation occur \cite{CruzOsorio2023JCAP, Donmez2024arXiv240216707D}. This mode, confined to a specific region, causes oscillation in a regular manner. These oscillation frequencies can be revealed in observational studies. In this context, we are examining the behavior of shock cones resulting from BHL accretion to understand the physical mechanism behind the observed data. Due to the change in pressure of matter around the non-rotating Horndeski black hole, four different frequencies can occur, depending on the physical structure of the cone. These are the azimuthal oscillation frequency, $f_{sh}$, inside the shock cone; the frequency of oscillation of matter confined between the potential barrier defined by the Horndeski hair parameter and the stagnation point, $f_h$; the oscillation frequency $f_{eh}$ between the stagnation point and the black hole horizon; and finally, the bow shock frequency that can occur as a result of the interaction of the shock cone with the strong gravitational field of the black hole and the scalar hair parameter. The pressure mode confined within this shock can create the frequency, $f_{bow}$ \cite{CruzOsorio2023JCAP, Donmez2024arXiv240216707D}.

\begin{figure*}
  \vspace{1cm}
  \center
     \psfig{file=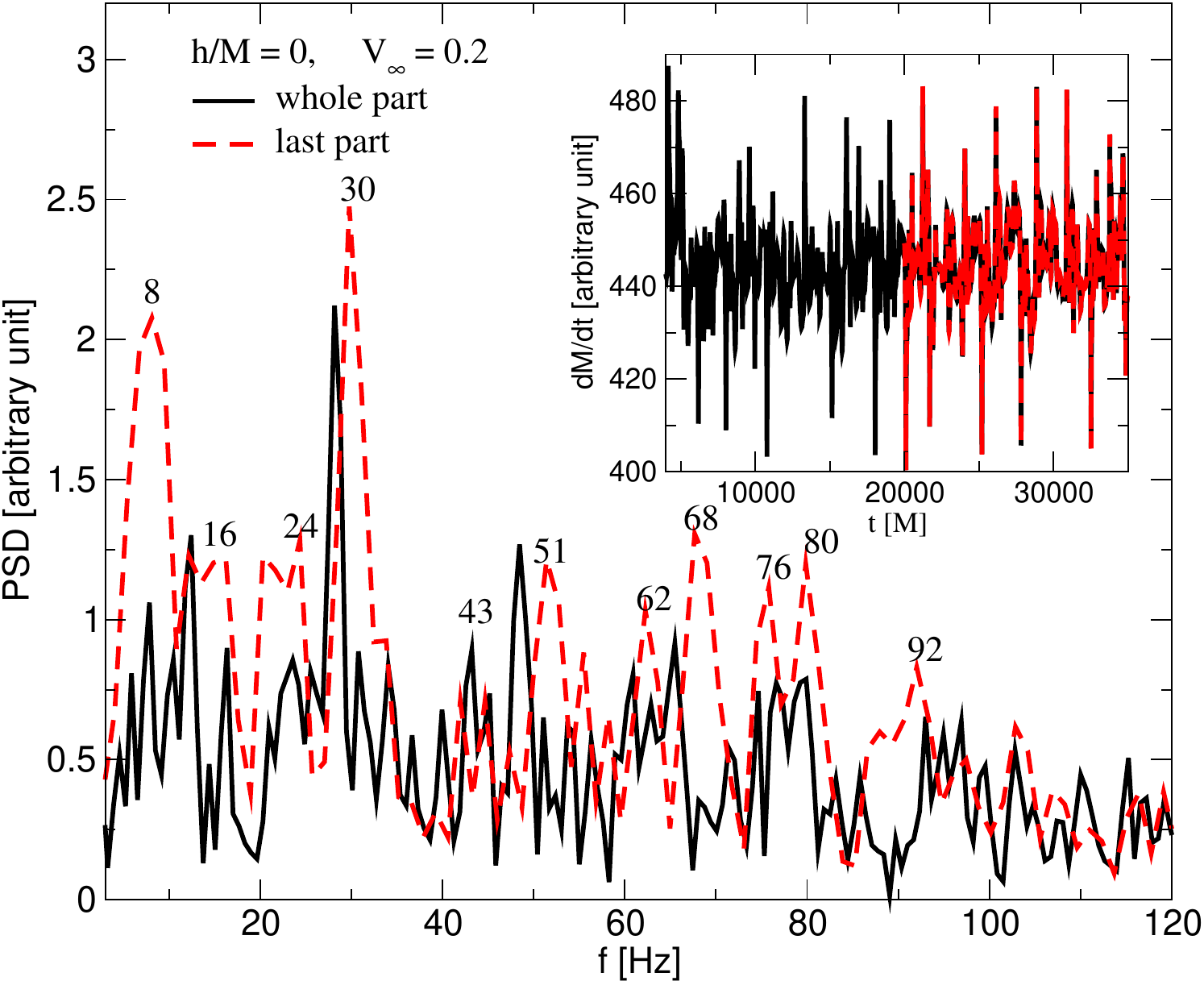,width=7.0cm}\hspace*{0.15cm}
     \psfig{file=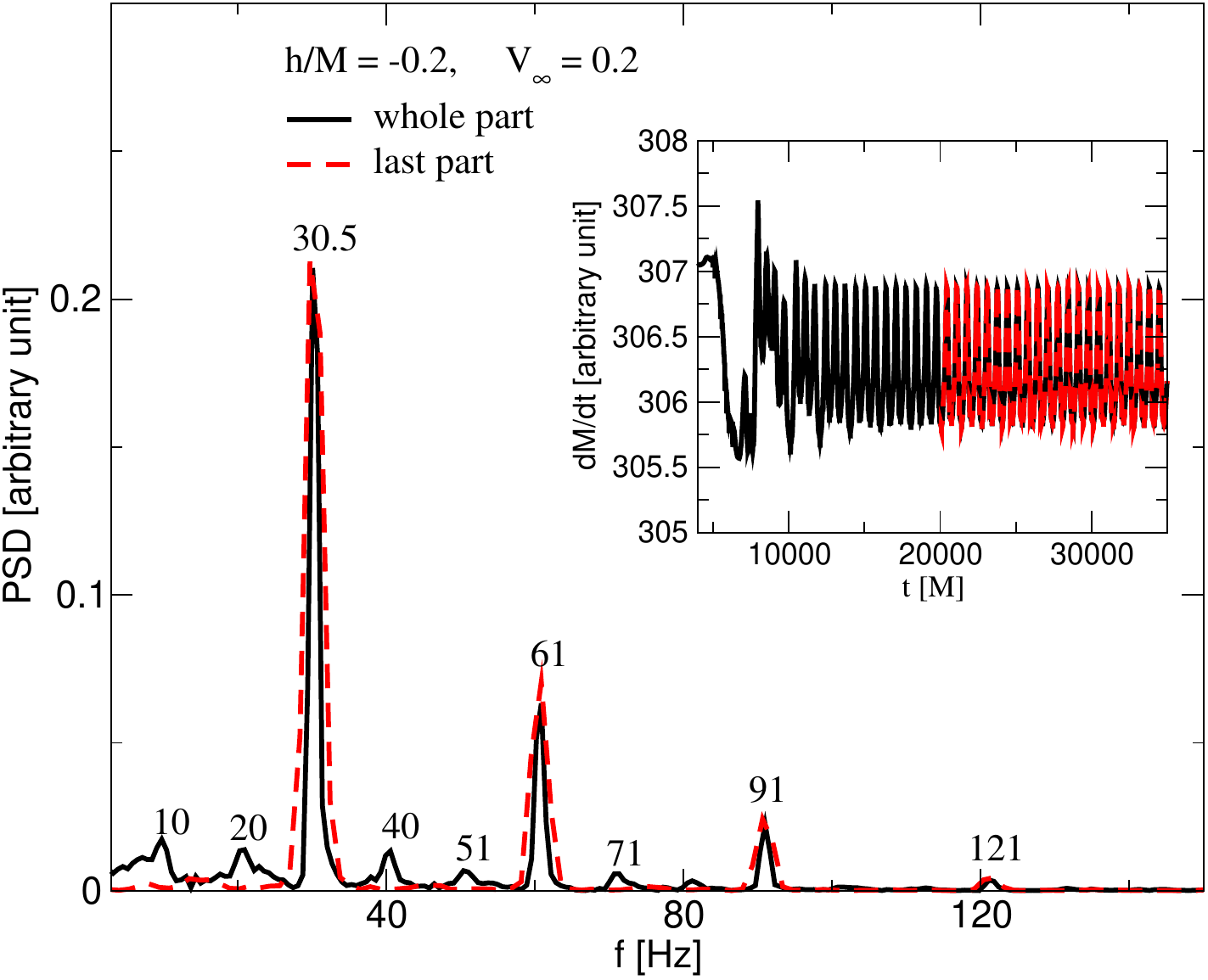,width=7.0cm}\\
    \vspace*{0.5cm} 
    \psfig{file=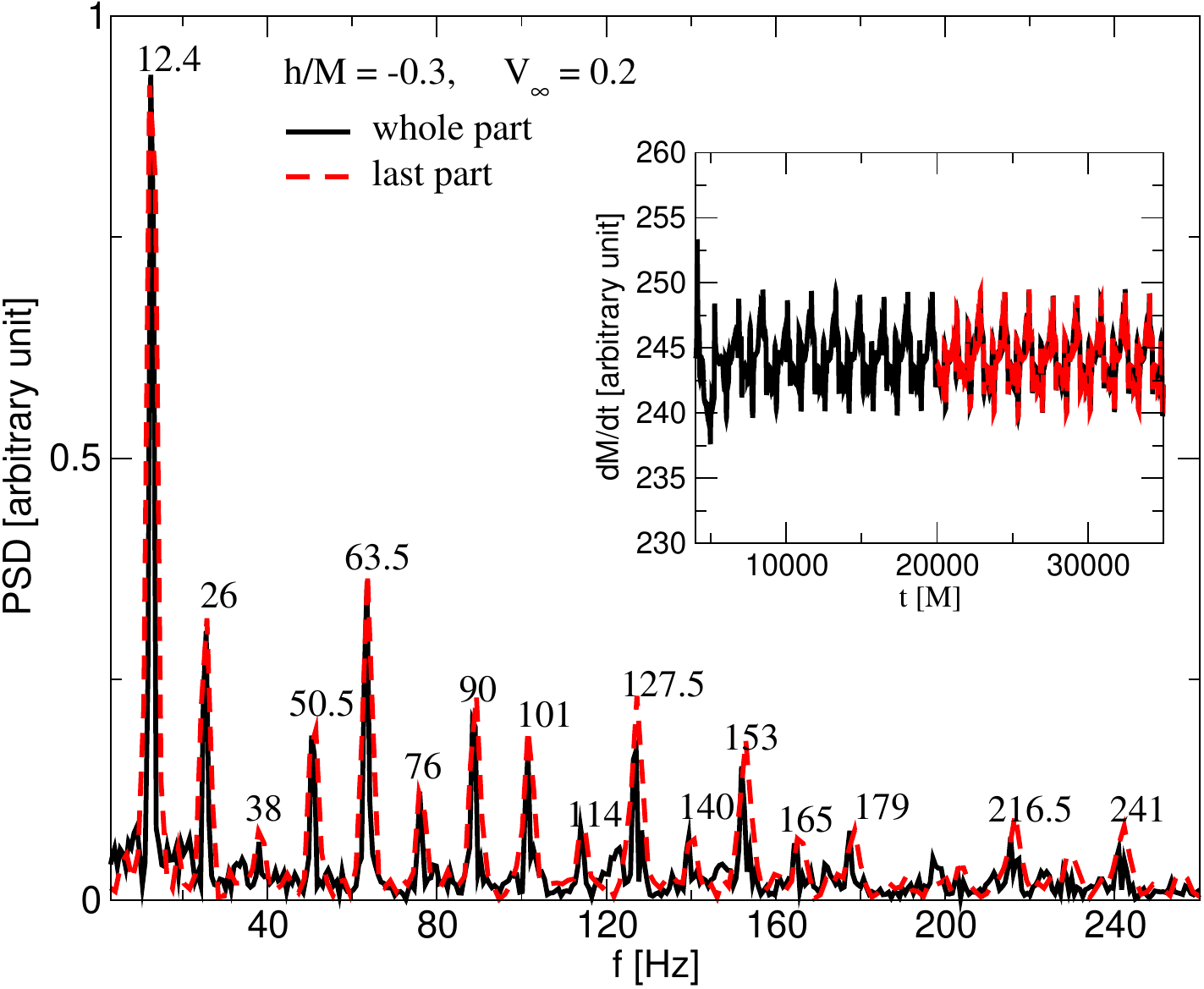,width=7.0cm}\hspace*{0.15cm}
    \psfig{file=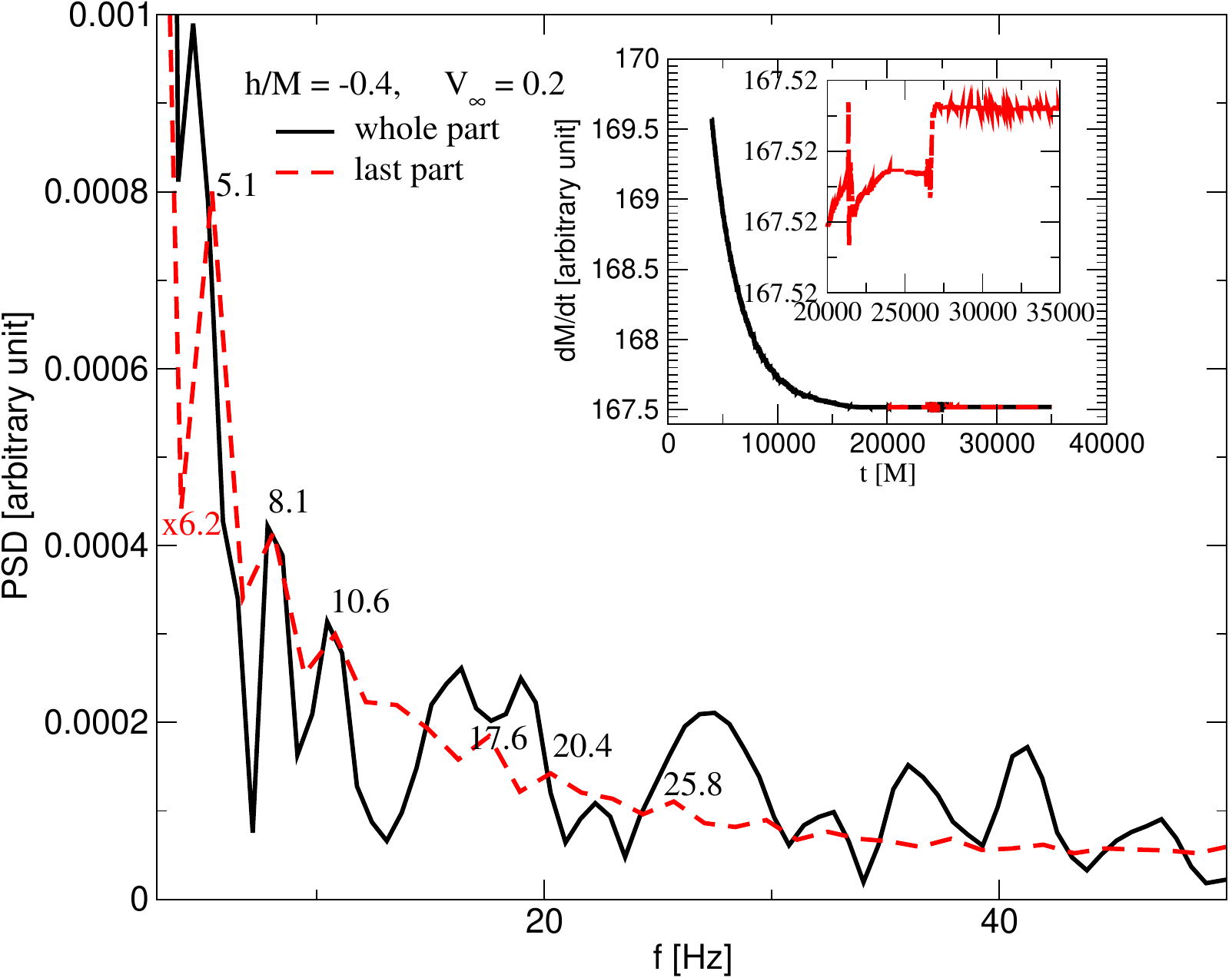,width=7.2cm}\\
    \vspace*{0.5cm}
    \psfig{file=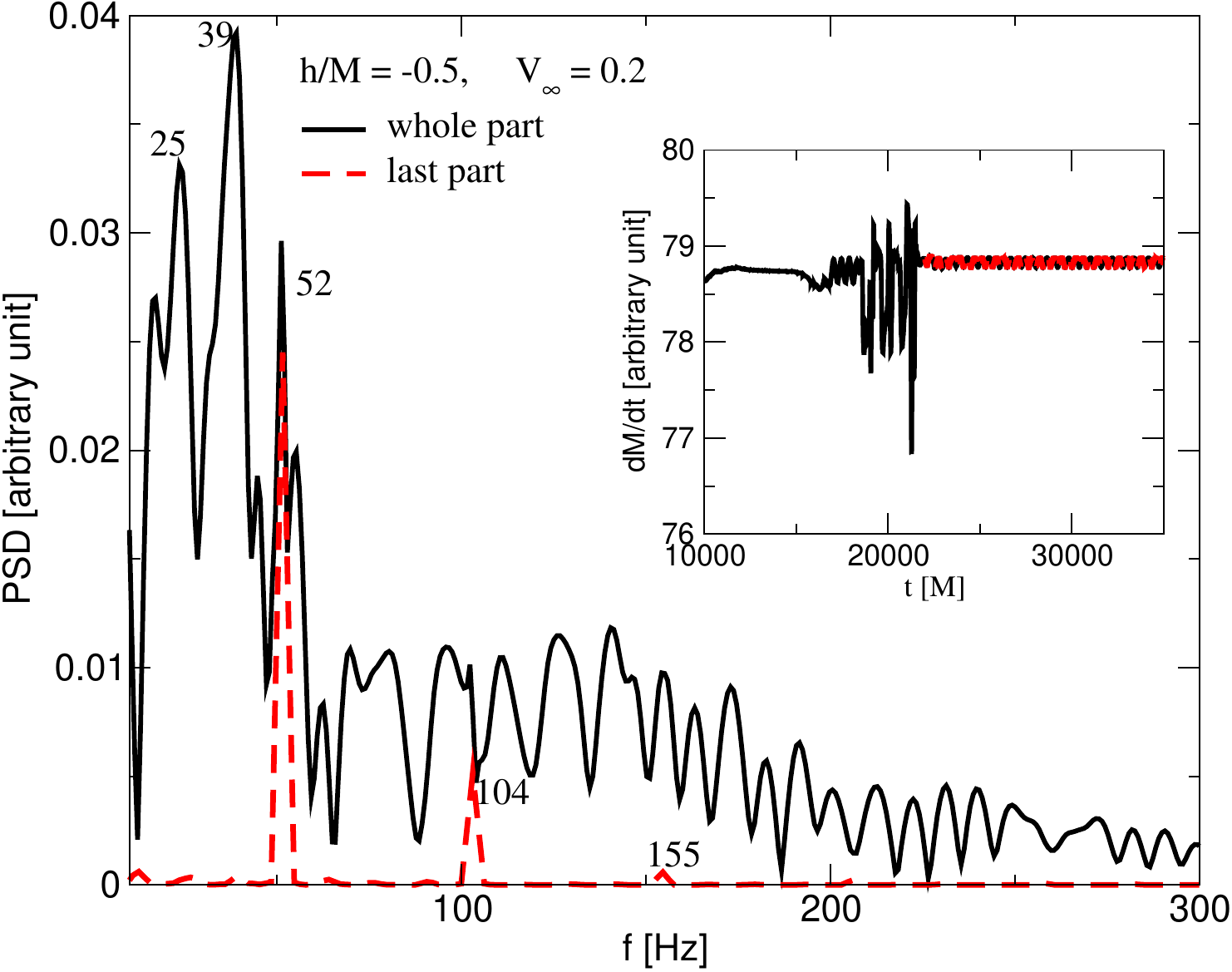,width=7.0cm}\hspace*{0.15cm}    
    \psfig{file=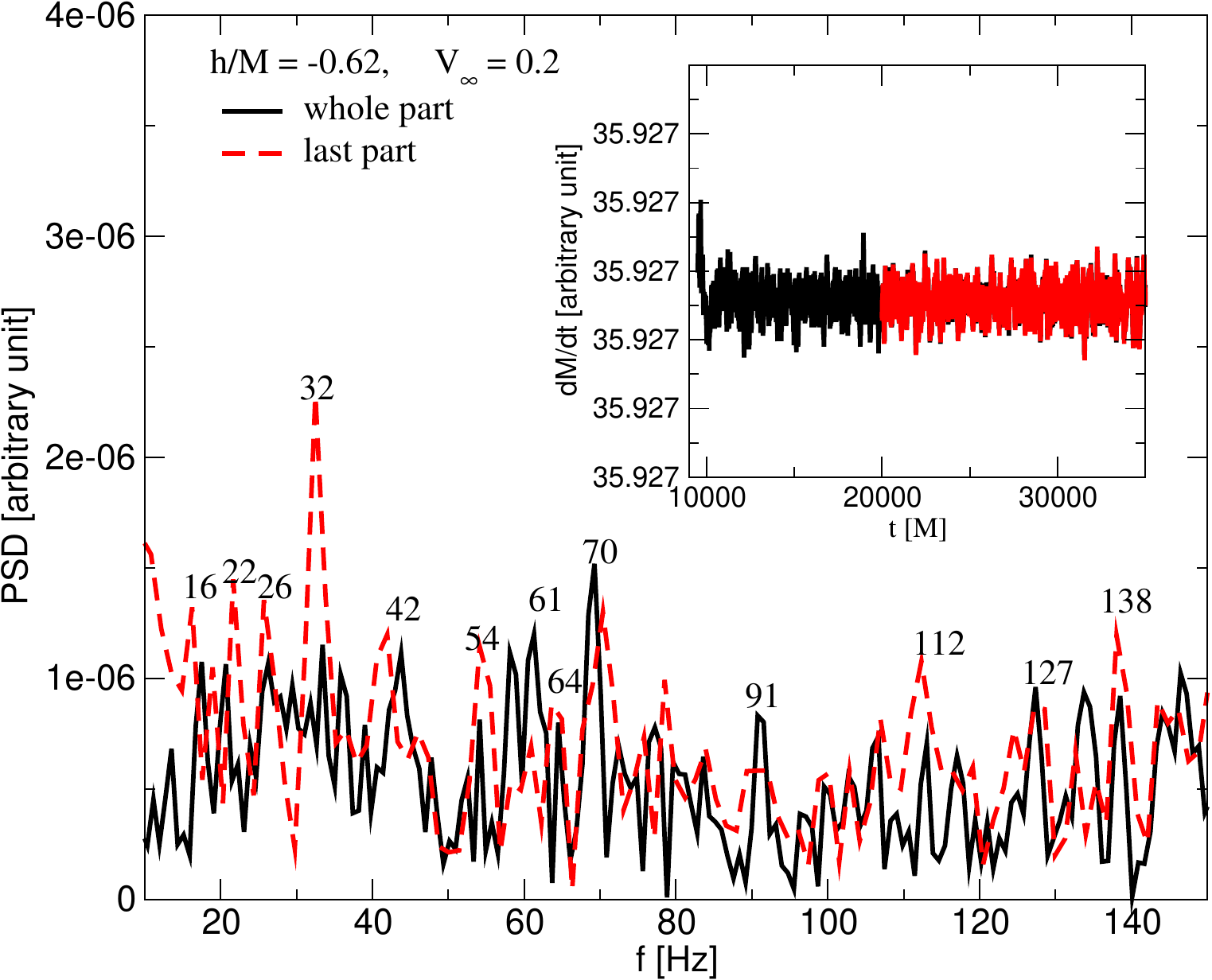,width=7.0cm} 
    \caption{For different scalar hair parameters, and in the case of $V_{\infty}/c=0.2$, PSD analyses have been computed from the mass accretion rates. The mass accretion rates for each $h/M$ value are presented in the graphs within the PSD analyses. Two different PSD analyses have been conducted. The PSD spectrum represented by the solid black line shows the QPO frequencies that occurred from the time the shocks stabilized until $\rm{t_{max}}=35000M$, while the dashed red line represents the QPO frequencies occurring from $t=20000M$ to $\rm{t_{max}}=35000M$. This approach has helped determine whether the QPOs are transient or persistent.
    }
\vspace{1cm}
\label{PSDV02_1}
\end{figure*}

\begin{figure*}
  \vspace{1cm}
  \center
    \psfig{file=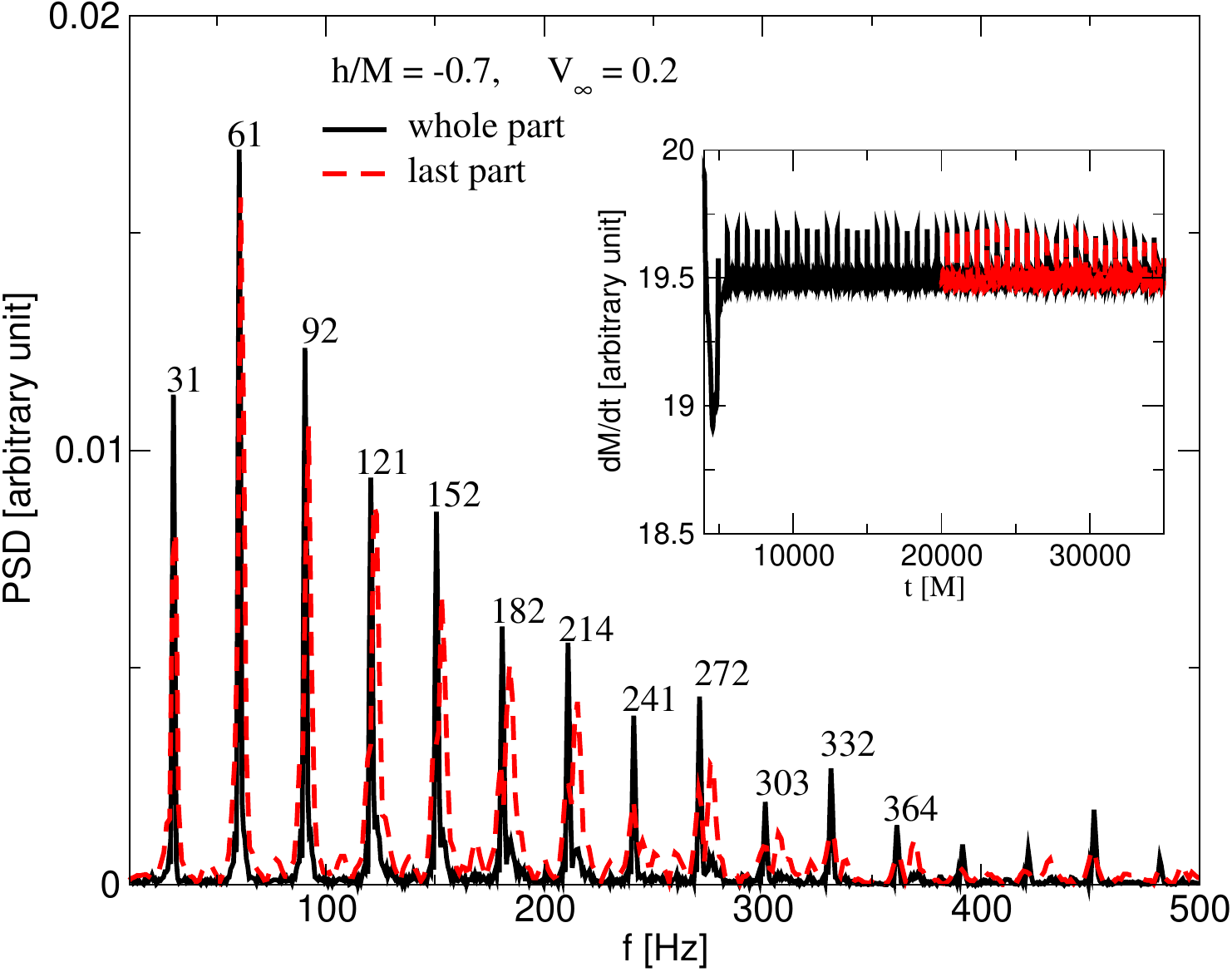,width=7.0cm}\hspace*{0.15cm}
    \psfig{file=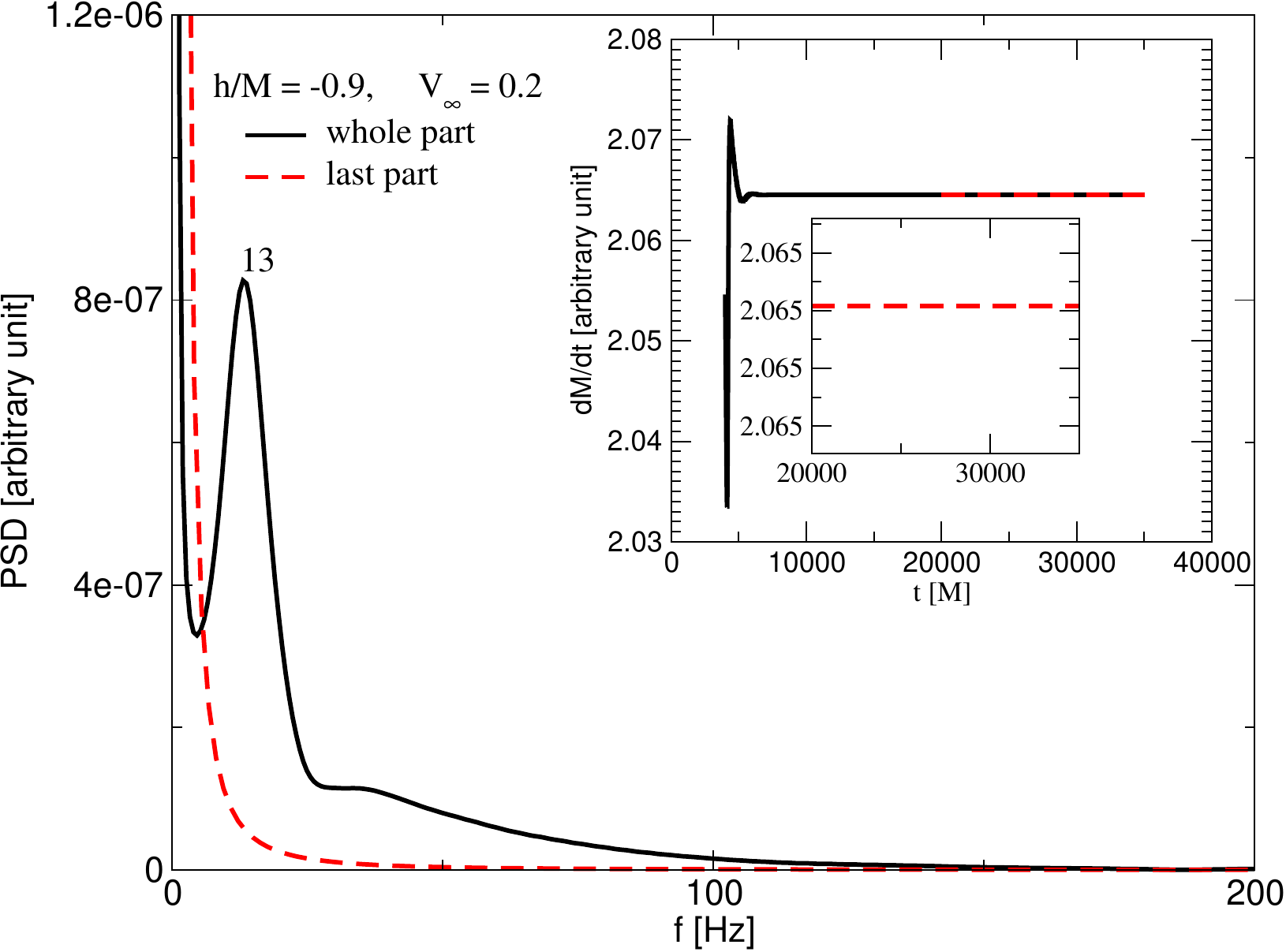,width=7.2cm} \\
    \vspace*{0.5cm}
    \psfig{file=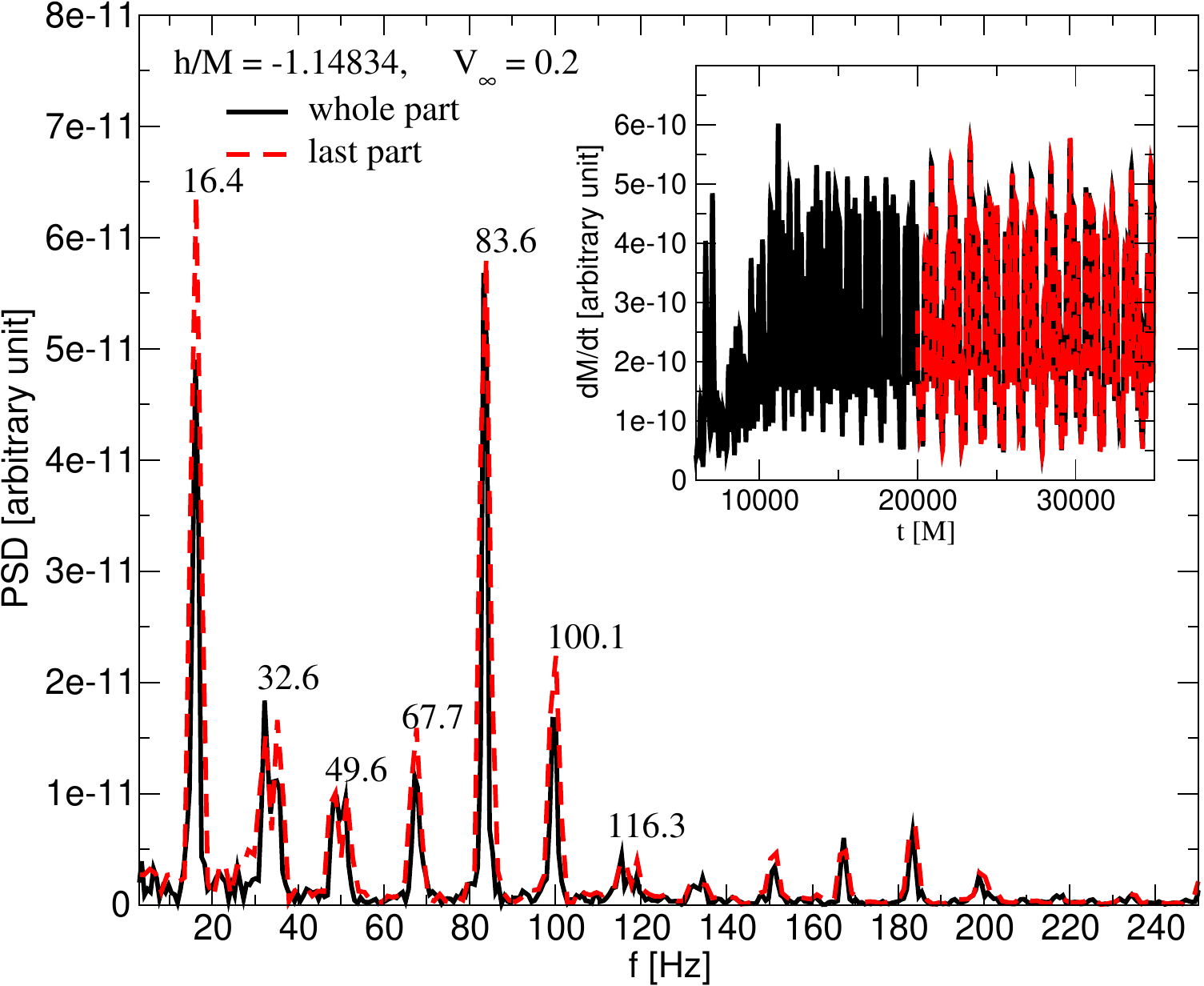,width=7.0cm}\hspace*{0.15cm}
    \psfig{file=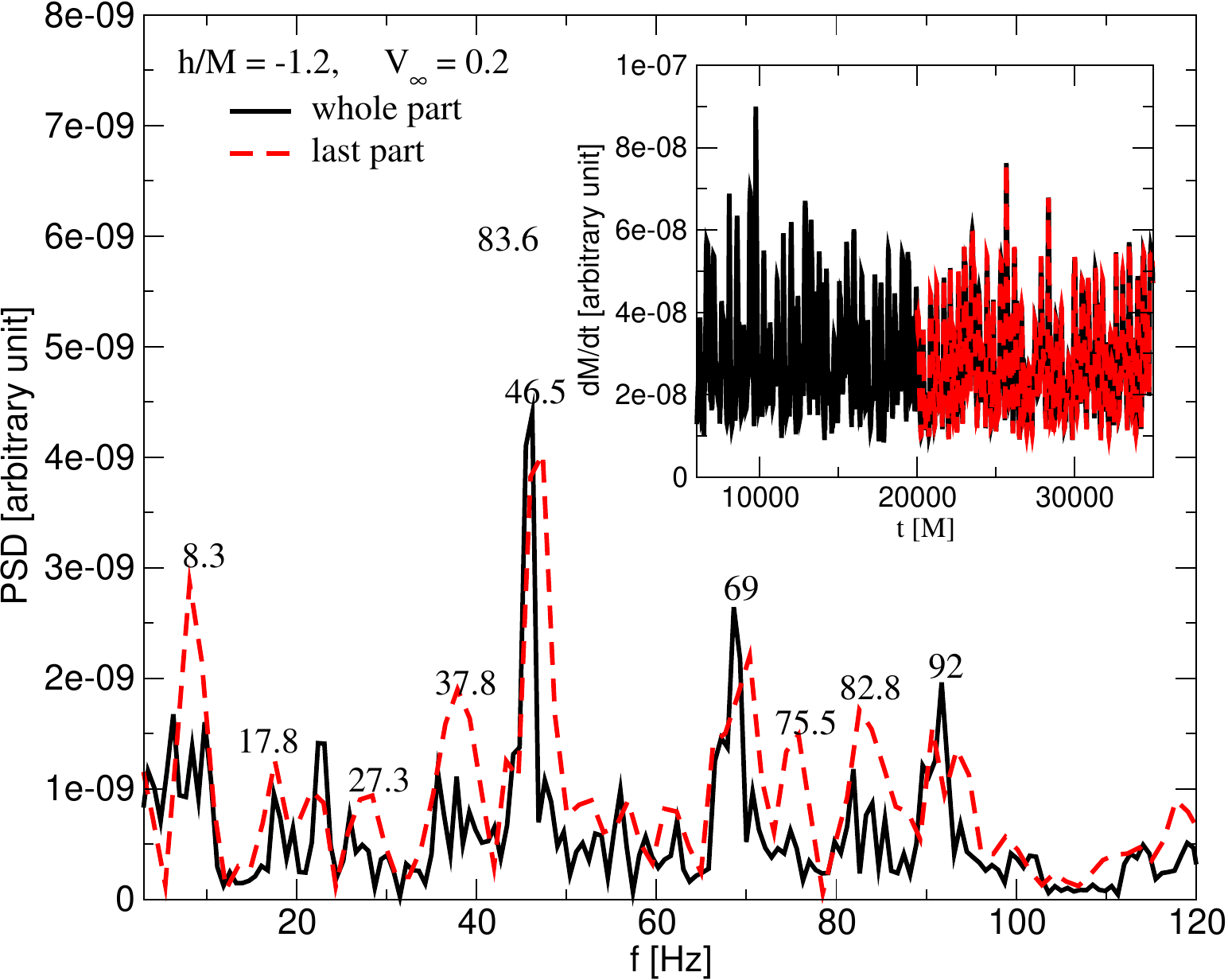,width=7.2cm} 
    \caption{Same as Fig. \ref{PSDV02_1}, but for smaller values of $h/M$.
}
\vspace{1cm}
\label{PSDV02_2}
\end{figure*}

The physical structure of the shock cone formed by the BHL accretion around the Horndeski black hole with a scalar hair parameter undergoes a significant change, as previously explained. This change results from an increase in the absolute value of the hair parameter $h/M$, leading to a stronger scalar field that modifies the spacetime around the black hole. This modification significantly affects the accretion mechanism around the black hole, as discussed earlier. Changes in the structure of the shock cone can lead to alterations or complete disappearance of the frequencies of pressure modes trapped within the cones. In Figs. \ref{PSDV02_1} and \ref{PSDV02_2}, the results of PSD analyses are presented for different hair parameter scenarios, calculated from the mass accretion rate ($\frac{dM}{dt}$) at the inner radius of the computational domain, $r=2.3M$. The mass accretion rates for which these QPOs are calculated are also presented within the same figures. It is crucial to determine whether the QPOs obtained from the numerical models represent continuously generated modes or temporary modes. The persistent modes from numerical calculations can then be compared with observations, aiding in the understanding of the physical properties of the black hole at the center. To identify continuously generated modes, two PSD analyses are conducted in each case: the solid black line represents the mass accretion rate from the shock cone reaching the steady-state, up to $\rm{t_{max}}=35000M$, while the dashed red line covers the interval from $20000M$ to $\rm{t_{max}}$. The times to reach a steady state for each model are given in Table \ref{Inital_Con}. Upon examining Figs. \ref{PSDV02_1} and \ref{PSDV02_2}, we observe that the frequency modes, across various values of $h/M$, consistently persist.

Before delving into detailed explanations, it is necessary to understand the nomenclature, coloring, and symbols used in Figs. \ref{PSDV02_1}, \ref{PSDV02_2}, and \ref{PSDV04_1} to make the graphs more comprehensible. The larger graphs display the PSD analyses for each $h/M$ case, while the inset graphs show the variation in the mass accretion rate on which the PSD is conducted. As seen in most of these figures, two different graphs have been drawn: the solid black line represents the behavior of the material falling into the black hole from the time it reaches the steady-state until $\rm{t_{max}}=35000M$, while the dashed red line shows the behavior from $t=20000M$ to $\rm{t_{max}}=35000M$ again. 

The top-left PSD graph in Fig.\ref{PSDV02_1} shows the frequencies of pressure modes trapped inside the shock cone around the Schwarzschild black hole. As seen here, the PSD has many rich peaks and exhibits a highly chaotic oscillation. In this model, all modes have formed except for the $f_{bow}$ mode. Because, as seen in Fig.\ref{colV02_1}, in the rest-mass density graph, a bow shock has not formed. Thus, the QPO frequency that could occur due to this shock has not emerged. The peaks represented by the dashed red line are all persistents because they have been observed in both short and long-term oscillations. The first peaks seen in this top-left graph at $8$, $24$, and $30$ Hz are the fundamental modes $f_{sh}$, $f_{eh}$, and $f_h$. The others are entirely due to the nonlinear couplings formed either within these modes themselves or with each other \cite{Donmez6, Landau1976, Zanotti2005MNRAS}. For example, these three fundamental modes, by coupling, have generated the peak occurring at $62$ Hz. Furthermore, even in the same graph, the occurrence at $16$ Hz is the double frequency of the $8$ Hz fundamental mode, indicating nonlinear coupling. In this model, both low- and high-frequency QPO formations have been observed.

As explained above, the hair parameter significantly influences the formation mechanism of the shock cone. Therefore, in the PSD analyses presented in Figs. \ref{PSDV02_1} and \ref{PSDV04_1}, it has been observed that both the fundamental modes and their nonlinear couplings undergo significant changes. While no substantial difference is noted in the rest-mass density graph in Fig. \ref{colV02_1} when comparing $h/M=-0.2$ with $h/M=0$, the top right graph of Fig. \ref{PSDV02_1} reveals that the PSD analysis at $h/M=-0.2$ produces very different QPO behaviors and frequencies. In the $h/M=-0.2$ case, as seen in the inner graph, after reaching the steady state, the accreted matter around the black hole and the formed shock cone exhibit a different instability, resulting in QPOs that are present in the solid black line but not in the dashed red line. Consequently, the initially produced QPOs are transient. Specifically, the QPOs represented by the solid black line, calculated until the system reaches the steady state and until the end of the simulation based on the mass accretion rate, are temporary. In comparison to the permanent QPOs, represented by the dashed red line, only a fundamental mode at $30.5$ Hz is observed, with $61$ Hz and $91$ Hz as its linear couplings.

In the $h/M=-0.3$ case of Fig. \ref{PSDV02_1}, the influence of the scalar field is strongly felt. Due to the narrowing of the opening angle of the shock cone and the approach of the stagnation point inside the cone to the black hole horizon, as seen in Fig. \ref{Stag_1}, changes in the frequencies observed in the Schwarzschild case have been noted. According to our estimations, in the $h/M=-0.3$ PSD analysis shown in the middle left of Fig. \ref{PSDV02_1}, the fundamental modes $f_{sh}$, $f_{eh}$, and $f_h$ occur at $12.4$, $26$, and $50.5$ Hz, respectively. Again, as seen in Fig. \ref{colV02_1}, no bow shock has formed, and therefore, the frequency that could have been generated by it has not been observed. Due to the modification of spacetime by the scalar hair parameter, we can assert that the frequencies shift from low to high. Similarly, other observed frequencies result from nonlinear couplings, for example, $38$ Hz could be three times $12.4$ Hz. Unlike in the Schwarzschild case, the frequencies in the $h/M=-0.3$ case are more distinct, regular, consecutive, and persistent, as all the frequencies given in the solid and dashed line scenarios have the same values. This can be attributed to a moderate scalar field defined by the hair parameter, leading to a reduction in the opening angle of the cone and a closer approach of the stagnation point to the black hole horizon. However, it is evident that as the scalar field value increases, the potential barrier becomes larger, affecting the state of matter falling due to BHL accretion and causing a significant change in the structure of the shock cone formed in the downstream region.

In the $h/M=-0.4$ case, as seen in Fig. \ref{colV02_1}, unlike the other three models, a less intense bow shock has formed in addition to the shock cone. Consequently, the time for the mass accretion and the forming shocks around the black hole to stabilize has extended to approximately $t=17000M$. As shown in the innermost drawing of the middle-right graph in Fig. \ref{PSDV02_1}, the formed shock mechanisms have not reached a fully stable state by $\rm{t_{max}}=35000M$. Therefore, to form a more definitive conclusion about the QPOs generated at $h/M=-0.4$, the model must be run for a much longer period. Ignoring this detail, in Fig. \ref{PSDV02_1}, the QPO frequencies found at $h/M=-0.4$ are represented as solid black lines and dashed red lines. Due to the matter not reaching a completely steady-state around the black hole, it has been shown that some of the low-frequency ( below $30$ Hz) and most of the high-frequency QPOs seen in the solid black line are transient. However, even though we cannot assert that the QPOs observed in the dashed red line are definitively permanent, the fact that they are observed in both PSD analyses strengthens the possibility of their persistence. Specifically, the frequencies of $5.1$, $8.1$, $10.6$, and $17.6$ Hz observed in the dashed red line are fundamental modes that occur in both situations. Unlike in the $h/M>-0.4$ scenarios, an additional fundamental mode has emerged here due to the bow shock. The other modes that form besides the fundamental modes are nonlinear couplings of the fundamental modes. For example, $20.4 \sim 2 \times 10.6$ and $25.8 \sim 2 \times 10.6 + 5.1$. The differences that occur are entirely due to numerical errors.  Theoretically, all pressure-based QPO modes that could arise from BHL accretion have been observed in this model. Nevertheless, nonlinear coupling frequencies observed in the dashed red line are weak in magnitude, likely because these couplings necessitate running the code for the $h/M=-0.4$ model for a longer duration to become more pronounced.

The bottom-left graph of Fig. \ref{PSDV02_1} displays the $h/M=-0.5$ condition, showing the mass accretion and PSD analysis. In this scenario, the intensity of the scalar field has increased significantly, greatly affecting the BHL mechanism, as previously discussed. Additionally, a significant change in the structure of the shock cone is observed. Compared to the $h/M>-0.4$ cases, the cone around the black hole reached a steady state later, and even after reaching stability, it exhibited instability between $t=17000M$ and $t=22000M$, as evident in the mass accretion graph in Fig. \ref{PSDV02_1}. The PSD analyses reveal that the frequencies obtained from long-term mass accretion (represented by the solid black line) are not very clear, due to the instability that occurs after the shock cone reaches the steady-state, significantly impacting the fundamental modes. Consequently, the numerical observability of the fundamental modes becomes challenging. However, the PSD analysis indicated by the dashed red line shows that only one fundamental mode occurs, at $52$ Hz. Based on our experiments, we estimate that this frequency corresponds to $f_{sh}$, because as long as the shock cone exists, the azimuthal fundamental mode would always be present. However, the $f_h$ and $f_{eh}$ modes are not excited due to the closer approach of the stagnation point to the black hole horizon. In the case of $h/M=-0.5$, as seen on the far left of Fig. \ref{colV02_1}, the presence of a bow shock is evident. Due to the complete displacement of this shock away from the black hole, meaning it does not occur in the strong gravitational field, and its significant loss of intensity, it has not been numerically observed that the bow shock excites any mode. As previously mentioned, only a fundamental mode at $52$ Hz and its linear couplings, resulting in peaks at $104$ and $155$ Hz, have been observed. We believe that in the case of $h/M=-0.5$, the distinctive behavior of QPO formation, as compared to previous and subsequent $h/M$ values, may be attributed to the bow shock formed at $h/M=-0.4$ being pushed far from the black hole, nearly exiting the computational domain. We hypothesize that the presence of the bow shock within the computational domain causes a shift in the resulting QPO modes to a different phase, thus leading to the distinct behavior observed.

So far, QPO analyses have shown that different values of $h/M$ cause significant changes in QPOs. As expected, since the potential of the scalar field modifies the spacetime around the Schwarzschild black hole, different shock mechanisms and QPOs have been observed. The motion of a test particle around the non-rotating Horndeski black hole has been theoretically investigated, with initial QPO oscillations presented by Ref.\cite{Rayimbaev2021}. The calculations have found that at $h/M=-0.62$, the ISCO occurs at a minimum point, namely at $r=5.785M$. In the $h/M=-0.62$ scenario, we specifically modeled the $h/M=-0.62$ case to uncover the behavior of BHL accretion and the resulting QPOs. As seen in the bottom right of Fig. \ref{PSDV02_1},  the observed instability, while exhibiting QPO behaviors similar to the Schwarzschild case, has produced peaks with much lower amplitude. In the case of $h/M=-0.62$, if we assume that the oscillation induces QPO modes, they occur at $22$, $26$, and $32$ Hz. Then, in this strong gravitational field due to strong non-linear oscillations, other new peaks arise from nonlinear couplings. As seen in Fig.\ref{colV02_1} for the case of $h/M=-0.62$, the formed shock cone and the widening bow shock angle do not possess sufficient intensity. Therefore, the QPO amplitudes in this case are much lower than those obtained for $h/M>-0.5$. Consequently, it is not very feasible to distinguish whether these peaks are genuinely induced modes or just background noise. Therefore, the likelihood of observing these peaks with detectors is very low.

In the $h/M=-0.7$ case, depicted in the top-left of Fig. \ref{PSDV02_2} where the scalar field is getting stronger, it is observed that due to the narrowing of the opening angle of the cone and the closer approach of the stagnation point to the black hole horizon (Fig. \ref{Stag_1}), the $f_{eh}$ and $f_h$ modes completely disappear, leaving only the azimuthally oscillating pressure mode. As expected, the frequencies of this mode have shifted towards higher frequencies due to the reduction in the opening angle, with $f_{sh} = 31$ Hz forming the fundamental mode. The other peaks, as clearly seen, result from the linear coupling of this fundamental mode, indicating that the observed fundamental mode and its various couplings are all persistent.

In Fig. \ref{colV02_2}, the $h/M=-0.9$ case shows a significant change in the structure of the shock cone compared to all models in Fig. \ref{colV02_1}, resulting in the complete ejection of the bow shock. Once the bow shock disappears, the material in the region of the shock cone, including the material inside the cone, is expelled from the computational domain. This is because as $h/M$ decreases, the intensity of the scalar potential increases, preventing the material from falling into the black hole, leading to a continuous and significant change in the physical mechanisms around the black hole, as detailed in Section \ref{Result1}. The $h/M=-0.9$ case in Fig. \ref{colV02_2}, compared to other values of $h/M$ in our models, indicates the transition parameter between the complete dumping of the shock cone into the black hole and its expulsion primarily out of the computational domain, followed by cavity formation. As seen in the mass accretion rate and PSD analysis graphs in the upper right of Fig. \ref{PSDV02_2}, no persistent QPO frequency has emerged in the $h/M=-0.9$ case, because the material and shock cone falling towards the black hole have reached a strong stability state.

For the non-rotating Horndeski black hole, the motion of a test particle has been found to be the same as the ISCO Schwarzschild solution for $h/M=-1.14834$ \cite{Rayimbaev2021}. Therefore, it has been demonstrated that the QPOs of the test particle in this scalar hair parameter mimic the Schwarzschild solution. On the other hand, in our numerical study, for $h/M=-1.14834$, the model showes that the scalar potential around the Horndeski black hole not only prevents the matter falling towards the black hole due to BHL accretion but also lead to the complete disappearance of the shock cone and the formation of the cavity in that region. For $h/M=-1.14834$, as observed in numerical simulation, the QPO frequencies generated by different physical mechanisms and their excitations are different, leading to the conclusion that it does not produce results similar to the Schwarzschild solution as expected in the theoretical study of the test particle  \cite{Rayimbaev2021}. This is because the QPO frequencies at $h/M=0$ and their behaviors are entirely different from those at $h/M=-1.14834$, as seen in Figs.\ref{PSDV02_1} and \ref{PSDV02_2}. As seen in Fig.\ref{colV02_2} for  $h/M=-1.14834$, the increasing scalar hair parameter not only causes the matter flowing towards the black hole to be pushed away from it, but also leads to the complete disappearance of the shock cone. The matter inside the cone is predominantly sent away from the black hole, while some falls towards it. Thus, in the region where the shock cone existed, a cavity forms. As seen in the bottom-left graph of Fig.\ref{PSDV02_2} for $h/M=-1.14834$, by exciting the trapped mode within the cavity, it generates QPO frequencies. Only one geniue mode is excited which is $16.4$ Hz. Similarly, the other peaks are formed as a result of linear coupling at $16.4$ Hz.

In the case of $h/M=-1.2$, as seen in Figs. \ref{colV02_2} and \ref{PSDV02_2}, we observe how the enlargement of the cavity affects the change in QPO frequencies. For $h/M=-1.2$, Fig. \ref{PSDV02_2} models the excitation of the mode trapped within the cavity and its nonlinear couplings. The fundamental frequency occurs at $8.3$ Hz, summarized in Table \ref{geniue_mode}, with others forming as multiples of this $8.3$ Hz. That is, a single fundamental mode has formed, and other QPO frequencies have arisen due to nonlinear couplings. At $h/M=-1.2$, the QPO frequencies and their behavior exhibit some differences compared to $h/M=-1.14834$, which can be attributed to the larger cavity size in the $h/M=-1.2$ case, as also seen in Table \ref{Inital_Con}, where the stagnation point is closer to the outer boundary of the computation domain. This is because as $h/M$ decreases, it is predicted that the cavity will transform into a channel extending from the horizon to the outer boundary of the calculation domain.

Finally, we discuss the effect of the cavity in generating QPOs. In Fig. \ref{PSDV02_2}, as seen in the cases of $h/M=-1.14834$ and $h=-1.2$, only a single fundamental mode has occurred, which is distinct from the modes we have discussed earlier and has been formed as a result of the excitation of the sound wave trapped entirely within the cavity. Other modes, such as $f_{sh}$, $f_h$, $f_{eh}$, and $f_{bow}$, have not formed because the physical mechanisms that produce them have completely vanished. Or if they have formed, they have not been observed here because the frequency generated by the azimuthal oscillation modes of the sound wave trapped inside the cavity has a dominant magnitude compared to them.

%%%%%%%%%%%%%%%%%%%%%%%%%%%%%%%%%%%%%%%%%%%%%%%%%%%%%%%%%%%%%%%%%%%%%%%%%%%%%%%%%%%
\subsection{The Case of $V_{\infty}/c=0.4$ and Comparison with  $V_{\infty}/c=0.2$ }
\label{QPOs3}

\begin{figure*}
  \vspace{1cm}
  \center
  \psfig{file=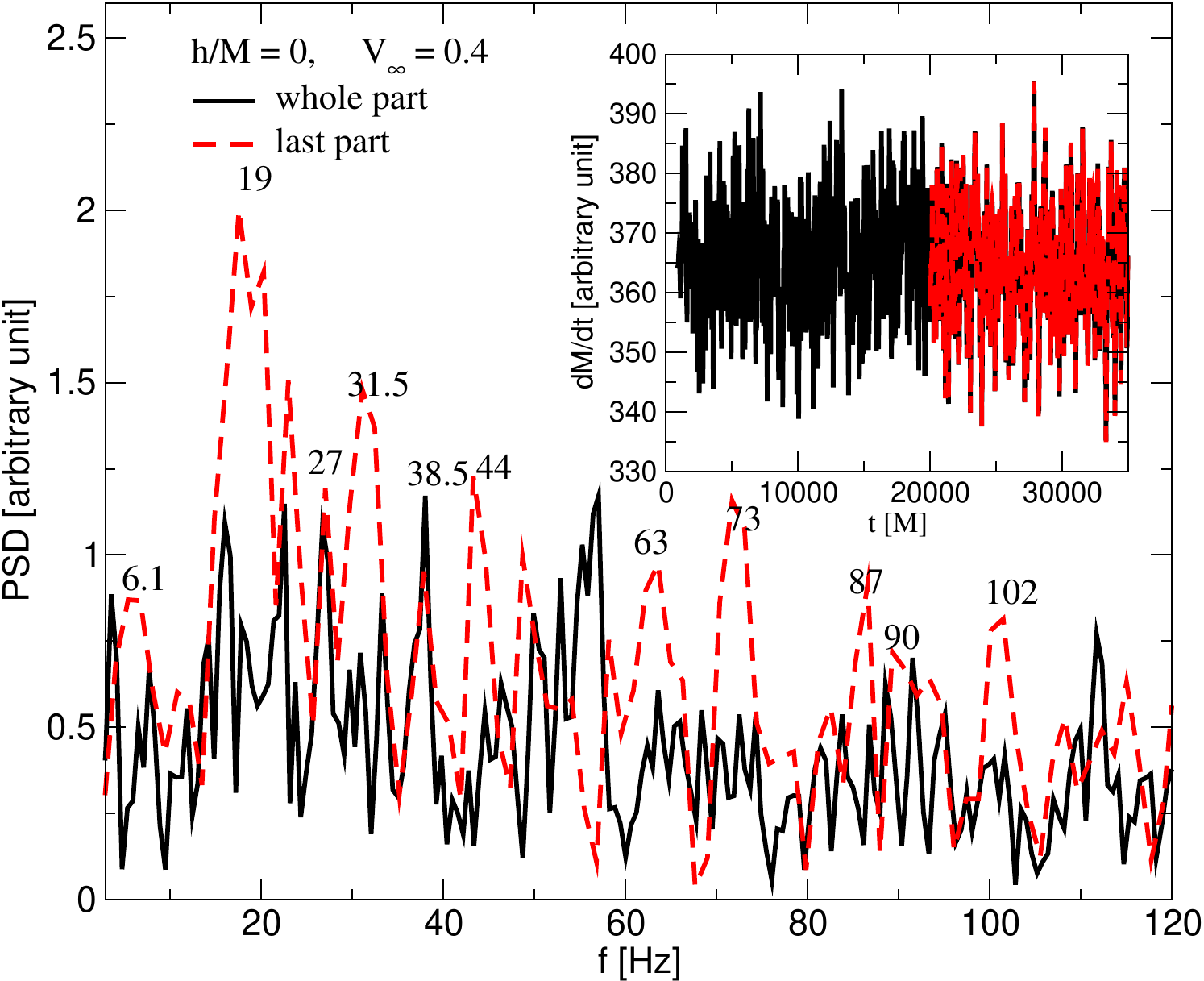,width=9.0cm, height=7.0cm} \\
      \vspace*{0.5cm}
   \psfig{file=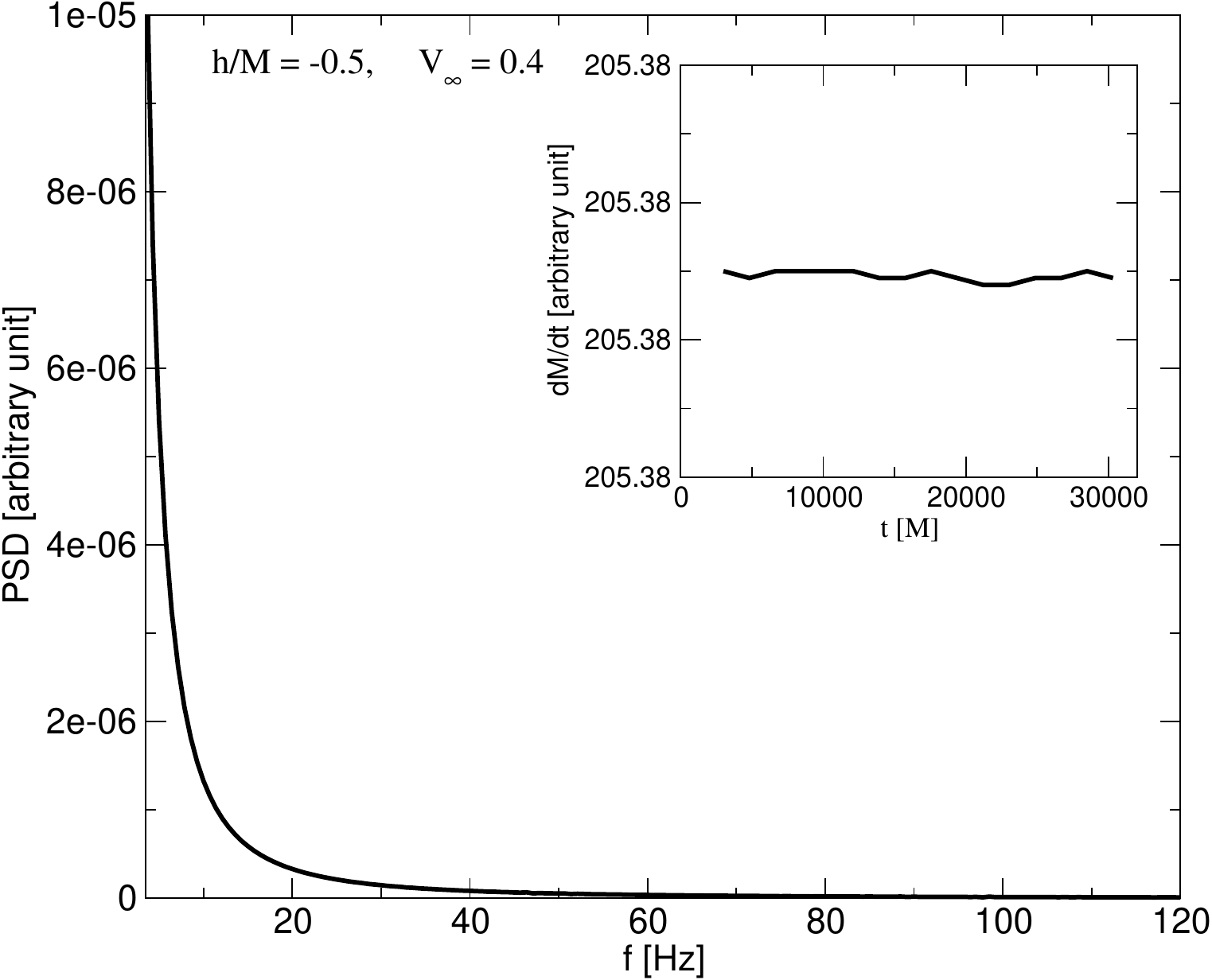,width=9.0cm, height=7.0cm} \\
      \vspace*{0.5cm}
  \psfig{file=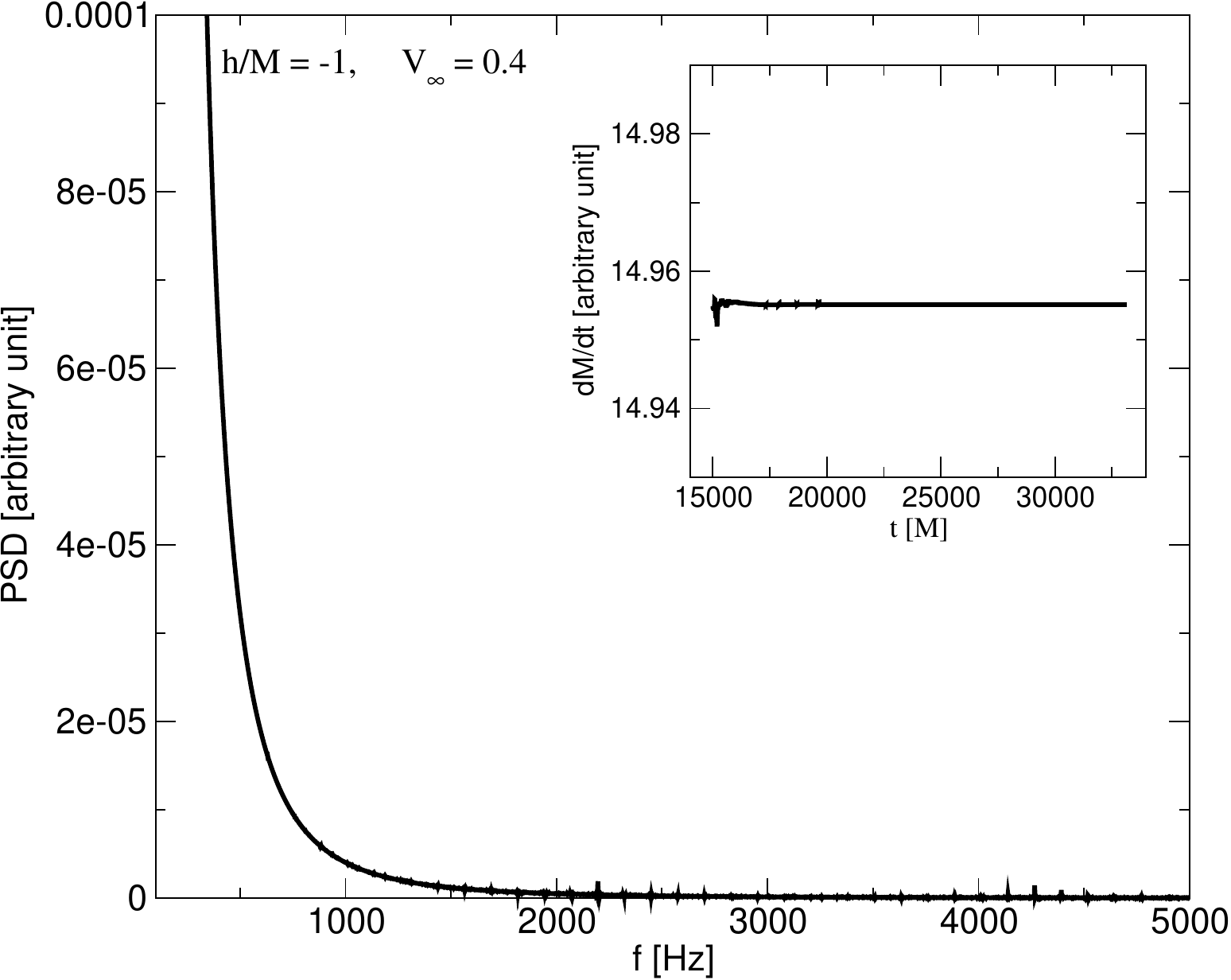,width=9.0cm, height=7.0cm}
    \caption{Same as Fig. \ref{PSDV02_1}, but for $V_{\infty}/c=0.4$.
}
\vspace{1cm}
\label{PSDV04_1}
\end{figure*}

Unlike Section \ref{QPOs2}, in this section, we perform QPO analysis at an asymptotic speed of $V_{\infty}/c=0.4$, but only for three different scalar hair parameters. We reveal the QPOs likely to be excited by physical mechanisms in response to this speed for the matter sent from the upstream region, which may form a shock cone, bow shock, and cavity around the black hole. As numerically observed in Fig. \ref{colV04_1}, the shock cone has formed in $h/M=0$, and the PSD analysis has been conducted using the mass accretion rate given in Fig. \ref{PSDV04_1}. This mass accretion rate in Fig. \ref{PSDV04_1} shows instability, leading to a rich set of QPO frequencies. As seen in the PSD analysis, most frequencies appear to be persistent. The analysis indicates, as summarized in Table \ref{geniue_mode}, that fundamental modes at $19$, $27$, and $31.5$ Hz have been established, with other peaks resulting from the nonlinear coupling of these fundamental modes in different combinations. On the other hand, although LFQPOs and HFQPOs are observed for $V_{\infty}/c=0.4$ with $h/M=0$, given the LFQPOs observed in the source discussed in Section \ref{QPOs4}, we can suggest that the $V_{\infty}/c=0.2$ case generates QPOs capable of explaining the observational results.

For the same $V_{\infty}/c=0.4$, but now at $h/M=-0.5$, we conducted a PSD analysis. Although the structure of the shock cone around the black hole and the contour plot appear dynamically similar to the $h/M=0$ case, as seen in Fig.\ref{colV04_1}, the PSD analysis in Fig.\ref{PSDV04_1} reveals that at $h/M=-0.5$, the matter reached a steady state and the formed shock cone did not exhibit any instability throughout the simulation, resulting in the absence of observed QPO frequencies. This situation is also summarized in Table \ref{geniue_mode}. As observed in the color distribution in Fig. \ref{colV04_1}, the only difference is that while the matter is densely gathered near the horizon of the black hole at $h/M=0$, at $h=0.5$, it is spread over a wider area around the black hole. Due to this distribution, the shock cone may have achieved strong stability.

In the case of the same asymptotic speed, we performed our PSD analysis at $h/M=-1$. As seen in Fig.\ref{colV04_1}, despite the complex physical mechanisms formed around the black hole, no instability in the mass accretion rate is observed, as shown in Fig.\ref{PSDV04_1}. Therefore, similar to the $h/M=-0.5$ case, no QPO frequencies are observed. Consequently, according to our initial model setups, the physical mechanisms and the resultant QPOs at $V_{\infty}/c=0.2$ are such that they can explain the observational results. This issue is discussed in detail in Sections \ref{QPOs4} and \ref{physical_mech}.

As a result, with the increase in asymptotic speed, changes in the nature of the accretion mechanism are observed, as well as differences in the oscillation characteristics. Indeed, this is an expected situation. When Fig.\ref{colV04_1} is compared with Figs.\ref{colV02_1} and \ref{colV02_2}, it is observed that the accreting mass reaches the black hole more quickly, creating physical mechanisms like the shock cone, and also, the formed bow shock appears more intense. These changes would undoubtedly affect how the formed mechanisms excite QPO modes. The results in the $V_{\infty}/c=0.4$ case have shown that in Horndeski gravity, no instability has occurred around the black hole. However, it is not possible to draw a general conclusion from the existing model results. Therefore, it is necessary to increase the number of models for a broader range of $h/M$ parameters to determine whether QPOs form in the $V_{\infty}/c=0.4$ case.

%%%%%%%%%%%%%%%%%%%%%%%%%%%%%%%%%%%%%%%%%%%%%%%%%%%%%%%%%%%%%%%%%%%%%%%%%%%%%%%%%%%%%%%%%%%%%
%%%%%%%%%%%%%%%%%%%%%%%%%%%%%%%%%%%%%%%%%%%%%%%%%%%%%%%%%%%%%%%%%%%%%%%%%%%%%%%%%%%%%%%%%%%%%%%%%%%%%%%%%%%%%
\section{Comparison with Observational Data: GRS 1915+105}
\label{QPOs4}

In numerical calculations, the instabilities that appeared after the systems reached the steady-state can be analyzed with PSD analyses, which provide information about the resolution of the calculated frequencies and whether these frequencies are persistent or transient. Therefore, in numerical simulations, models must be run for a sufficient duration to achieve this resolution. Thus, the obtained QPO frequencies can be compared with observational QPOs to make predictions about the physical properties of different astrophysical sources. In this study, each model was run up to $\rm{t_{max}}=35000M$. Considering the locations of the frequencies that occurred for the GRS 1915+105 source, as summarized in Table 3 and graphically shown in Figures 6 and 8 of Ref.\cite{Dhaka_2023}, and in Table 2 and graphically shown in Figure 4 of Ref.\cite{Liu_2021}, we believe that the maximum duration in numerical calculations is sufficient for the frequencies to be persistent. According to the results in Ref.\cite{Liu_2021, Dhaka_2023}, while the HFQPOs occur around $r=3.38M$, the situation for the LFQPOs takes place between the black hole horizon and $r=36.3M$. Considering the running time of each model, while the Keplerian period $T=13.79M$ at $r=3.38M$, it is $T=1373M$ at $r=36.3M$. Therefore, based on the total running time of each model, at $r=3.38M$ the matter orbits around the black hole approximately 2538 times, and 26 times at $r=36.3M$. This demonstrates that the resolution of the numerically obtained frequencies is sufficient and that the frequencies are persistent.

GRS 1915+105, a binary system that emits uniquely and violently varying X-ray radiation, contains a black hole at its center. Observational data indicate that this system has a complex accretion mechanism. Particularly, due to its ability to produce significant changes in X-ray luminosities within a few seconds, it has been accepted as an important source in spectral studies in the literature. The mass of the black hole at the center of this binary system is $12.4M_{\odot}$ \cite{Reid2014}. According to the source, 14 different X-ray classes have been observed based on the X-ray flux, color-color diagram, and hardness ratio \cite{Klein-WoltMNRAS2002, Hannikainen2005A&A}. From the observed X-ray data of this source, QPO behaviors have been revealed through Fourier analyses. As a result of the PSD analyses, it has been found that it exhibits different behaviors and has produced sharp peaks \cite{Belloni_2002}. The QPO frequencies of this source range from approximately mHz to 74 Hz \cite{Belloni2013MNRAS, Sreehari_2020}. In certain spectral states and transitions, the centroid frequencies of these QPOs are known to be related to the physical mechanisms occurring around the black hole. The frequency of LFQPOs is less than approximately 30 Hz, while the centroid frequency of HFQPOs is greater than approximately 60 Hz, and HFQPOs can reach several hundred Hz \cite{Belloni2013MNRAS, Majumder_2022, Sreehari_2020, Dhaka_2023, Belloni_2009, Motta2023}.

It is numerically observed that the shock cones formed around the non-rotating Horndeski black hole, due to BHL accretion, trap and subsequently excite pressure-based QPO modes, as described in Section \ref{QPOs1}. As seen in Figs. \ref{PSDV02_1} and \ref{PSDV02_2}, the numerically observed lowest frequency for a black hole with $M=10M_{\odot}$ is around $8$ Hz. Recalculating this for GRS 1915+105, with a mass of $M=12.4M_{\odot}$, yields a frequency of $6.45$ Hz. Furthermore, in each PSD analysis, the lowest frequency increases as $h/M$ decreases. Considering that LFQPO frequencies are typically below $30$ Hz, the trapping of modes by the shock cone around the black hole contributes to LFQPO formation. However, a comparison of the numerically obtained QPO frequencies with the observational data in Ref.\cite{Liu_2021, Dhaka_2023} reveals inconsistencies. These references indicate that LFQPOs for GRS 1915+105 occur between $2.01$ Hz and $5.41$ Hz. In contrast, the HFQPOs in Ref.\cite{Sreehari_2020, Dhaka_2023} align with the numerical results. Notably, the QPO frequency found in numerical simulations for $h/M=-0.4$ shows the lowest QPO for a black hole with $M=10M_{\odot}$ observed at $5.1$ Hz. When recalculated for GRS 1915+105, it amounts to $4.11$ Hz, consistent with the LFQPO frequencies observed in these studies. Therefore, the formation of the bow shock alongside the shock cone at $h/M=-0.4$ may explain such low frequencies. Thus, the bow shock could be the physical mechanism necessary to explain the LFQPOs occurring around the non-rotating Horndeski black hole. However, as mentioned in Section \ref{QPOs1}, a longer simulation duration is necessary to definitively conclude that the $h/M=-0.4$ case produces these QPOs. Although most low frequencies identified here do not exactly match the LFQPOs, they are close to the upper limit values of the observational results.

\begin{figure*}
  \vspace{1cm}
  \center
     \psfig{file=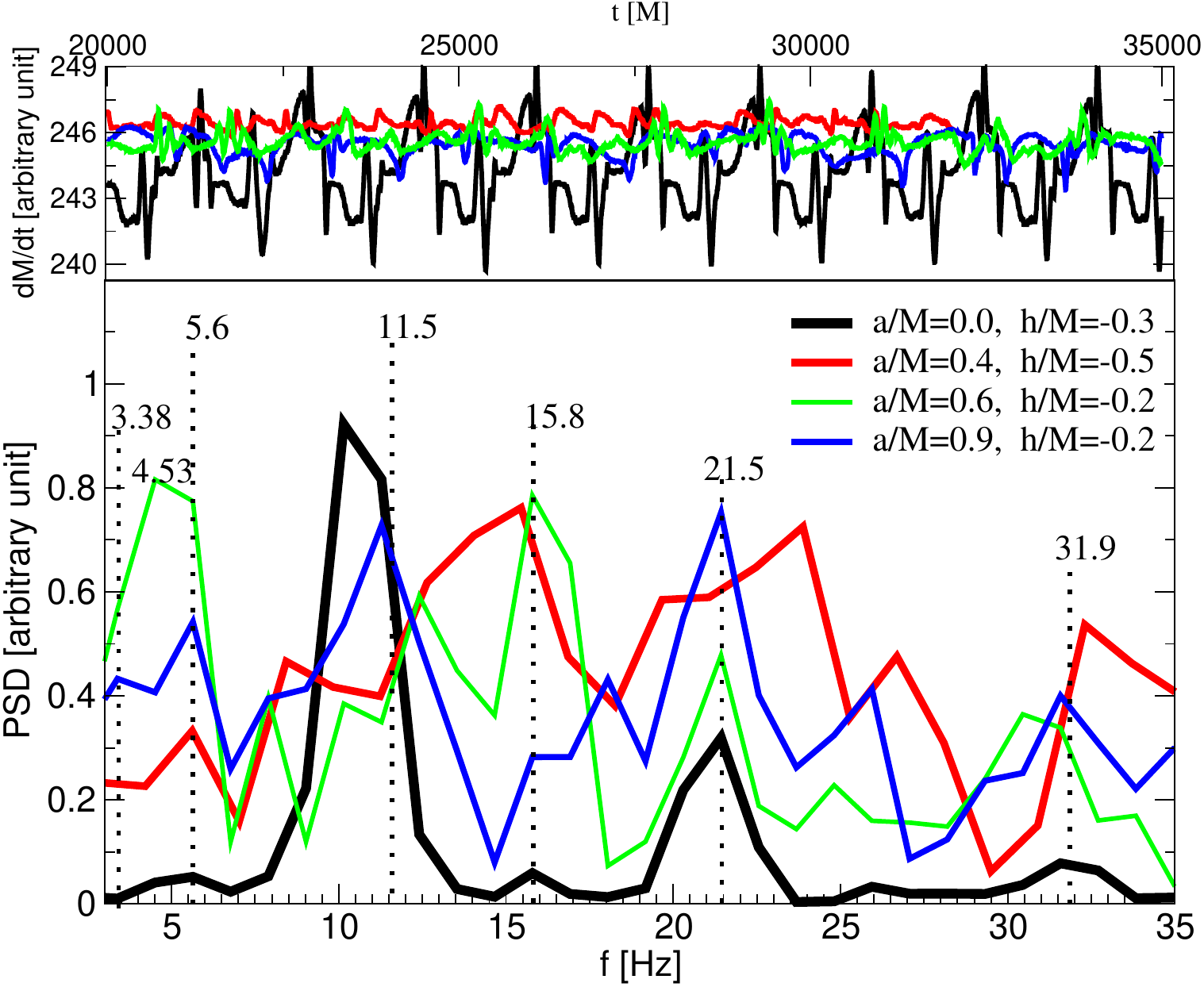,width=14.0cm}
     \caption{In the top row, the mass accretion rate is shown, while the bottom row displays the PSD analyses of the mass accretion rate for $M=12.4M_{\odot}$.
  The amplitudes of the peaks obtained from the PSD analyses of different parameters are different. To compare them, those with lower amplitudes are multiplied by a certain factor to achieve similar amplitude values. Thus, QPO frequencies under different $h/M$ and $a/M$ could be compared more easily.
QPO modes trapped within the shock cone around both the rotating and non-rotating hairy Horndeski black holes are shown. The results for the rotating Horndeski black hole from Ref.\cite{Donmez2024arXiv240216707D} are compared with those for the non-rotating Horndeski black hole found in this article.
}
\vspace{1cm}
\label{non-rotat_BH}
\end{figure*}

In conclusion, while the numerical models for the non-rotating Horndeski black hole may not explain the LFQPOs of GRS 1915+105, they can provide an explanation for the HFQPOs. The shock cone or cavity around the non-rotating Horndeski black hole, as revealed through numerical modeling, can be proposed as a physical mechanism for explaining HFQPOs, which is discussed in more detail in Section \ref{physical_mech}. In our previous study \cite{Donmez2024arXiv240216707D}, we examined the behavior of the shock cone around the rotating Horndeski black hole. The PSD analyses in Figures 14 and 15 of Ref.\cite{Donmez2024arXiv240216707D} showed that lower frequencies were produced by the shock cone around the rotating Horndeski black hole. This observation is also seen in Fig. \ref{non-rotat_BH}, which displays the QPOs excited by the shock cones around both non-rotating and rotating Horndeski black holes. The frequency that falls within the range of the observed QPOs in Ref.\cite{Liu_2021, Dhaka_2023} is only present in the $a/M=0.9$ case in numerical simulations. Thus, the numerically calculated frequency for the rapidly rotating black hole, $f=3.38$ Hz, is within the $2.01$ Hz to $5.41$ Hz range for the LFQPOs observed in these studies, as listed for the GRS 1915+105 source.
In summary, while the rapidly rotating Horndeski black hole can explain both low and high frequencies observed from the GRS 1915+105 source, the non-rotating black hole can explain LFQPOs and HFQPOs with frequencies above approximately $6$ Hz. This does not fall within the range of low-frequency QPOs obtained from observations of the GRS 1915+105 source. In other words, we cannot explain the LFQPOs of the GRS 1915+105 source with a non-rotating black hole model. For this, as confirmed by observational results, the black hole must be rapidly rotating. On the other hand, in both rotating and non-rotating black hole models, the necessary physical mechanism for QPO formation is the shock cone. Therefore, we can say that the shock cone is a good mechanism for the trapping and excitation of QPO modes. It can be used to explain QPO frequencies observed from different sources.

After demonstrating that rapid rotation is necessary to explain the very low frequencies observed in LFQPOs around the black hole GRS 1915+105, we explore the possible range of the rotation parameter. Numerical studies, as calculated in Ref.\cite{Donmez2024arXiv240216707D} and illustrated in Fig. \ref{non-rotat_BH}, have shown that no LFQPOs formed around the non-rotating ($a/M=0$) or slowly rotating ($a/M=0.4$) hairy Horndeski black hole for the source GRS 1915+105, as listed in Ref.\cite{Dhaka_2023}. However, for both the non-rotating and the rotating black hole models, HFQPOs above $5.5$ Hz have been calculated numerically, independent of the parameters used in our numerical simulations (e.g., hair parameter and asymptotic velocity). As a result, the observed LFQPOs from different telescopes \cite{Liu_2021, Dhaka_2023, Athulya_2021, YUAN2023} are consistent with a rotation parameter of $a \ge 0.6$, as seen in Fig. \ref{non-rotat_BH}. Considering possible resolution-related errors in numerical calculations, we surmise that the lower limit of the rotation parameter could be greater than $0.6$. For the upper limit, we cannot be certain because we only have QPOs formed at $a=0.9$ for the rapidly rotating black hole. However, given that the frequency at $a=0.9$ is $3.28$ Hz and the lowest LFQPO limit according to Ref.\cite{Liu_2021, Dhaka_2023} should be $2.01$ Hz, it suggests that the black hole could rotate faster than $a=0.9$. This is because as the black hole rotation parameter increases, the observed lowest frequency decreases further. According to the PSD analyses conducted in Ref.\cite{Donmez2024arXiv240216707D} and in this study, the predicted rotation parameter value for the source GRS 1915+105 is consistent with the spin values estimated from observational data \cite{Liu_2021, Athulya_2021, YUAN2023}. Through extensive parameter analysis, Ref.\cite{Liu_2021, Athulya_2021, YUAN2023} have predicted that the spin parameter of the black hole could be between $a/M=0.8$ and $a/M=0.999$. These observational results are consistent with the spin values we have numerically predicted here. Finally, to numerically validate the range of the rotation parameter estimated from observational data, we aim to provide a more definitive prediction on the lower and upper limits of the rotation parameter. We plan to achieve this by simulating the scenario of the shock cone formation around the rotating Horndeski black hole at different rotation speeds in future studies.

%%%%%%%%%%%%%%%%%%%%%%%%%%%%%%%%%%%%%%%%%%%%%%%%%%%%%%%%%%%%%%%%%%%%%%%%%%%%%%%%%%%%%%%%%%%%%
%%%%%%%%%%%%%%%%%%%%%%%%%%%%%%%%%%%%%%%%%%%%%%%%%%%%%%%%%%%%%%%%%%%%%%%%%%%%%%%%%%%%%%%%%%%%%%%%%%%%%%%%%%%%%
\section{Physical Models Founded to Explain GRS 1915+105 Black Hole Observations.}
\label{physical_mech}

The X-ray emission of black hole binary systems is an important tool for understanding the physical mechanisms resulting from the interaction between black hole mass accretion and the surrounding environment. Observational analyses of these X-ray power spectral densities (PSDs) reveal how they change and elucidate the related QPO frequencies \cite{Klis1989ARA&A}. As mentioned earlier, the physical mechanism of the GRS 1915+105 source is related to the C-type QPO frequencies it produces. QPOs from PSD analyses form peaks with varying frequencies, often accompanied by a subharmonic and a second harmonic peak \cite{Casella_2005}. In the literature, there are predictions about the physical mechanisms causing C-type QPOs. These mechanisms are either related to instabilities in mass accretion \cite{CabanacMNRAS2010} or geometric effects like Lense-Thirring (LT) precession around the black hole \cite{Schnittman_2006, IngramMNRAS2009}. Additionally, variations in the amplitude of these QPOs have been found to depend on the disk's inclination angle relative to the observer \cite{Heilmnras2015}, and it has been shown that QPOs are modulated by the iron line and central energy \cite{Ingram2016mnras}. Beyond these models, others have been developed, including the accretion-ejection instability model in the magnetized disk \cite{Tagger1999A&A}, the two-component advection flow model \cite{Chakrabartimnras2009}, corrugation modes which are transverse standing waves in the disk \cite{Kato2001}, and the magneto-rotational instability \cite{Tagger_2006}. Despite these studies, the physical mechanisms behind the wide spectrum of frequencies (low to high) observed from the GRS 1915+105 source are not fully understood. To contribute to the literature, we propose a physical mechanism, the shock cone, that can produce QPOs, after demonstrating the consistency of the QPO frequencies obtained from numerical simulations with those observed from the GRS 1915+105 source.

Observations in the X-ray spectrum of GRS 1915+105 have shown dramatic variability \cite{Athulya_2021, Shi2023MNRAS}, leading to a wide range of behavioral changes and disk transitions \cite{Markwardt_1999, Belloni2000A&A}. This continuous change and instability have made it a focal point for researchers for many years. Observational results confirm that data obtained from the black hole source vary between hard and soft X-rays \cite{Belloni2024MNRAS, Chauhan2024MNRAS, Boked2024MNRAS}. Therefore, understanding the matter accreting around this source and the physical mechanisms formed is crucial to explain the observed QPOs and reveal the black hole's physical characteristics. Here, we present the interaction of matter falling towards the black hole via BHL accretion with the central black hole, leading to the formation of shock cones, bow shock, and cavities depending on parameters such as scalar hair and asymptotic velocity. These mechanisms can explain the disk transitions and the sudden changes in GRS 1915+105, including LFQPOs and HFQPOs. Such interactions are known to produce C-QPOs \cite{Mastichiadis2022}.

In this study, we numerically demonstrated the physical mechanisms, namely the shock cones, bow shock, and the cavities, formed around the Horndeski black hole with BHL accretion. As seen in Figs. \ref{colV02_1} and \ref{colV02_2}, for $h/M=0$ (Schwarzschild black hole), the shock cone forms on the downstream side. The matter trapped in this cone oscillates due to the pressure mode and generates QPOs, as revealed in PSD analyses. Subsequently, we created different models for various values of the hair parameter $h/M$, defining the scalar field around the Horndeski black hole, to uncover its effect on the formed shock cone mechanism and thus on the induced QPOs, as presented in Table \ref{Inital_Con}. In numerical calculations, as $h/M$ decreases, changes in the shock cone mechanism begin to occur, significantly affecting the cone structure. With the decreasing scalar field parameter, the cone’s opening angle narrows, and the stagnation point approaches the black hole horizon. As the scalar field intensity increases, the dynamic structure of the shock cone is disrupted, and at values of $h/M$ below $-0.9$, the matter inside the shock cone starts being pushed out by the scalar field. Then, a cavity forms downstream of the black hole at the location of the shock cone. Although at some $h/M$ values, the shock cones or cavities appear to have a similar physical dynamic structure, PSD analyses continuously change at each $h/M$ value, affecting both the formation of the fundamental modes and the other peaks created through linear or nonlinear couplings. With changes in the scalar field parameter, the continuous change demonstrated in Figs. \ref{PSDV02_1} and \ref{PSDV02_2} in the QPOs is consistent with the observed X-ray variations of GRS 1915+105. The physical mechanism we propose here can explain the behavior of the GRS 1915+105 source, which we have demonstrated numerically and continue to reveal through the analysis of observational data. The shock cone mechanism we numerically presented, its changing structure over time, and the formation of the cavity by pushing the matter inside the cone away from the black hole, can be predicted due to the presence and temporal change of a scalar field defined in the spacetime of the black hole. This could result from dark matter or energy, as explained in the introduction and literature \cite{Comelli_2003, BERTACCA_2007, Chung_2007, de_la_Macorra_2008, Arbuzov_2022}. Changes in the scalar field around black holes can also occur due to various reasons, including changes in the physical properties of the black hole (such as mass and spin), in the dynamics of the field or energy in the expanding universe, and due to the gravitational interaction of the black hole with its surroundings, as well as alterations of the existing scalar field due to the changes caused by the matter falling towards the black hole \cite{Jetzer_1997, Hegade2022PhysRevD, Pombo:2023lxg}. In conclusion, the shock cones, bow shock, and cavities formed around the non-rotating and rotating hairy Horndeski black holes are important mechanisms that can explain the continuous spectral variability of the GRS 1915+105 source. For explaining most of the observed LFQPOs (around $4$ Hz) with the non-rotating Horndeski black hole model, the most suitable mechanism is the bow shock. Bow shock has been numerically observed in the $h/M=-0.4$ case, and LFQPOs have been encountered in its PSD analyses. These mechanisms not only explain the observed LFQPOs and HFQPOs but also demonstrate the compatibility of the possible rotation parameters of the black hole with the literature, as seen in Section \ref{QPOs4}.

%%%%%%%%%%%%%%%%%%%%%%%%%%%%%%%%%%%%%%%%%%%%%%%%%%%%%%%%%%%%%%%%%%%%%%%%%%%%%%%%%%%%%%%%%%%%%%%%%%%%%%%%%%%%%
%%%%%%%%%%%%%%%%%%%%%%%%%%%%%%%%%%%%%%%%%%%%%%%%%%%%%%%%%%%%%%%%%%%%%%%%%%%%%%%%%%%%%%%%%%%%%
%%%%%%%%%%%%%%%%%%%%%%%%%%%%%%%%%%%%%%%%%%%%%%%%%%%%%%%%%%%%%%%%%%%%%%%%%%%%%%%%%%%%%%%%%%%%%%%%%%%%%%%%%%%%%
\section{Conclusions}
\label{conc}

In this article, we numerically model and reveal how the shock cone, bow shock, and cavity forming around the non-rotating Horndeski black hole via BHL accretion vary with the scalar hair parameter, $h/M$. We detail the changes in the physical mechanisms across a wide range of $h/M$ values, the transitions between these mechanisms, and the excitation of LFQPOs and HFQPOs within them. The QPO frequencies obtained from our numerical simulations align with the observational results of the GRS 1915+105 source, thereby proposing physical mechanisms for such dynamically varying sources \cite{Motta2023}.

When matter accretes from the upstream region towards the Horndeski black hole via the BHL mechanism, the  shock cone is observed to form on the downstream side. With a decrease in the scalar hair parameter, $h/M$, the opening angle of the shock cone narrows, and the stagnation point within the cone moves closer to the black hole horizon. These changes result in noticeable alterations to the physical structure of the shock cone. As the hair parameter decreases, there is a reduction in the amount of matter accreting towards the black hole. This reduction is attributed to the barrier formed by the scalar field's potential, which slows the matter flowing in BHL accretion and causes it to scatter towards points far from the black hole. Furthermore, the movement of the stagnation point closer to the black hole pushes the matter within the cone outwards. At the same time, the scalar field propels the surrounding matter away from the black hole. Consequently, the intensity of the shock cone gradually decreases. Additionally, an increase in the scalar potential repels the accreting matter, leading to the formation of a bow shock. This bow shock, driven by the increasing scalar potential, moves towards the incoming matter and disappears as $h/M$ continues to decrease. Lastly, the decreasing $h/M$, coupled with the increasing scalar potential, leads to the formation of a cavity on the side where the shock cone is created. As this cavity encircles the black hole and expands with the decreasing $h/M$, it fosters the development of a distinct physical mechanism around the black hole.

At the same time, to measure the effect of the asymptotic velocity of matter sent from the outer boundary on the physical mechanisms, and consequently on the QPOs, we considered cases with $V_{\infty}/c=0.2$ and $V_{\infty}/c=0.4$. As expected, in the $V_{\infty}/c=0.4$ case, since the matter falls towards the black hole at a much faster rate, the formation time of the shock cone and the time to reach the steady-state are shorter compared to the $V_{\infty}/c=0.2$ case. Additionally, at $V_{\infty}/c=0.4$, the opening angle of the cone is smaller, and the stagnation point forms closer to the horizon. Besides these, similar physical mechanisms are observed. However, the significant differences between these two asymptotic velocities manifest in the formation and excitation of QPO frequencies. In the $V_{\infty}/c=0.4$ case, QPOs are observed only for $h/M=0$, that is, in the case of a Schwarzschild black hole. It is understood that the cone does not show instability for the other two values of $h/M$ in the case of $V_{\infty}/c=0.4$, and therefore, QPOs do not form.

In the PSD analyses, it has been observed that the physical mechanism changes with different values of $h/M$, while LFQPOs and HFQPOs have been detected across a broad frequency range \cite{Banerjee2021ApJ}. The fundamental QPO modes observed in numerical simulations for various models are presented in Table \ref{geniue_mode}. The increase in $\lvert h/M \rvert$ significantly impacts the fundamental modes and other peaks. As $\lvert h/M \rvert$ increases, both the fundamental frequencies and their couplings shift from low to high frequencies. In the context of observational results for a non-rotating black hole, using a Horndeski black hole model suggests that low-frequency QPOs should correspond to $h/M$ values close to zero (equivalent to a Schwarzschild black hole) or near zero. For high-frequency QPO oscillations, larger negative values of $h/M$ are preferable.

HFQPOs arise from rapid variations in brightness as matter descends into a black hole. These oscillations exhibit frequencies ranging from the low twenties to several hundred Hertz, often maintaining stable frequency ratios like $3:2$ \cite{Pasham_2015, Varniere_2018}. Such ratios, including $3:2$, $5:3$, and $2:1$, represent resonance conditions observed in different sources. In our case, when we examine the QPO frequencies calculated numerically, it is observed that these ratios occur in almost every model. According to the numerical results, the observation of such ratios is possible not only for HFQPOs but also for LFQPOs.

GRS 1915+105 is a binary system that emits $X$-rays with varying intensity, attracting considerable attention due to its constant activity and variability. In this context, the numerically obtained QPOs are compared with the observed QPO frequencies for the source GRS 1915+105. In the PSD analyses, we have generated low- and high-frequency QPOs. Considering the mass of the GRS 1915+105 source, the lowest frequency obtained in our numerical calculations is approximately $6.45$ Hz, which coincides with the upper limit of some observed LFQPO values \cite{Liu_2021, Dhaka_2023}. However, the HFQPOs found through our numerical calculations can explain the observational results for this source. Additionally, in this paper, we use the results of QPOs around a rotating Horndeski black hole from our previous work to compare rotating and non-rotating black hole models and the QPOs they produce. Particularly, to explain the QPOs observed in Ref.\cite{Liu_2021, Dhaka_2023}, which occur between $2.01$ Hz and $5.41$ Hz, it has been concluded that the black hole must be rotating. The calculations have shown that the spin parameter $a/M$ must be at least $0.6$. In this case, the physical mechanisms identified here can explain all LFQPOs around $3$ Hz.

On the other hand, for the GRS 1915+105 or the microquasars which produces similar observed frequencies, the physical mechanisms found here, particularly the shock cone, can be suitable for capturing and exciting QPO modes. As a result of our analyses in this study, the black hole must be spinning to explain the LFQPO and HFQPO observed in the GRS 1915+105. In this case, as seen in Figure 1 of Ref.\cite{Donmez2024arXiv240216707D}, there is a lower limit for $h/M$ depending on the spin parameter. Furthermore, both the numerical simulations in Figure 1 and the numerical calculations made in this article for a non-spinning black hole confirm that there must be a lower limit for $h/M$. As seen in Table \ref{geniue_mode} and Figures \ref{colV02_1} and \ref{PSDV02_2}, we can explain the QPO frequencies observed in GRS 1915+105 or most other microquasars with the QPOs formed in cases where the hair parameter is $h/M > -0.5$. Indeed, as seen in Fig.\ref{non-rotat_BH} and discussed in Section \ref{QPOs4}, the values $h/M > -0.5$ are appropriate hair parameters for the QPOs observed in the GRS 1915+105 source.

Finally, since the LFQPOs and HFQPOs observed for GRS 1915+105 are consistent with our numerically calculated QPOs, we propose these numerical QPOs' generating mechanisms as the physical mechanisms for the GRS 1915+105 source. This suggests that the existence of a shock cone, bow shock, and cavity around the GRS 1915+105 source warrants discussion. GRS 1915+105 is known for its high activity and continuously changing QPOs, implying that the scalar field may adapt over time. The changes in the black hole’s physical properties (such as mass and spin), the dynamics of the field or energy in the expanding universe, the gravitational interaction of the black hole with its environment, and the alteration of the existing scalar field due to accreting matter, could influence the scalar field \cite{Jetzer_1997, Hegade2022PhysRevD, Pombo:2023lxg}. Therefore, the ongoing changes in the GRS 1915+105 source can be elucidated by our model. Our numerical PSD analyses demonstrate that the behavior of matter around the black hole and the resulting QPOs vary with each $h/M$ value. Thus, we conclude that the observed changes may be attributed to the continuous transformation of the scalar field around the GRS 1915+105 source.

%%%%%%%%%%%%%%%%%%%%%%%%%%%%%%%%%%%%%%%%%%%%%%%%%%%%%%%%%%%%%%%%%%%%%%%%%%%%%%%%%%%%%%%%%%%%%
\section*{Acknowledgments}
All simulations were performed using the Phoenix  High Performance Computing facility at the American University of the Middle East (AUM), Kuwait. I would like to thank the reviewer very much for their contributions to enhancing the readability and enriching the content of the paper.\\

%%%%%%%%%%%%%%%%%%%%%%%%%%%%%%%%%%%%%%%%%%%%%%%%%%
\section*{Data Availability}

The results obtained in the article were obtained by numerically solving the general
relativistic hydrodynamic equations on a High-Performance Computing (HPC) system.
The total disk space covered by each model is approximately 5GB, and these are raw
data. The results obtained from each model are processed with different approaches
to generate new data. Together, they total 7GB. In this article, the total disk
space covered by 13 different models is around 91GB. These data can be provided
to you whenever needed.


\begin{thebibliography}{10}

\bibitem{Pasham2018ALQ}
  Pasham D.~R.,  et~al.
\newblock  {\em   Science}, 363, 531, 2018.

\bibitem{Smith_2021}
  Smith K.~L.,  Tandon C.~R.,   Wagoner R.~V.
\newblock  {\em    The Astrophysical
  Journal}, 906, 92, 2021.

\bibitem{Singh2022MNRAS}
  {Singh} C.~B.,  {Mondal} S.,   {Garofalo} D.
  \newblock  {\em   MNRAS} {510, 807}, 2022.

\bibitem{charles2003optical}
  Charles P.~A.,  Coe M.~J.
  \newblock  {\em  arXiv astro-ph/0308020}, 2003.

  \bibitem{Fragile_2016}
  Fragile P.~C.,  Straub O.,   Blaes O.
  Monthly Notices of the
\newblock  {\em     Royal Astronomical Society}, 461, 1356–1362, 2016.

 \bibitem{Walter_2015}
  Walter R.,  Lutovinov A.~A.,  Bozzo E.,   Tsygankov S.~S.
\newblock  {\em     The
  Astronomy and Astrophysics Review}, 23, 2015.

\bibitem{Ackermann2015ApJ}
  {Ackermann} M.,  et~al., 2015.
\newblock  {\em  ApJ} {813, L41}, 2015.

\bibitem{Zhang_2021}
  Zhang P.,  Wang Z.
  \newblock  {\em    The Astrophysical Journal}, 914, 1, 2021.

\bibitem{Homan2001ApJS}
{Homan} J.,  {Wijnands} R.,  {van der Klis} M.,  {Belloni} T.,  {van Paradijs}
J.,  {Klein-Wolt} M.,  {Fender} R.,   {M{\'e}ndez} M.
\newblock  {\em ApJS} {132, 377}, 2001.

 \bibitem{belloni2011black}
Belloni T.~M.,  Motta S.~E.,   Muñoz-Darias T.
\newblock  {\em arXiv} {1109.3388}, 2011.

\bibitem{Belloni2024MNRAS}
  {Belloni} T.~M.,  {M{\'e}ndez} M.,  {Garc{\'\i}a} F.,   {Bhattacharya} D.
\newblock  {\em Monthly Notices of the
  Royal Astronomical Society} {527, 7136}, 2024.

\bibitem{Motta_2012}
{Motta} S.,  {Homan} J.,  {Mu{\~n}oz Darias} T.,  {Casella} P.,  {Belloni}
T.~M.,  {Hiemstra} B.,   {M{\'e}ndez} M.
\newblock  {\em MNRAS} {427, 595}, 2012.

\bibitem{Motta2016AN}
  {Motta} S.~E.
   \newblock  {\em    Astronomische Nachrichten} {337, 398}, 2016.

\bibitem{Liu_2021}
Liu H.,  Ji L.,  Bambi C.,  Jain P.,  Misra R.,  Rawat D.,  Yadav J.~S.,
Zhang Y.
 \newblock  {\em  The Astrophysical Journal}, 909, 63, 2021.

\bibitem{Belloni2013MNRAS}
  {Belloni} T.~M.,  {Altamirano} D..
  \newblock  {\em    MNRAS} {432, 19}, 2013
  
\bibitem{Misra2020ApJ}
  {Misra} R.,  {Rawat} D.,  {Yadav} J.~S.,   {Jain} P.
   \newblock  {\em    ApJ} {889, L36}, 2020.

\bibitem{wang20242022}
Wang J.,  et~al., 2024, The 2022 Outburst of IGR J17091-3624: Connecting the
exotic GRS 1915+105 to standard black hole X-ray binaries
\newblock  {\em  arXiv:2401.10192}, 2022.

\bibitem{Chauhan2024MNRAS}
{Chauhan} J.,  {Bharali} P.,  {Lohfink} A.,  {Abdulghani} Y.,   {Davidson} E.
\newblock  {\em    MNRAS} {527, 11801}, 2024.

\bibitem{Pasham_2015}
Pasham D.~R.,  Cenko S.~B.,  Zoghbi A.,  Mushotzky R.~F.,  Miller J.,   Tombesi
F.
\newblock  {\em The Astrophysical Journal},
  811, L11, 2015.

\bibitem{Varniere_2018}
  {Varniere} P.,  {Rodriguez} J.
  \newblock  {\em    ApJ} {865, 113}, 2018.

\bibitem{DonmezMPLA24}
  {Donmez} O.
  \newblock  {\em Modern Physics Letters A} {39(16), 2450076-665}, 2024

\bibitem{DonmezEPJC24}
    {Donmez} O.
    \newblock  {\em   European Physical Journal C} {84(5), 524}, 2024

    

\bibitem{Strohmayer2001ApJ}
  {Strohmayer} T.~E.
  \newblock  {\em    ApJ} {554, L169}, 2001.

\bibitem{Remillard2006ARA&A}
  {Remillard} R.~A.,  {McClintock} J.~E.
  \newblock  {\em   ARA\&A} {44, 49}, 2006.

 \bibitem{BelloniMNRAS2012}
  Belloni T.~M.,  Sanna A.,   Méndez M.
\newblock  {\em Monthly Notices of the
  Royal Astronomical Society}, 426, 1701, 2012

\bibitem{Majumder_2022}
  Majumder S.,  Sreehari H.,  Aftab N.,  Katoch T.,  Das S.,   Nandi A.
  \newblock  {\em   Monthly Notices of the Royal Astronomical Society}, 512, 2508–2524, 2022.

\bibitem{Sreehari_2020}
Sreehari H.,  Nandi A.,  Das S.,  Agrawal V.~K.,  Mandal S.,  Ramadevi M.~C.,
Katoch T. M.
\newblock  {\em   onthly Notices of the Royal Astronomical Society}, 499, 5891–5901, 2020.
\bibitem{Ingram2019}
  {Ingram} A.~R.,  {Motta} S.~E.
  \newblock  {\em   NewAR} {85, 101524}, 2019.

\bibitem{Bondi1}
  {Bondi} H.,  {Hoyle} F.
\newblock  {\em Monthly Notices of the
  Royal Astronomical Society} {104, 273}, 1944.

\bibitem{Bondi1952MNRAS}
  {Bondi} H.
\newblock  {\em Monthly Notices of the
  Royal Astronomical Society}  {528, 7016}, 2024.
  {112, 195}, 1952.

\bibitem{Edgar1}
  {Edgar} R.
\newblock  {\em   NewAR} {48, 843}, 2004.

\bibitem{Donmez6}
  {D{\"o}nmez} O.,  {Zanotti} O.,   {Rezzolla} L.
  \newblock  {\em  MNRAS} {412, 1659}, 2011.
  
\bibitem{Zanotti1}
  {Zanotti} O.,  {Roedig} C.,  {Rezzolla} L.,   {Del Zanna} L.
  \newblock  {\em  MNRAS} {417, 2899}, 2011.

\bibitem{Penner2}
  Penner A.~J.
\newblock  {\em    Monthly Notices of the Royal Astronomical
  Society}, 428, 2171, 2012.

\bibitem{Donmez5}
  {D{\"o}nmez} O.
  \newblock  {\em  MNRAS} {426, 1533}, 2012.

 \bibitem{Koyuncu1}
  {Koyuncu} F.,  {D{\"o}nmez} O.
  \newblock  {\em    Modern Physics Letters A} {29, 1450115}, 2014.

 \bibitem{DonmezUniverse2022}
  {Donmez} O.,  {Dogan} F.,   {Sahin} T.
  \newblock  {\em   Universe} {8(9), 458}, 2022.

 \bibitem{CruzOsorio2023JCAP}
{Cruz-Osorio} A.,  {Rezzolla} L.,  {Lora-Clavijo} F.~D.,  {Font} J.~A.,
{Herdeiro} C.,   {Radu} E.
\newblock  {\em   JCAP} {2023, 057}, 2023.

\bibitem{Donmez2024arXiv240216707D}
{Donmez} O.
\newblock  {\em arXiv240216707D} {p. arXiv:2402.16707}, 2024.

\bibitem{Ruffert2}
  {Ruffert} M.,  {Arnett} D.
 \newblock  {\em  ApJ} {427, 351}, 1994.

\bibitem{Foglizzo1}
  {Foglizzo} T.,  {Galletti} P.,   {Ruffert} M.
  \newblock  {\em    A\&A} {435, 397}, 2005.

\bibitem{Ohsugi1}
  Ohsugi Y.
\newblock  {\em   Astronomy and Computing}, 25, 44, 2018.

\bibitem{Wenrui1}
  {Xu} W.,  {Stone} J.~M.
 \newblock  {\em  MNRAS} {488, 5162}, 2019.

\bibitem{Donmez20}
  Donmez O.,  Dogan F.
  \newblock  {\em  Universe}, 10, 2024.

\bibitem{Donmez_2024ES}
  Donmez O.,  Dogan F.
  \newblock  {\em arXiv:2407.01478}, 2024

\bibitem{Thomson2014PhDT}
  {Thomson} B.~W. PhD thesis, University of North Dakota, 2014.

\bibitem{Comelli_2003}
  Comelli D.,  Pietroni M.,   Riotto A.
\newblock  {\em    Physics Letters B}, 571, 115–120, 2003.   
 
\bibitem{BERTACCA_2007}
  BERTACCA D.,  MATARRESE S.,   PIETRONI M.
  \newblock  {\em Modern Physics  Letters A}, 22, 2893–2907, 2007

 \bibitem{Chung_2007}
  Chung D. J.~H.,  Everett L.~L.,   Matchev K.~T.
\newblock  {\em   Physical
  Review D}, 76, 2007. 
 
\bibitem{de_la_Macorra_2008}
  de~la Macorra A.
  \newblock  {\em     Journal of Cosmology and Astroparticle
    Physics}, 2008, 030, 2008.

\bibitem{Errehymy2024EPJC}
 Errehymy A., Banerjee A., Hansraj S., Donmez O., Nisar K. S., Abdel-Aty A.-H., 
 \newblock  {\em European Physical Journal C}, 84, 573, 2024
  
\bibitem{Maurya2024ptep}
Maurya S. K., Errehymy A., Mustafa G., Donmez O.,  Nisar K. S., Abdel-Aty A.-H.
\newblock  {\em Progress of Theoretical and Experimental Physics}, 2024, 043E02, 2024

   
\bibitem{Arbuzov_2022}
  Arbuzov A.,  Latosh B.,   Nikitenko A.
  \newblock  {\em  Classical and Quantum   Gravity}, 39, 055003, 2022.

 \bibitem{martens2020dark}
Martens N. C.~M.,  Lehmkuhl D. Dark Matter = Modified Gravity?
  Scrutinising the spacetime-matter distinction through the modified gravity/
  dark matter lens
    \newblock  {\em    arXiv:2009.03890}, 2020.

\bibitem{Hu_2000}
  Hu W.,  Barkana R.,   Gruzinov A.
 \newblock  {\em   Physical Review Letters}, 85, 1158–1161, 2000.
    
  \bibitem{Hui_2017}
  Hui L.,  Ostriker J.~P.,  Tremaine S.,   Witten E.
  Physical
  \newblock  {\em  Review D}, 95, 2017.

 \bibitem{Wang_2023}
  Wang D.,  Koussour M.,  Malik A.,  Myrzakulov N.,   Mustafa G.
\newblock  {\em     The European Physical Journal C}, 83, 2023. 

\bibitem{Font2000LRR}
  {Font} J.~A.
  \newblock  {\em     Living Reviews in Relativity} {3, 2}, 2000.

 \bibitem{Donmez1}
  {D{\"o}nmez} O.
\newblock  {\em   Ap\&SS} {293, 323}, 2004.

\bibitem{Donmez2}
  {Donmez} O.
  \newblock  {\em    AM\&C},  {181, 256}, 2006.

 \bibitem{Donmez_EGB_Rot}
  {Donmez} O.
\newblock  {\em   Physics Letters B} {827, 136997}, 2022.

\bibitem{Donmez2023arXiv231013847D}
  {Donmez} O.
\newblock  {\em    Research in Astronomy and Astrophysics} {24(8), 085001}, 2024.

\bibitem{Horndeski1974IJTP}
  {Horndeski} G.~W.
\newblock  {\em    International Journal of Theoretical
  Physics} {10, 363}, 1974.

 \bibitem{Heydari-Fard2023arXiv}
  {Heydari-Fard} M.,  {Heydari-Fard} M.,   {Riazi} N.
  \newblock  {\em   arXiv231112393} {p. arXiv:2311.12393}, 2023.

  \bibitem{Badia2017EPJC}
  {Bad{\'\i}a} J.,  {Eiroa} E.~F.
  \newblock  {\em European Physical Journal C}, {77, 779}, 2017.

 \bibitem{Esteban2021arXiv}
  {Esteban Perez Bergliaffa} S.,  {Maier} R.,   {de Oliveira Silvano} N.
  \newblock  {\em   arXiv210707839E} {p. arXiv:2107.07839}, 2021.

 \bibitem{Walia2022EPJC}
  {Walia} R.~K.,  {Maharaj} S.~D.,   {Ghosh} S.~G.
\newblock  {\em      European
  Physical Journal C} {82, 547}, 2022.

\bibitem{Kerr1963PhRvL}
  {Kerr} R.~P.
 \newblock  {\em   PhRvL} {11, 237}, 1963.

\bibitem{Rayimbaev2023EPJC}
  {Rayimbaev} J.,  {Dialektopoulos} K.~F.,  {Sarikulov} F.,   {Abdujabbarov} A.
  \newblock  {\em   European Physical Journal C} {83, 572}, 2023.

 \bibitem{Misner1977}
  {Misner} C.~W.,  {Thorne} K.~S.,   {Wheeler} J.~A.
  \newblock  {\em      Gravitation}, Volume 1, 1977.

 \bibitem{Tzikas2021PhLB}
  {Tzikas} A.~G.
  \newblock  {\em     PhLB} {819, 136426}, 2021.

  \bibitem{Yang2019CoTPh} {Rayimbaev}, {Tadjimuratov}, {Abdujabbarov},
  {Ahmedov}  \& {Khudoyberdieva},
  \newblock  {\em   CoTPh}, 2019.

 \bibitem{Yalinewich2018MNRAS}
{Yalinewich} A.,  {Sari} R.,  {Generozov} A.,  {Stone} N.~C.,   {Metzger}
B.~D.
\newblock  {\em  MNRAS} {479, 4778}, 2018.

\bibitem{VarniERe:2015kda}
Varni\`ERe P.,  Tagger M.,   Rodriguez J. in {13th Marcel Grossmann
  Meeting on Recent Developments in Theoretical and Experimental General
  Relativity, Astrophysics, and Relativistic Field Theories}. pp 2398--2400,
\newblock  {\em     10.1142/9789814623995\_0450}, 2015.

\bibitem{Landau1976}
  {Landau} L.~D.,  {Lifshitz} E.~M.~a.
 \newblock  {\em   Mechanics}, Vol.1, 1976.

\bibitem{Zanotti2005MNRAS}
  {Zanotti} O.,  {Font} J.~A.,  {Rezzolla} L.,   {Montero} P.~J.
 \newblock  {\em MNRAS} {356, 1371}, 2005.

 \bibitem{Rayimbaev2021}
{Rayimbaev} J.,  {Tadjimuratov} P.,  {Abdujabbarov} A.,  {Ahmedov} B.,
{Khudoyberdieva} M.
\newblock  {\em  Galaxies} {9, 75}, 2021.

 \bibitem{Dhaka_2023}
  Dhaka R.,  Misra R.,  Yadav J.~S.,   Jain P.
\newblock  {\em  Monthly Notices
  of the Royal Astronomical Society}, 524, 2721–2732, 2023.

\bibitem{Reid2014}
{Reid} M.~J.,  {McClintock} J.~E.,  {Steiner} J.~F.,  {Steeghs} D.,
{Remillard} R.~A.,  {Dhawan} V.,   {Narayan} R.
\newblock  {\em ApJ} {796, 2}, 2014.

\bibitem{Klein-WoltMNRAS2002}
Klein-Wolt M.,  Fender R.~P.,  Pooley G.~G.,  Belloni T.,  Migliari S.,  Morgan
E.~H.,   van~der Klis M.
 \newblock  {\em  Monthly Notices of the Royal
   Astronomical Society}, 331, 745, 2002.

\bibitem{Belloni_2002}
Belloni T.,  Psaltis D.,   van~der Klis M.
\newblock  {\em   The Astrophysical Journal }, 572, 392–406, 2002. 

 \bibitem{Hannikainen2005A&A}
  {Hannikainen} D.~C.,  et~al.
\newblock  {\em     A\&A} {435, 995}, 2005.

\bibitem{Belloni_2009}
T.~Belloni, \emph{States and transitions in black hole binaries},  in
  \emph{Lecture Notes in Physics}, p.~53–84, Springer Berlin Heidelberg
  (2009), \href{https://doi.org/10.1007/978-3-540-76937-8_3}{DOI}.
  
\bibitem{Motta2023}
  {Motta} S.~E.,  {Belloni} T.~M.
  \newblock  {\em   arXiv230700867} {p. arXiv:2307.00867}, 2023.
  
\bibitem{Athulya_2021}
Athulya M.~P.,  Radhika D.,  Agrawal V.~K.,  Ravishankar B.~T.,  Naik S.,
Mandal S.,   Nandi A.
\newblock  {\em Monthly Notices of the Royal Astronomical Society}, 510, 3019–3038, 2021.

\bibitem{YUAN2023}
Yuan W.,  Li-jun G.,  Xue-shan Z.,  Ye F.,  Nan J.,  Zhen-xuan L.,   Yu-feng
L.
\newblock  {\em Chinese Astronomy and Astrophysics}, 47, 625, 2023.
  
 
\bibitem{Klis1989ARA&A}
  {van der Klis} M.
  \newblock  {\em    ARA\&A} {27, 517}, 1989.

 \bibitem{Casella_2005}
  Casella P.,  Belloni T.,   Stella L.
\newblock  {The Astrophysical
   Journal}, 629, 403, 2005. 

\bibitem{CabanacMNRAS2010}
Cabanac C.,  Henri G.,  Petrucci P.-O.,  Malzac J.,  Ferreira J.,   Belloni
T.~M.
\newblock  {\em Monthly Notices of the
  Royal Astronomical Society} , 404, 738, 2010.

\bibitem{Schnittman_2006}
  Schnittman J.~D.,  Homan J.,   Miller J.~M.
\newblock  {\em  The Astrophysical
  Journal}, 642, 420, 2006.

\bibitem{IngramMNRAS2009}
  Ingram A.,  Done C.,   Fragile P.~C.
\newblock  {\em    Monthly Notices of the
  Royal Astronomical Society: Letters}, 397,
L101, 2009.

\bibitem{Heilmnras2015}
  Heil L.~M.,  Uttley P.,   Klein-Wolt M.
 \newblock  {\em  Monthly Notices of the
   Royal Astronomical Society}, 448, 3348, 2015.

 \bibitem{Ingram2016mnras}
Ingram A.,  van~der Klis M.,  Middleton M.,  Done C.,  Altamirano D.,  Heil L.,
Uttley P.,   Axelsson M.
\newblock  {\em  Monthly Notices of the Royal
  Astronomical Society}, 461, 1967, 2016.

\bibitem{Tagger1999A&A}
  {Tagger} M.,  {Pellat} R.
  \newblock  {\em    A\&A} {349, 1003}, 1999.

 \bibitem{Chakrabartimnras2009}
  Chakrabarti S.~K.,  Dutta B.~G.,   Pal P.~S.
\newblock  {\em  Monthly Notices
  of the Royal Astronomical Society}, 394,
1463, 2009.

\bibitem{Kato2001}
  Kato S.
  \newblock  {\em   Publications of the Astronomical Society of Japan}, 53, 1, 2001

\bibitem{Tagger_2006}
  Tagger M.,  Varniere P.
  \newblock  {\em    The Astrophysical Journal}, 652, 1457–1465, 2006.

\bibitem{Shi2023MNRAS}
  {Shi} Z.,  {Wu} Q.,  {Yan} Z.,  {Lyu} B.,   {Liu} H.
\newblock  {\em    MNRAS} {525, 1431}, 2023.
   
\bibitem{Markwardt_1999}
  Markwardt C.~B.,  Swank J.~H.,   Taam R.~E.
   \newblock  {\em   The Astrophysical
  Journal}, 513, L37, 1999.

\bibitem{Belloni2000A&A}
{Belloni} T.,  {Klein-Wolt} M.,  {M{\'e}ndez} M.,  {van der Klis} M.,   {van
  Paradijs} J.
\newblock  {\em  A\&A...355..271B} {355, 271}, 2000.

\bibitem{Boked2024MNRAS}
{Boked} S.,  {Maqbool} B.,  {V} J.,  {Misra} R.,  {Bhat} N.~I.,   {Bhulla} Y.
\newblock  {\em Monthly Notices of the
  Royal Astronomical Society}  {528, 7016}, 2024.

\bibitem{Mastichiadis2022}
  {Mastichiadis} A.,  {Petropoulou} M.,   {Kylafis} N.~D.
  \newblock  {\em    A\&A} {662, A118}, 2022.
  
\bibitem{Jetzer_1997}
  Jetzer P.,  Scialom D.
 \newblock  {\em   Physical Review D}, 55, 7440–7450, 1997.

  \bibitem{Hegade2022PhysRevD}
  Hegade K.~R. A.,  Most E.~R.,  Noronha J.,  Witek H.,   Yunes N.
\newblock  {\em   Phys. Rev. D}, 105, 064041, 2022.

\bibitem{Pombo:2023lxg}
  Pombo A.~M.,  Doneva D.~D.
\newblock  {\em     Phys. Rev. D}, 108, 124068, 2023.
  
\bibitem{Banerjee2021ApJ}
{Banerjee} A.,  {Bhattacharjee} A.,  {Chatterjee} D.,  {Debnath} D.,
{Chakrabarti} S.~K.,  {Katoch} T.,   {Antia} H.~M.
  \newblock  {\em ApJ} {916, 68}, 2021.




\end{thebibliography}
\end{document}